\documentclass[%Wo
 aip,
 amsmath,amssymb,
 reprint,%
]{revtex4-1}

\usepackage{graphicx}% Include figure files
\usepackage{dcolumn}% Align table columns on decimal point
\usepackage{bm}% bold math
\usepackage[utf8]{inputenc}
\usepackage[T1]{fontenc} % Use modern font encodings
\usepackage{mathptmx}
\usepackage{etoolbox}

\usepackage[version=3]{mhchem} % Formula subscripts using \ce{}    
\usepackage{comment}

\usepackage[separate-uncertainty = true,multi-part-units=single]{siunitx}
\DeclareSIUnit{\wtpercent}{wt.\%}
\DeclareSIUnit{\Torr}{Torr}
\usepackage{mhchem}
\usepackage{booktabs}
\usepackage{svg}
\usepackage{natbib}
\usepackage{physics}

\usepackage{hyperref}

\usepackage{cleveref}
\crefname{figure}{Fig.\!}{Figs.\!}
\Crefname{figure}{Figure}{Figures}
\crefname{table}{Table~}{Tables~}

\usepackage{braket} % to add | > ket

\makeatletter
\def\@email#1#2{%
 \endgroup
 \patchcmd{\titleblock@produce}
  {\frontmatter@RRAPformat}
  {\frontmatter@RRAPformat{\produce@RRAP{*#1\href{mailto:#2}{#2}}}\frontmatter@RRAPformat}
  {}{}
}%
\makeatother
\begin{document}

\title[TaO2]{Synthesis of epitaxial \ce{TaO2} thin films on \ce{Al2O3} by suboxide molecular-beam epitaxy and thermal laser epitaxy}

%% Candidates for the title

\begin{comment}
Stabilizing a 5d1 Oxide: Challenges and Pathways in the Epitaxy of Metastable TaO2
Single-domain rutile TaO2 on $r$-plane Sapphire: Direct epitaxy without buffer layers

%% Acknowledging the theory part more
“Competing Phases in TaO₂: Theoretical Landscape and Epitaxial Realization of a Metastable Oxide”
“Epitaxy of Metastable Rutile TaO₂ via suboxide molecular-beam epitaxy and thermal laser epitaxy: A Candidate for a Metal-Insulator Transition”

“Metastable TaO₂ via Dual-Route Epitaxy: Insights into Phase Stability and the Potential for a Metal-Insulator Transition”

“Accessing Rutile TaO₂ via Epitaxy: Thermodynamic and Kinetic Growth of a Non-Ground-State Phase with Metal-Insulator Potential”

“Epitaxial Growth of Rutile TaO₂ via Thermodynamic and Kinetic Routes: Stabilization of a Non-Ground-State Phase with Metal-Insulator Transition Potential”

“Epitaxial Stabilization of Rutile TaO₂: Thermodynamic and Kinetic Growth Pathways and Proximity to a Metal-Insulator Transition”

\end{comment}

\author{Yorick A. Birkhölzer}
\homepage{contributed equally}
\affiliation{Department of Materials Science and Engineering, Cornell University, Ithaca, New York 14853, USA}
\email{y.birkholzer@cornell.edu}

\author{Anna S. Park}
\homepage{contributed equally}
\affiliation{Department of Materials Science and Engineering, Cornell University, Ithaca, New York 14853, USA}
\affiliation{Platform for the Accelerated Realization, Analysis, and Discovery of Interface Materials (PARADIM), Cornell University, Ithaca, New York 14853, USA}
\author{Noah Schnitzer}
\affiliation{Department of Materials Science and Engineering, Cornell University, Ithaca, New York 14853, USA}
\affiliation{Kavli Institute at Cornell for Nanoscale Science, Ithaca, New York 14853, USA}

\author{Jeffrey Z. Kaaret}
\affiliation{School of Applied and Engineering Physics, Cornell University, Ithaca, New York 14853, USA}
\author{Benjamin Z. Gregory}
\affiliation{Department of Materials Science and Engineering, Cornell University, Ithaca, New York 14853, USA}

\author{Tomas A. Kraay}
\affiliation{Department of Materials Science and Engineering, Cornell University, Ithaca, New York 14853, USA}

\author{Tobias Schwaigert}
\affiliation{Department of Materials Science and Engineering, Cornell University, Ithaca, New York 14853, USA}
\affiliation{Platform for the Accelerated Realization, Analysis, and Discovery of Interface Materials (PARADIM), Cornell University, Ithaca, New York 14853, USA}

\author{Matthew R. Barone}
\affiliation{Department of Materials Science and Engineering, Cornell University, Ithaca, New York 14853, USA}
\affiliation{Platform for the Accelerated Realization, Analysis, and Discovery of Interface Materials (PARADIM), Cornell University, Ithaca, New York 14853, USA}

\author{Brendan D. Faeth}
\affiliation{Platform for the Accelerated Realization, Analysis, and Discovery of Interface Materials (PARADIM), Cornell University, Ithaca, New York 14853, USA}

%%%%%%%
\author{Felix V.E. Hensling}
\affiliation{Max Planck Institute for Solid State Research, 70569 Stuttgart, Germany}

\author{Iris C.G. van den Bosch}
\affiliation{MESA+ Institute for Nanotechnology, University of Twente, 7500 AE, Enschede, The Netherlands}
\author{Ellen M. Kiens}
\affiliation{MESA+ Institute for Nanotechnology, University of Twente, 7500 AE, Enschede, The Netherlands}
%\author{Jelle R.H. Ruiters}
%\affiliation{MESA+ Institute for Nanotechnology, University of Twente, 7500 AE, Enschede, The Netherlands}
\author{Christoph Baeumer}
\affiliation{MESA+ Institute for Nanotechnology, University of Twente, 7500 AE, Enschede, The Netherlands}

%

%%%%%
\author{Enrico Bergamasco}
\affiliation{Institute of Physics II, University of Cologne, 50937 Cologne, Germany}

\author{Markus Grüninger}
\affiliation{Institute of Physics II, University of Cologne, 50937 Cologne, Germany}

%%%%%%%

\author{Alexander Bordovalos}
\affiliation{Department of Physics and Astronomy, University of Toledo, Toledo, Ohio 43606, USA}
\affiliation{Wright Center for Photovoltaics Innovation and Commercialization, University of Toledo, Toledo, Ohio 43606, USA}

\author{Suresh Chaulagain}
\affiliation{Department of Physics and Astronomy, University of Toledo, Toledo, Ohio 43606, USA}
\affiliation{Wright Center for Photovoltaics Innovation and Commercialization, University of Toledo, Toledo, Ohio 43606, USA}

\author{Nikolas J. Podraza}
\affiliation{Department of Physics and Astronomy, University of Toledo, Toledo, Ohio 43606, USA}
\affiliation{Wright Center for Photovoltaics Innovation and Commercialization, University of Toledo, Toledo, Ohio 43606, USA}

%%%%%
\author{Waldemar Tokarz}
\affiliation{Faculty of Physics and Applied Computer Science, AGH University of Krakow, 30-059, Krakow, Poland}

\author{Wojciech Tabis}
\affiliation{Faculty of Physics and Applied Computer Science, AGH University of Krakow, 30-059, Krakow, Poland}
%%%%%%

\author{Matthew J. Wahila}
\affiliation{Analytical and Diagnostics Lab, Binghamton University, State University of New York,  Binghamton, New York 13902, USA}
%%%%
\author{Suchismita Sarker}
\affiliation{Cornell High Energy Synchrotron Source, Wilson Laboratory, Cornell University, Ithaca, New York 14853, USA}

\author{Christopher J. Pollock}
\affiliation{Cornell High Energy Synchrotron Source, Wilson Laboratory, Cornell University, Ithaca, New York 14853, USA}
%%%%

\author{Shun-Li Shang}
\affiliation{Department of Materials Science and Engineering, The Pennsylvania State University, University Park, Pennsylvania 16802, USA}

\author{Zi-Kui Liu}
\affiliation{Department of Materials Science and Engineering, The Pennsylvania State University, University Park, Pennsylvania 16802, USA}

\author{Nongnuch Artrith}
\affiliation{Debye Institute for Nanomaterials Science, Utrecht University, 3584 CG, Utrecht, The Netherlands}

\author{Frank M.F. de Groot}
\affiliation{Debye Institute for Nanomaterials Science, Utrecht University, 3584 CG, Utrecht, The Netherlands}

\author{Nicole A. Benedek}
\affiliation{Department of Materials Science and Engineering, Cornell University, Ithaca, New York 14853, USA}

\author{Andrej Singer}
\affiliation{Department of Materials Science and Engineering, Cornell University, Ithaca, New York 14853, USA}

\author{David A. Muller}
\affiliation{Platform for the Accelerated Realization, Analysis, and Discovery of Interface Materials (PARADIM), Cornell University, Ithaca, New York 14853, USA}
\affiliation{Kavli Institute at Cornell for Nanoscale Science, Ithaca, New York 14853, USA}
\affiliation{School of Applied and Engineering Physics, Cornell University, Ithaca, New York 14853, USA}

\author{Darrell G. Schlom}
\email{schlom@cornell.edu}
\affiliation{Department of Materials Science and Engineering, Cornell University, Ithaca, New York 14853, USA}
\affiliation{Platform for the Accelerated Realization, Analysis, and Discovery of Interface Materials (PARADIM), Cornell University, Ithaca, New York 14853, USA}
\affiliation{Kavli Institute at Cornell for Nanoscale Science, Ithaca, New York 14853, USA}
\affiliation{Leibniz-Institut für Kristallzüchtung, Max-Born-Str. 2, 12489 Berlin, Germany}

%Lines 

\date{12 May 2026}% It is always \today, today,

\begin{abstract}
Tantalum dioxide (\ce{TaO2}) is a metastable tantalum compound. Here, we report the epitaxial stabilization of \ce{TaO2} on \ce{Al2O3} (1$\bar{1}$02) ($r$-plane sapphire) substrates using suboxide molecular-beam epitaxy (MBE) and thermal laser epitaxy (TLE), demonstrating single-oriented, monodomain growth of anisotropically strained thin films.
Microstructural investigation is performed using synchrotron X-ray diffraction and scanning transmission electron microscopy.
The tetravalent oxidation state of tantalum is confirmed using X-ray absorption and photoemission spectroscopy as well as electron energy-loss spectroscopy. Optical properties are investigated via spectroscopic ellipsometry and reveal a \SI{0.3}{\electronvolt} Mott gap of the tantalum $5d$ electrons.
Density-functional theory and group theoretical arguments are used to evaluate the limited stability of the rutile phase and reveal the potential to unlock a hidden metal-insulator transition concomitant with a structural phase transition to a distorted rutile phase, akin to \ce{NbO2}. 
Our work expands the understanding of tantalum oxides and paves the way for their integration into next-generation electronic and photonic devices.
\end{abstract}

\maketitle

%%%%%%%%%%%%%%%%%%%%%%%%%%%%%%%%%%%%%%%
%Introduction
\section{Introduction}

New materials with tunable electrical and optical conductivities can be a tremendous boon to the development of next-generation fast and energy-efficient nanoelectronics. Such materials may exhibit substantial changes between high-conductivity (``on") and low-conductivity (``off") states in response to thermodynamic control parameters, such as strain or temperature. Unfortunately, most existing materials are either metallic or insulating, with limited ability to modify their conductivity. 
%For the realization of the next generation of fast and energy-efficient nanoelectronics, there is a great need for new materials whose electrical and optical conductivities can be sensitively tuned between high (on) and low (off) states by altering a thermodynamic control parameter, e.g., strain or temperature. Unfortunately, most materials are either metallic or insulating and their conductivities cannot be changed substantially. 
Historically, advancements in microelectronics have relied on tuning semiconductor conductivity through chemical doping and the electric-field effect. 
%The evolution of microelectronics has hinged on the ability to tune the conductivity of semiconductors like silicon using chemical doping and the electric-field effect.
Materials undergoing a thermodynamic phase transition offer an exciting alternative to conventional semiconductors, enabling potential breakthroughs in device design.
Among these, materials exhibiting a metal-insulator transition (MIT) are particularly intriguing.\cite{Imada1998}
Yet, such materials are rare, and even fewer exhibit an MIT above room temperature, limiting their integration in practical devices.\cite{Noskin2017}
Expanding the library of MIT materials is therefore essential to broadening their applications.
%Materials exhibiting a metal-insulator transition (MIT) are rare, and among that small class even fewer exhibit an MIT above room temperature. This small design space of MIT materials limits their applicability in devices. It is therefore of great interest to discover more materials that display an MIT.

The archetypical compound exhibiting an MIT is \ce{VO2}, with a transition temperature of \SI{65}{\celsius}. Shortly after the discovery of the MIT in $3d^1$ \ce{VO2},\cite{Morin1959} a similar effect was discovered in $4d^1$ \ce{NbO2} albeit at a much higher temperature of \SI{807}{\celsius}.\cite{Janninck1966, Belanger1974} Thus far, high-quality samples of the $5d^1$ analog \ce{TaO2} have remained elusive. The $d^1$ oxides \ce{VO2}, \ce{NbO2}, and \ce{TaO2} crystallize in the rutile structure (space group $P4_2/mnm$, \# 136) at high temperature.
Interest in \ce{TaO2} is also motivated by the desire to compare it with its isoelectronic sulfide analog, \ce{TaS2}, which exhibits  greater thermodynamic stability. Unlike the oxide, \ce{TaS2} crystallizes in layered polymorphs, referred to as 1T (trigonal), 2H (hexagonal), and 3R (rhombohedral), similar to \ce{MoS2}. 
\ce{TaS2} is renowned for its exotic electronic properties including an MIT, charge-density wave, \cite{Hart2023} and superconductivity. \cite{Navarro2016, Sipos2008, Maaren1967}
Furthermore, the conductivity can be tuned dynamically by electrical pulses \cite{Venturini2022, Huber2025} as well as by ultrashort light pulses.\cite{Sun2018} 

Beyond its immediate relevance to MIT materials, studying \ce{TaO2} could inform advancements in quantum information processing. Tantalum is a key material in the fabrication of superconducting quantum bits,\cite{Place2021, Ganjam2024, Bland2025}
and understanding its oxides may help mitigate decoherence from surface defects caused by uncontrolled oxidation.\cite{Oh2024, Majer2024} Furthermore, tantalum monoxide (\ce{TaO}) has been identified as a superconductor with a critical temperature surpassing that of tantalum metal,\cite{Zhang2023, Cao2024} further underscoring the need to explore tantalum oxides comprehensively.
Last but not least, sub-stoichiometric tantalum oxides (Ta$_2$O$_{5-x}$, TaO$_{2-x}$) are technologically relevant to resistive switching memories.
The resistive switching phenomenon in these devices is attributed to the reversible formation and rupture of conductive filaments within the tantalum oxide layer, typically associated with the movement of oxygen vacancies.\cite{Lee2011}

The inherent tendency of tantalum oxides to form oxygen vacancies and exhibit multiple valence states presents significant challenges in stabilizing intermediate oxidation states, particularly \ce{Ta^{4+}} in \ce{TaO2}.
Reports of the synthesis of bulk \ce{TaO2} are scarce. Historically, some researchers focused on the partial reduction of \ce{Ta2O5}, \cite{Schafer1952}
% Schäfer and Breil  refer to papers that are even older, but I cannot access those to verify 
% they also mention reproducibility issues and contradicting results
while others attempted the oxidation of TaC \cite{Schoenberg1954} or experimented with redox reactions of tantalum with a different metal oxide.\cite{Terao1967} None of these approaches resulted in the realization of phase-pure \ce{TaO2}.
\citet{Syono1983} used shock reduction at $\approx$\SI{60}{\giga\pascal} to dissociate \ce{Ta2O5} under high pressure and high temperature and obtained a specimen with the rutile structure -- albeit not phase-pure either (maximum yield $\approx$ \SI{70}{\percent}). Subsequent annealing of the material in an oxygen atmosphere resulted in a rutile structure with reduced lattice parameters, which the authors referred to as \ce{Ta_{0.8}O2}.

We are only aware of one report of the growth of epitaxial \ce{TaO2} thin films.
\citet{Muraoka2016} utilized pulsed-laser deposition and employed a buffer layer of \ce{NbO2} to epitaxially stabilize \ce{TaO2} on \ce{Al2O3} (0001) ($c$-plane sapphire), achieving an orientation relationship of \ce{TaO2} (100) on \ce{NbO2} (110) on \ce{Al2O3} (0001). Their attempts to grow \ce{TaO2} directly on $c$-plane sapphire were unsuccessful.

In this work, we demonstrate the buffer-free epitaxial growth of anisotropically strained \ce{TaO2} on $r$-plane sapphire (\ce{Al2O3} (1$\bar{1}$02)) using suboxide molecular-beam epitaxy ($S$-MBE) and thermal laser epitaxy (TLE).
%% new
Most of the characterization presented in this article has been performed on MBE-grown samples.

The remainder of this article is organized as follows. Section II outlines the synthesis of \ce{TaO2} thin films and the experimental techniques employed for their characterization. The experimental results and their discussion are presented in Section III. Section IV provides a summary of the first-principles calculations, which are described in detail in the supplementary material. Finally, Section V presents the conclusions and outlook.

\section{Experiment}
\subsection{Suboxide molecular-beam epitaxy ($S$-MBE)}
%Here, we demonstrate for the first time the growth of untwinned epitaxial thin films of phase-pure $5d^1$ \ce{TaO2} (the rutile-like polymorph) using suboxide molecular-beam epitaxy (MBE). 
A molecular beam of \ce{TaO2} is achieved by heating an iridium crucible containing \ce{Ta2O5} in an effusion cell heated to a temperature of about \SI{1700}{\celsius}. \cite{Schwaigert2023} 
The calculated vapor pressures of species over \ce{Ta2O5} shows that the beam is dominated by the incongruent evaporation of the suboxide \ce{TaO2} (see Figs. \ref{Fig_VaporPressure_Ta_Ta2O5} and \ref{Fig_VaporPressureComponents_Ta2O5}). 
The $S$-MBE approach avoids the notoriously unstable electron-beam evaporation of tantalum metal in conventional MBE and the need for subsequent oxidation using a background gas or plasma. The latter is particularly challenging to control in the quest for \ce{TaO2} as the stable bulk phase of tantalum oxide is the $5d^0$ compound \ce{Ta2O5}, a band insulator without an MIT, similar to the case of $3d^0$ \ce{V2O5} and $4d^0$ \ce{Nb2O5}. In the $S$-MBE approach,  \ce{Ta^4+} is delivered to the substrate from a pre-oxidized molecular beam of \ce{TaO2} and no additional oxidant such as oxygen or ozone background gas is required.

We have found that $S$-MBE of tantalum oxide benefits from the preparation of the oxide source material in the form of densely compacted pellets, as opposed to loose powders. \cite{Schwaigert2023} As a starting material, we purchased \ce{Ta2O5} powder from Alfa Aesar with a purity of \SI{99.993}{\percent}.
The high source temperatures (above \SI{1700}{\celsius}) required to create a suitable flux of \ce{TaO2} pose stringent demands on the crucible material. Here, we use iridium crucibles known for their thermal stability and resistance to oxidation. At \SI{1750}{\celsius}, we achieve a growth rate around \SI{10}{\nano\meter\per\hour}. Unfortunately, such extreme temperatures significantly reduce the lifetime of high-temperature effusion cells.
The chamber background pressure of the Veeco Gen10 system used for this study was $\approx$ \SI{5E-8}{\Torr} during \ce{TaO2} film growth and consisted mostly of oxygen (emanating from the \ce{Ta2O5} source). %turbomolecular and cryogenic pumps.

\ce{Ta2O5} thin films were epitaxially grown on (111)-oriented yttria-stabilized zirconia (YSZ) substrates to act as a pentavalent tantalum oxide reference for X-ray and electron spectroscopy experiments. For \ce{Ta2O5} growth via $S$-MBE, distilled ozone was supplied to the center of the front side of the substrate via a water-cooled stainless steel nozzle and the background chamber pressure was kept at \SI{1E-6}{\Torr}.

\begin{figure}[tb]
	\centering
	\includegraphics[width=\columnwidth]{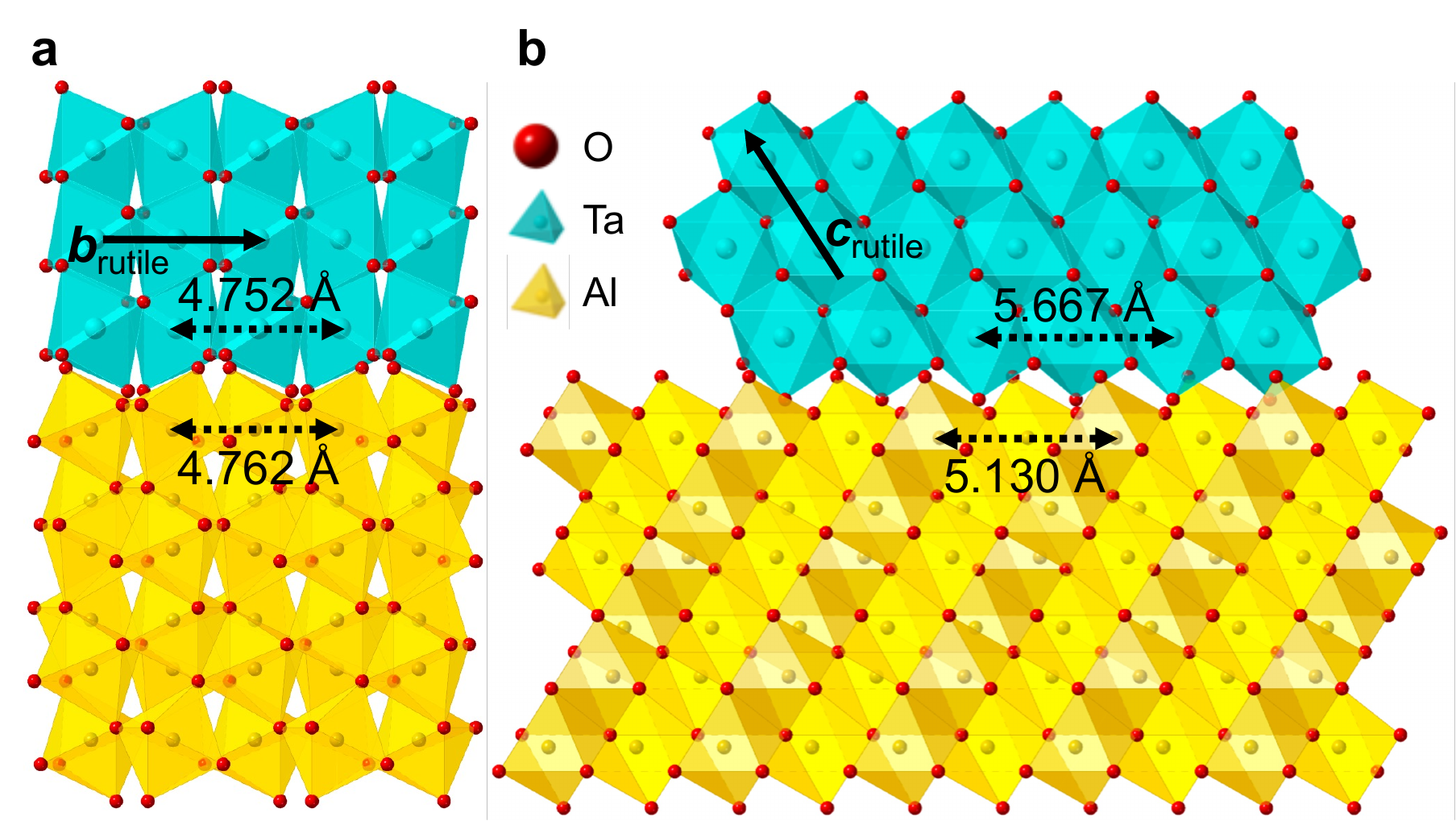}
	\caption{Cross-sectional schematics of \ce{TaO2} (101) (blue) on \ce{Al2O3} ($1\bar{1}2$) (yellow) perpendicular to the rutile $b$ and $c$-axes are shown in (a) and (b), respectively. The view direction into the plane for panel (a) is \ce{TaO2} [$\bar{1}01$] $\parallel$ \ce{Al2O3} [$\bar{1}101$] and for panel (b) is \ce{TaO2} [$010$] $\parallel$ \ce{Al2O3} [$11\bar{2}0$].}
	\label{Fig_Cartoon}
\end{figure}

We selected $r$-plane sapphire as the most promising substrate for \ce{TaO2} film growth because it is stable under ultrahigh vacuum (UHV) conditions even at elevated substrate temperatures ($T_\text{sub}$) around \SI{1000}{\celsius} that were found necessary to crystallize \ce{TaO2}. Figure \ref{Fig_Cartoon} highlights the oxygen octahedral network of the film and substrate that continues across the heteroepitaxial interface. While one in-plane direction has an ideal fit along the rutile $b$-axis with only $\approx$ \SI{0.2}{\percent} epitaxial strain, the orthogonal in-plane direction unfortunately suffers from a large misfit of $\approx$ \SI{9.5}{\percent}.
See supplementary material \cref{sec:misfit} for calculation and discussion as well as a topview schematic (\cref{Fig_Cartoon_Topview}).
%The epitaxial synthesis of rutile \ce{TiO2} on $r$-plane sapphire was published by Fukushima, Takaoka, and Yamada \cite{Fukushima1993} and we observe the same epitaxial relationship for the growth of rutile \ce{TaO2} on $r$-plane sapphire as they reported for rutile \ce{TiO2} on $r$-plane sapphire.
We observe an epitaxial relationship for the growth of rutile \ce{TaO2} on $r$-plane sapphire in the same way as \citet{Fukushima1993} reported it for rutile \ce{TiO2} on $r$-plane sapphire.

Of course, the ideal substrate for \ce{TaO2} would have the rutile crystal structure itself. Unfortunately, both commercially available substrates with the rutile structure (\ce{TiO2} and \ce{MgF2}) proved unsuitable for the synthesis of \ce{TaO2}. Unlike its 3$d^1$ analog \ce{VO2}, \ce{TaO2} cannot conveniently be grown on abundantly available \ce{TiO2} substrates due to the high $T_\text{sub}$ required, around \SI{1000}{\celsius}, which promotes the reduction of  \ce{Ti^{4+}}. Attempts to grow \ce{TaO2} on \ce{TiO2} substrates led exclusively to \ce{Ta2O5} formation.
This is consistent with Ellingham diagrams showing that at the oxygen partial pressure needed to stabilize \ce{Ta2O5}, \ce{Ti2O3} (and not \ce{TiO2}) is stable.\cite{Shang2024}
This implies the reaction

\begin{align}
    2~ \ce{TiO2} (s) +  2~ \ce{TaO2}(s) & \rightarrow 2~  \ce{Ti2O3} (s) + \ce{Ta2O5}(s).
\end{align}

%We reason that this is due to the ease of oxygen vacancy creation in the \ce{TiO2} substrate, and the oxygen then diffuses into the film and oxidizes the Ta to \ce{Ta^{5+}}.

Attempts to grow \ce{TaO2} on \ce{MgF2} were also unsuccessful due to the decomposition of the substrate at temperatures around \SI{600}{\celsius}, well below the crystallization temperature of \ce{TaO2} that was found to be around \SI{1000}{\celsius}.
The propensity of \ce{MgF2} to oxidize and form a volatile fluoride gas during the growth of a metal oxide poses an additional challenge.\cite{Kubovsky2025}

Preliminary attempts to grow \ce{TaO2} on \ce{Mg2SiO4} (010) -- a new substrate for rutile oxide heteroepitaxy that we recently introduced \cite{Kubovsky2025} -- resulted in the formation of \ce{MgTa2O6} and rough interfaces  at substrate temperatures in the range of 1000-\SI{1200}{\celsius}.
The proposed reaction is
\begin{align}
    \ce{Mg2SiO4} (s) + 4~ \ce{TaO2} (s) + \frac{1}{2} \ce{O2} (g) & \rightarrow 2 ~\ce{MgTa2O6} (s) + \ce{SiO} (g). 
\end{align}

%Another possibility is
%\begin{align}
%    \ce{Mg2SiO4} (s) + 2~ \ce{TaO2} (s) & \rightarrow  ~\ce{MgTa2O6} (s) + \ce{MgO} (s) + \ce{SiO} (g) 
%\end{align}

The formation of \ce{MgTa2O6} resembles the findings of
\citet{Sun2004} who reported the formation of \ce{MgTa2O6} when subjecting MgO (001) substrates to Ta-O vapors (from electron-beam evaporation of a \ce{Ta2O5} powder charge) at $T_\text{sub}$ between 600-\SI{1000}{\celsius} and $p$\ce{O2} = \SI{7.5E-5}{\Torr}. 

X-ray photoelectron spectroscopy (XPS) and scanning transmission electron microscopy (STEM) analysis showed that magnesium diffused from the substrate into the film and that tantalum was oxidized to its preferred pentavalent oxidation state. This observation motivates further research into the potential use of \ce{MgTa2O6} and related trirutiles as new substrates for the growth of rutile thin films with large lattice parameters.

As \ce{TaO2} is not known to be a stable phase in the bulk, we initially capped films \textit{in situ} with an amorphous layer of \ce{SiO_x} at room temperature in ultrahigh vacuum before exposing the samples to ambient air. To this end, we used a pre-oxidized beam of silicon oxide emanating from a SiO charge (Alfa Aesar, \SI{99.99}{\percent} purity). \cite{Azizie2023}
In later experiments, we found that \ce{TaO2} films remain stable in air at room temperature for at least 1 year even without a capping layer such as amorphous \ce{SiO_x}. 
All of the samples in this article were grown without a capping layer.

Before film growth, all substrates were sequentially cleaned in an ultrasonic bath using a series of solutions: a detergent solution (``Micro-90" (International Products Corporation (IPC))\footnote{\url{https://ipcol.com/cleaners/micro-90}} in deionized water, pH $\approx$ 9.5), followed by acetone, and finally isopropyl alcohol, with each cleaning step lasting several minutes.

To heat to $T_\text{sub}$ around \SI{1000}{\celsius} during growth, we utilize a mid-infrared (\SI{10.6}{\micro\meter}) \ce{CO2} laser-heating system built by epiray GmbH.
The substrate temperature is monitored and controlled during growth via an on-axis infrared (\SI{7.5}{\micro\meter}) pyrometer. The pyrometer focus is projected through a ZnSe beamsplitter in the \ce{CO2} shaping optics to target the bare substrate backside -- the same incident surface as the \ce{CO2} heating laser.
%This approach provides accurate temperature sensing across the full range from room temperature to \SI{2000}{\celsius} with a single pyrometer.
The pyrometer reading is then fed into a PID loop for temperature control of the \ce{CO2} laser output, resulting in a tight, stable laser output at fixed target temperatures across the full range from room temperature to \SI{2000}{\celsius} both with and without continuous substrate rotation.
To increase the accuracy of the temperature measurements despite partial attenuation of the \SI{7.5}{\micro\meter} radiation through the laser beamline optics, we calibrate the system by intentionally melting the backside of a test \ce{Al2O3} substrate. This provides a known melting point (\SI{2054\pm6}{\celsius})\cite{Schneider1970} and substrate emissivity at \SI{7.5}{\micro\meter} ($\approx$ 1).\cite{Sova1998}
Assuming a true emissivity of 1, the ratio between the measured \SI{7.5}{\micro\meter} radiation density and the known value for \ce{Al2O3} at the melting point represents the degree of \SI{7.5}{\micro\meter}  attenuation in the beam optics, which we find in our case to be $\approx$ 0.55.
Setting the pyrometer onboard electronics to an emissivity value matching this ratio then effectively compensates for the optical losses and provides substrate backside temperature readings across the full temperature range.

\subsection{Thermal laser epitaxy (TLE)}
Given the demanding deposition parameters required for the growth of epitaxial \ce{TaO2} by MBE, alternative approaches merit consideration. In TLE, elemental sources are heated by continuous-wave lasers with a wavelength of $\approx$ \SI{1}{\micro\meter} that is efficiently absorbed by most elements. Depending on the thermal properties of the element, the source material is either free-standing and self-contained or, if need be, placed in a crucible, from which it sublimates or evaporates to enable material transfer onto the substrate. \cite{Braun2019, Smart2021} The TLE system used in this work is equipped with the same \ce{CO2}-laser substrate heater as the aforementioned MBE system. 
Due to the unique mechanism of laser-based source heating and laser-based substrate heating, TLE offers a significantly broader process window than conventional epitaxial growth techniques. \cite{Hensling2024}
This is particularly advantageous given the narrow and challenging conditions identified for the suboxide MBE growth of epitaxial \ce{TaO2}. Notably, tantalum has been established as a well-suited, free-standing source for TLE, and high deposition rates are readily attainable. \cite{Majer2024}
%new
The objective of this work is not to compare different growth techniques, but rather to demonstrate that rutile \ce{TaO2} can be stabilized through multiple synthesis routes. To this end, we include a limited set of TLE-grown samples alongside those synthesized by $S$-MBE.

\begin{figure}
    \centering
    \includegraphics[width=\linewidth]{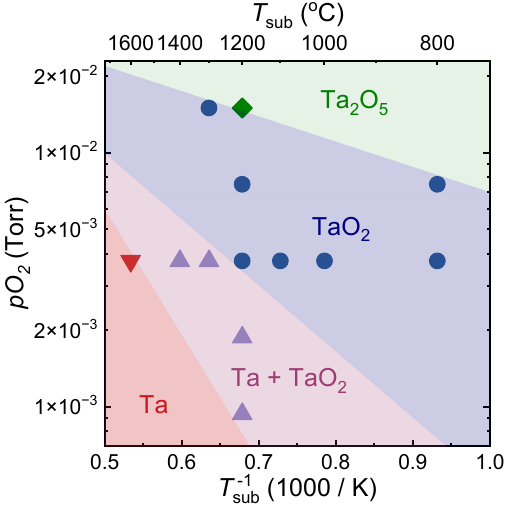}
    \caption{Growth diagram summarizing the oxygen partial pressure ($p\text{O}_2$) and substrate temperature ($T_\text{sub}$) dependence of the TLE growth of tantalum oxide. Phase-pure \ce{TaO2} was obtained in the blue-shaded region of the diagram between 800 - \SI{1300}{\celsius}. Scatter markers denote experimental data, colored backgrounds serve as guide to the eye.}
    \label{fig:TLE_GrowthDiagram}
\end{figure}

To explore the parameter space in which \ce{TaO2} may be grown by TLE, we applied a fixed source-laser power of \SI{300}{\watt}. This resulted in a growth rate of  $\approx$ \SI{10}{\nano\meter\per\minute} for the parameter space shown in Fig. \ref{fig:TLE_GrowthDiagram}, grown on $c$-plane and $r$-plane sapphire. Note that this growth rate exceeds the suboxide MBE growth rate by over an order of magnitude.
This allowed us to explore a wide range of $T_\text{sub}$ and pressures for the growth of tantalum oxides. In the remainder of the article we focus on samples grown on $r$-plane sapphire, which enables untwinned growth of rutile, unlike $c$-plane sapphire.

\Cref{fig:TLE_GrowthDiagram} summarizes the results from pressure and $T_\text{sub}$ sweeps during TLE growth parameter optimization.
At an oxygen partial pressure of \SI{1.5E-2}{\Torr}, we identify a transition from the growth of epitaxial \ce{TaO2} to \ce{Ta2O5}. 
The formation of the desired epitaxial \ce{TaO2} is observed for $T_\text{sub}$ below \SI{1200}{\celsius} and pressures between \SI{3.75E-3}{\Torr} $<p \text{O}_2$ < \SI{1.5E-2}{\Torr}. 
Among the \ce{TaO2} samples, the XRD rocking curve full width at half maximum is narrowest for the film grown at $T_\text{sub} =$ \SI{1200}{\celsius} and widest for the sample grown at $T_\text{sub} =$ \SI{800}{\celsius}.
At $T_\text{sub} >$ \SI{1200}{\celsius}, the films start to show metallic tantalum until $T_\text{sub}$ > \SI{1400}{\celsius} where only tantalum metal is observed. 
Conversely, at low $T_\text{sub}$ and low oxygen pressures, the films become partially amorphous. 
The growth diagram in Fig. \ref{fig:TLE_GrowthDiagram} indicates that the high tantalum flux achieved in TLE promotes the stabilization of the intermediate \ce{Ta^{4+}} oxidation state. 
While the fully oxidized \ce{Ta^{5+}} state is accessible under elevated oxygen partial pressures, the high tantalum flux in TLE likely exceeds the available oxidant supply within the regime where \ce{TaO2} is formed. This imbalance between metal and oxygen flux limits complete oxidation, favoring the formation of the intermediate \ce{Ta^{4+}} state. The emergence of metallic tantalum at higher $T_\text{sub}$ further supports this interpretation, as the diminished oxygen chemical potential at elevated $T_\text{sub}$ reduces the system's capacity to fully oxidize the incoming metal flux.

Taken together, these observations underscore the delicate interplay between thermodynamic and kinetic factors in TLE. Nevertheless, we reproducibly identify a relatively broad window for the epitaxial growth of \ce{TaO2} thin films using this technique.
%new
Supporting XRD data is provided in the Supplementary Material. 
\Cref{fig:XRD_TLE_c-plane} shows XRD of TLE-grown \ce{TaO2} (100) on $c$-plane sapphire. Notably, this was achieved without the use of a buffer layer in contrast to the work of \citet{Muraoka2016}.
\Cref{fig:XRD_MBE_TLE} presents a comparison of MBE- and TLE-grown \ce{TaO2} (101) on $r$-plane sapphire, indicating that both techniques can lead to \ce{TaO2} with a single orientation, i.e., free of twins.

\subsection{Characterization \label{sec:Characterization}}
X-ray diffraction and reflectivity (XRD and XRR) measurements were performed using a PANalytical Empyrean diffractometer equipped with a hybrid 2-bounce germanium monochromator and a PIXcel3D detector. High-resolution rocking curves were measured using a germanium analyzer crystal on the diffracted beam side and a proportional counter. $\theta-2\theta$ scans, rocking curves, and small two-dimensional (2D) reciprocal space maps (RSMs) were recorded using a 1/\SI{2}{\degree} divergence slit. For XRR, a narrow 1/\SI{32}{\degree} divergence slit was used.
Three-dimensional (3D) X-ray reciprocal space mapping was performed at the $QM^2$ beamline at the Cornell High Energy Synchrotron Source (CHESS) at a beam energy of \SI{15}{\kilo\electronvolt} using the protocol outlined by Wadehra \textit{et al.} \cite{Wadehra2025}
2D RSMs were extracted to show the film and substrate Bragg peaks in two representative orthogonal planes.

\begin{figure}[t]
	\centering
	\includegraphics[width=\columnwidth]{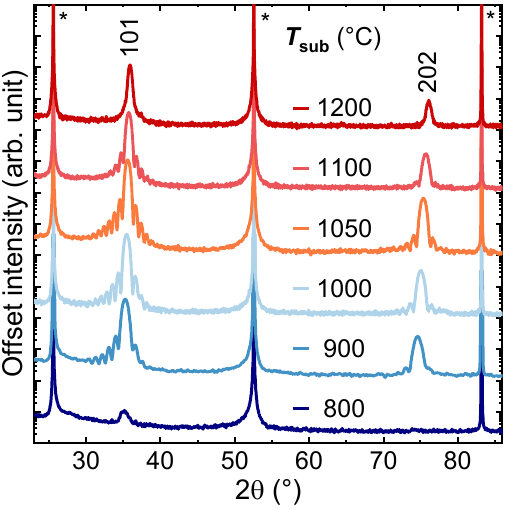}
	\caption{X-ray diffraction of MBE-grown \ce{TaO2} (101) thin film on \ce{Al2O3} ($1\bar{1}02$) scans showing pronounced Laue fringes for $T_\text{sub}$ between 900 and \SI{1100}{\celsius}. Peaks highlighted with an asterisk are due to the substrate.}
	\label{Fig_XRD_GrowthTemp}
\end{figure}

%%% Lab-source RSM
\begin{figure}[t]
	\centering
	\includegraphics[width=\linewidth]{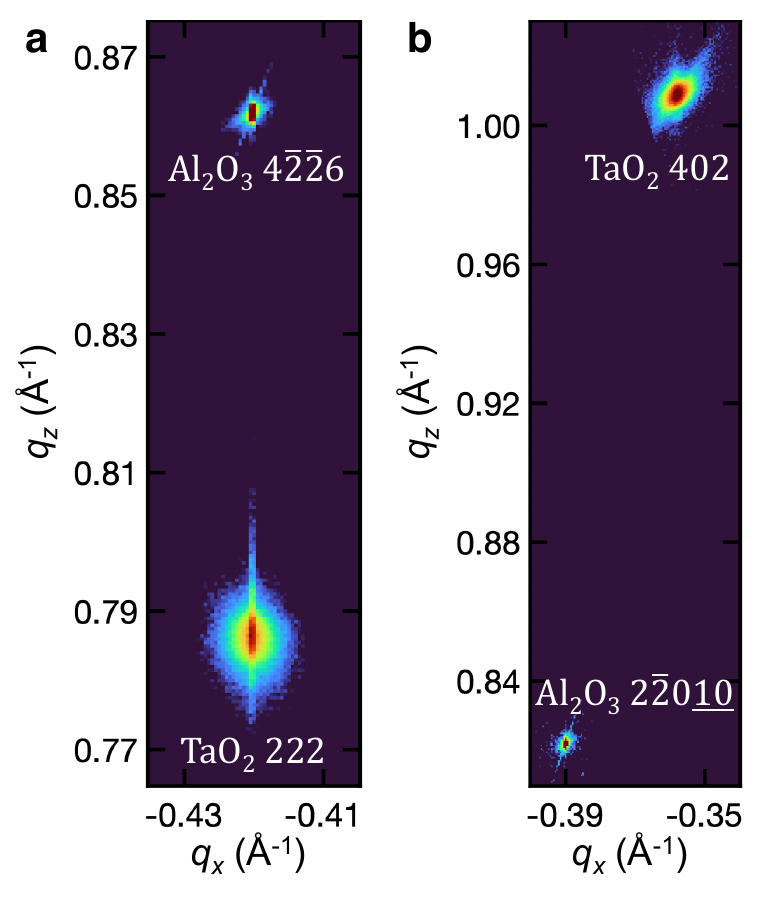}
	\caption{XRD reciprocal space map revealing the anisotropic strain state of MBE-grown \ce{TaO2} on \ce{Al2O3} ($1\bar{1}02$). Panel (a) shows that the film is commensurately strained to the substrate lattice parameter along the rutile $b$ direction (see \cref{Fig_Cartoon} for illustration); the film is fully relaxed in the orthogonal high-misfit direction (b).}
	\label{Fig_RSM_Lab}
\end{figure}	

Cross-sectional lamellae of MBE-grown \ce{TaO2} films for STEM imaging and electron energy-loss spectroscopy (EELS) measurements were prepared with a Thermo Fisher Helios G4 UX focused ion beam. High-angle annular dark-field (HAADF) STEM imaging was performed on an FEI Titan Themis operated at 300 kV with a 21.4 mrad probe convergence semi-angle and 50 pA probe current. For high-precision structural measurements, a series of 40 rapid-frame images (\SI{0.4}{\second} per frame) were acquired, aligned, and averaged with a rigid registration algorithm optimized to prevent lattice hops \cite{Savitzky2018}  to recover high signal, high fidelity atomic resolution images. EELS spectra were collected with a GIF Quantum ER spectrometer and a Gatan K2 Summit direct electron detector operated in counting mode.
Four-dimensional (4D) STEM datasets for multislice electron ptychography were collected at 300 kV with a 21.4 mrad probe convergence angle, 46 mrad collection angle, and 50 pA probe current, scanning 256 $\times$ 256 points at a step size of 0.44 Å and dwell time of \SI{100}{\micro\second} on an EMPAD-G2 detector.\cite{Philipp2022}
Multislice electron ptychography reconstructions were performed using the maximum likelihood algorithm \cite{Thibault2013} implemented in the fold-slice package \cite{Wakonig2020, Chen2021} using multiple probe modes to account for partial coherence. \cite{Chen2020, Thibault2013} Optimal reconstruction parameters were found with Bayesian optimization with an objective function minimizing the data error.\cite{Zhang2022}

\begin{figure*}[t]
	\centering
	\includegraphics[width=1.8\columnwidth]{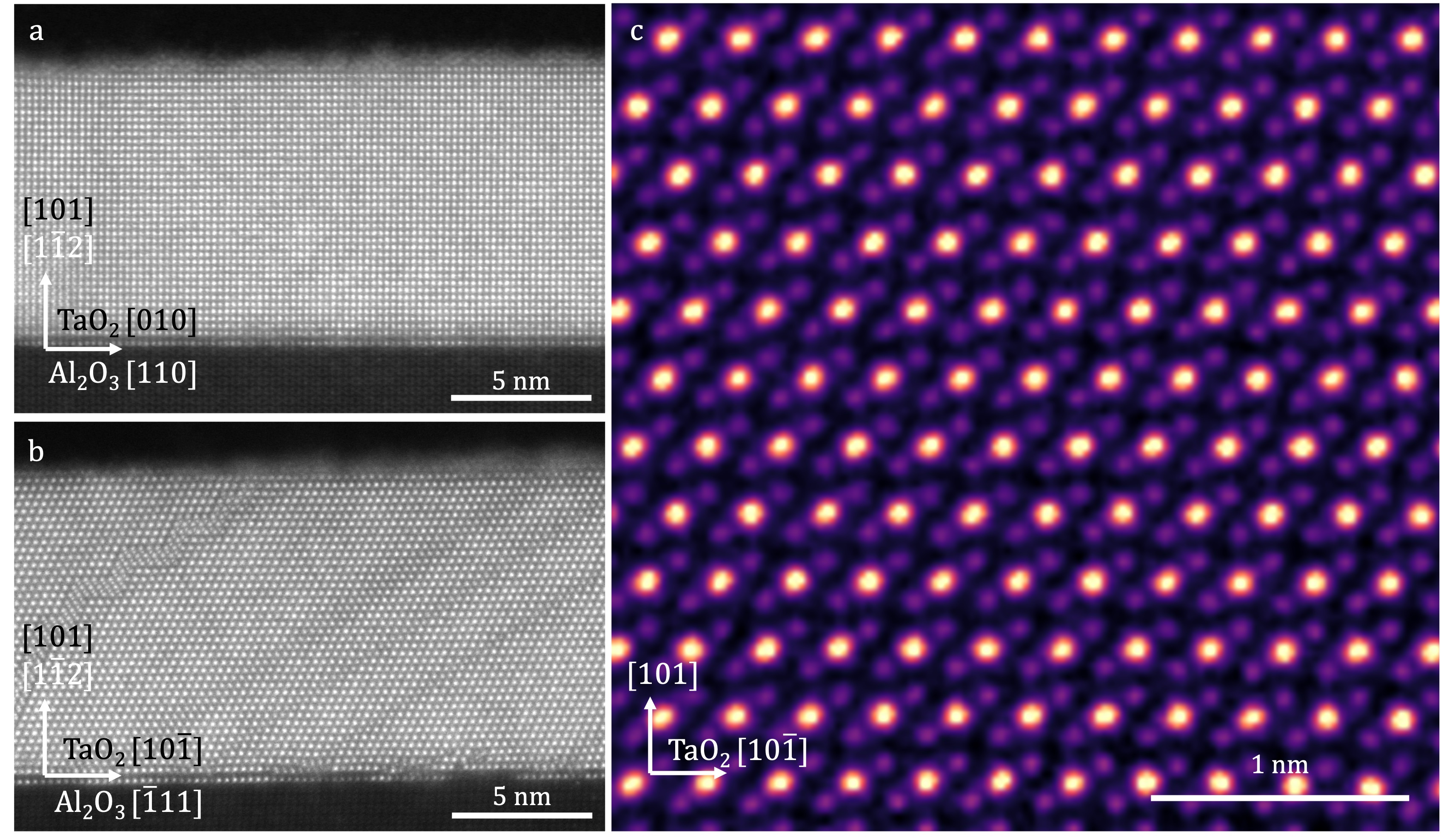}
    % redo figure with full res
	\caption{Cross-sectional HAADF-STEM images of MBE-grown \ce{TaO2} taken (a) perpendicular and (b) parallel to the rutile 
    $b$-axis. In panel (a), the film appears commensurately strained, as evidenced by the perfect alignment between tantalum ions in the film and aluminum ions in the substrate. In contrast, panel (b) reveals diagonal striations and an incommensurate interface marked by numerous dislocations. Panel (c) presents a magnified view of a defect-free region in the same orientation as (b), reconstructed using multislice electron ptychography to simultaneously resolve tantalum (depicted in yellow) and oxygen (purple) atomic columns.}
	\label{Fig_STEM}
\end{figure*}

\textit{In situ} non-monochromated XPS using a magnesium anode (\SI{1.3}{\kilo\electronvolt}) was performed at the PARADIM thin-film MBE-ARPES cluster at Cornell University. The sample was transferred from the growth chamber to the photoemission chamber at a pressure less than \SI{5E-8}{\Torr}.
Additional \textit{ex situ} XPS using a monochromated aluminum anode (\SI{1.5}{\kilo\electronvolt}) was performed using a ThermoFisher Nexsa G2 system with a \SI{400}{\micro\meter} diameter spot size and a low-energy electron flood gun for charge neutralization. The pressure during the analysis was $\approx$ \SI{1E-8}{\Torr}, mostly argon gas. High-resolution spectra were collected at a pass energy of \SI{40}{\electronvolt} and a step size of \SI{0.1}{\electronvolt}. Survey spectra were recorded at a pass energy of \SI{150}{\electronvolt} using a step size of \SI{0.4}{\electronvolt}. To avoid unintentional reduction, the samples were not sputter-cleaned before analysis. The XPS data were analyzed using CasaXPS (v. 2.3.25).

To gain information of the interior of the film, hard X-ray photoemission spectroscopy (HAXPES) was performed on the tantalum $3d$ and $4f$ peaks using two different X-ray energies. 
Variable-temperature operando HAXPES was performed using a SPECS system at the University of Twente in the Netherlands \cite{vandenBosch2025} equipped with a chromium anode (\SI{5.4}{\kilo\electronvolt}) for the tantalum $4f$ peak, and a Scienta Omicron system equipped with a gallium (\SI{9.3}{\kilo\electronvolt}) liquid-metal-jet anode at Binghamton University, New York, for the tantalum $3d$ peak.
Measurements of the tantalum $3d$ peaks show the same trends albeit less pronounced compared to the $4f$ peaks.

To investigate the stability of MBE-grown \ce{TaO2} at elevated temperatures in vacuum using operando HAXPES, the sample was heated using a near-infrared laser shining on the backside of a blackened flag-style sample holder. The temperature was monitored using a thermocouple spot-welded onto the sample plate. The pressure in the analysis chamber was on the order of \SI{1E-8}{\Torr} and spiked briefly during each heating step, likely due to the outgassing of the silver paint used to mount the sample on the flag-style holder. Residual gas analysis indicated that the background gas consisted mostly of \ce{H2O} and \ce{O2}.

%% XAS methods
Oxygen $K$-edge XAS was performed at the $Versox$ beamline at Diamond Light Source.\cite{Grinter2024} The signal was collected in total electron yield (TEY) mode by measuring the drain current from the sample holder. The highly insulating nature of film and substrate led to highly distorted spectra with curved backgrounds. Attempts to record and analyze the tantalum $N$-edges were unsuccessful due to their broad and weak nature.
Tantalum $L_3$ edge measurements were performed at the $PIPOXS$ beamline at CHESS in partial fluorescence yield (PFY) mode. 
The incident beam was energy selected using a Si(111) monochromator and reflected off two rhodium-coated mirrors set to 4 mrad for harmonic rejection.  Tantalum metal foil was measured in transmission geometry using \ce{N2}-filled ion chambers while \ce{TaO2} and \ce{Ta2O5} thin films were placed at \SI{45}{\degree} relative to the incident beam and signal detected using a four element Vortex detector.

%% RIXS methods
We studied resonant inelastic X-ray scattering (RIXS) at the tantalum $L_3$ edge at beamline ID20 of the European Synchrotron Radiation Facility. Photons from three consecutive U26 undulators were monochromatized by a Si(111) high-heat-load monochromator and a successive Si(311) channel cut monochromator. 
Using a spectrometer in Rowland geometry with a spherically shaped diced Si(660) analyzer crystal (1\,m radius of curvature), we achieved an overall energy resolution of 300\,meV.\@ 
All measurements were performed at 300 K.\@ RIXS is a photon-hungry technique. 
To enhance the RIXS signal for measurements on a TLE-grown thin film with a thickness of 43\,nm, we utilized a grazing incident angle $\theta$\,=\,1$^\circ$ that maximizes absorption within the thin film and a scattering angle $2\theta$ close to 90$^\circ$ that minimizes self-absorption.

%% Ellipso methods
Spectroscopic ellipsometry measurements were performed on MBE-grown epitaxial \ce{TaO2} ($\approx$ \SI{82}{\nano\meter}) on $r$-plane \ce{Al2O3} to determine the complex dielectric function ($\epsilon = \epsilon_1 + i \epsilon_2$) spectra of epitaxial \ce{TaO2}.
The thickness determined by modeling of the ellipsometry data was independently verified by fitting X-ray reflectivity measurements (not shown).
Reference measurements were performed on an uncoated $r$-plane sapphire substrate (see supplementary material). 
Generalized ellipsometric spectra were collected at 40, 45, and \SI{50}{\degree} angles of incidence in the 0.225 – \SI{0.50}{\electronvolt} photon energy range using a rotating compensator Fourier transform infrared (IR) ellipsometer (FTIR-VASE, J.A. Woollam Co.) \cite{Hilfiker2003} and in the 0.50 – \SI{4.13}{\electronvolt} photon energy range using a single-channel rotating-analyzer ellipsometer with an auto-retarder (V-VASE, J.A Woollam Co.). \cite{Johs1999, Woollam1999}
Generalized ellipsometric spectra are collected at four different sample orientations varying the azimuthal Euler angle $\phi$ with both ellipsometers to determine the structural and optical properties of the epitaxial \ce{TaO2} film.
\begin{comment}
The \ce{SiO_x} cap is modeled with a single Tauc-Lorentz oscillator, a Sellmeier expression, and a constant additive term to the real part of the complex dielectric function spectra, $\epsilon_\infty = 1$. 
This parameterization has been used to describe \ce{SiO_x} previously \cite{Hensling2024}. 
\end{comment}
The optical properties of \ce{Al2O3} are fixed from the measurement of the uncoated $r$-plane \ce{Al2O3}. 
The polar Euler angle $\theta$ for \ce{TaO2} is fixed to \SI{33.8}{\degree} which is the angle between the $c$-axis and the (101) surface plane of the \ce{TaO2} film.
Details are provided in Sec. \ref{sec:Ellipsometry} in the supplementary material.

Atomic force microscopy images were recorded \textit{ex situ} in air at room temperature using an Asylum Research Cypher S instrument with NanoWorld Arrow-UHFAuD-10 cantilever probes. Post-processing was performed using Gwyddion (v.2.67). \Cref{fig:AFM} shows a series of topography images. While the film appears to grow as small islands, the underlying step-and-terrace morphology of the $r$-plane sapphire substrate remains clearly visible.

%%%%%%%%%%%%%%%%%%%%%%%%%%%%%%%%%%%%%%%
\section{Results and discussion}
\subsection{X-ray diffraction}
% Al2O3         TaO2
% 5.13          5.667
% 4.762         4.752

XRD results in Fig. \ref{Fig_XRD_GrowthTemp} demonstrate that phase-pure \ce{TaO2} can be grown by suboxide MBE over a wide $T_\text{sub}$ range from 800-\SI{1200}{\celsius}. The exclusive presence of peaks arising from the (101) plane attests to the single out-of-plane orientation. The experimental lattice parameters are $a = 4.66(2)$, $b = 4.76(2)$, $c = 2.99(1)$ \SI{}{\angstrom}. 
%% new
Supplementary XRD data collected from MBE-grown \ce{TaO2} thin films of varying thicknesses is provided in \cref{fig:XRD_Thicknesses}. 
XRD $\phi$-scans (see Fig. \ref{fig:PhiScans}) confirm the untwinned nature of these films grown on $r$-plane sapphire.
The epitaxial growth is further corroborated by the streaky reflection high-energy electron diffraction (RHEED) pattern shown in Fig. \ref{fig:RHEED}.

\Cref{Fig_RSM_Lab} shows X-ray reciprocal space maps (RSMs) that probe two orthogonal in-plane directions, corresponding to the schematic depicted in \cref{Fig_Cartoon}.
The data reveals the highly anisotropic nature of \ce{TaO2} (101) on \ce{Al2O3} ($1\bar{1}02$). This is consistent with the surface unit cell of the substrate, see \cref{Fig_Cartoon_Topview}. The epitaxial mismatch along the $[010]_R$ direction (subscript $R$ denotes the rutile unit cell) is about \SI{0.2}{\percent} (tensile strain), whereas in the $[\bar{1}01]_R$ direction, it is about \SI{-9.5}{\percent} (compressive strain).
This results in the film being pseudomorphic only in the $[010]_R$ direction. In the orthogonal direction, the film is relaxed from the substrate-film interface and shows periodic defects. 
%From the weak satellite fringes in the $[\bar{1}01]$ direction, we deduce a value of \SI{7(1)}{\nano\meter} for the coherently scattering crystalline domain size.

% numbers from Ben from CHESS on sample YB240513A
% to be compared with sample 230410

% we are most sure about the b-axis because that's the one commensurate to Al2O3 [110]

Large field-of-view synchrotron X-ray RSMs are shown together with diffraction simulations in the supplementary material in Fig. \ref{Fig_RSM_overlay} and confirm the untwinned, single-oriented growth of rutile \ce{TaO2} on $r$-plane sapphire.
High-temperature XRD  data (Fig. \ref{Fig_XRD_Heating}) reveal that \ce{TaO2} irreversibly oxidizes to \ce{Ta2O5} around \SI{400}{\celsius} in air.

\subsection{Scanning transmission electron microscopy}

Atomic resolution HAADF-STEM images were collected on two orthogonal cross-sections of the film, the rutile $[\bar{1}01]$ and $[010]$ zone axes, revealing its anisotropic structure.
%updated
Along the $[\bar{1}01]$ zone axis, the tantalum sublattice visible with HAADF-STEM shows good agreement with a rutile polytype, and is fully strained along the $[010]$ direction with a low density of defects (\cref{Fig_STEM} a). On the other hand, along the $[010]$ zone axis a high density of diagonal striations perpendicular to the rutile $c$-axis is present (\cref{Fig_STEM} b). We attribute these striations to the formation of partial dislocations with a Burger’s vector $b=\frac{1}{2}[\bar{1}01]$ at the film-substrate interface. These dislocations drive a relaxation of the film along the $[\bar{1}01]$ direction accommodating the large misfit strain with the \ce{Al2O3} substrate. In addition to these striations, a lower density of intergrowths with non-rutile structure are also observed (also visible in \cref{Fig_STEM} b).
While conventional HAADF-STEM imaging readily resolves the film's cation sites and relevant symmetries and defects, it lacks the sensitivity to measure the arrangement of oxygen ions in the material. 

To confirm the film's rutile structure, multislice electron ptychography was used to fully characterize both the tantalum and oxygen sublattices, revealing the edge- and corner-sharing octahedra characteristic of the rutile structure, which is maintained even across the dense defects visible along the $[\bar{1}01]$ zone axis.
Thus, the combination of HAADF-STEM imaging and electron ptychography shows that the film exhibits a distorted rutile structure, with inequivalent $a$ and $b$ lattice parameters, and highly oriented defects associated with a partially relaxed, anisotropic epitaxial strain state.

\subsection{X-ray photoemission spectroscopy}
\begin{figure}[t]
	\centering
	\includegraphics[width=\columnwidth]{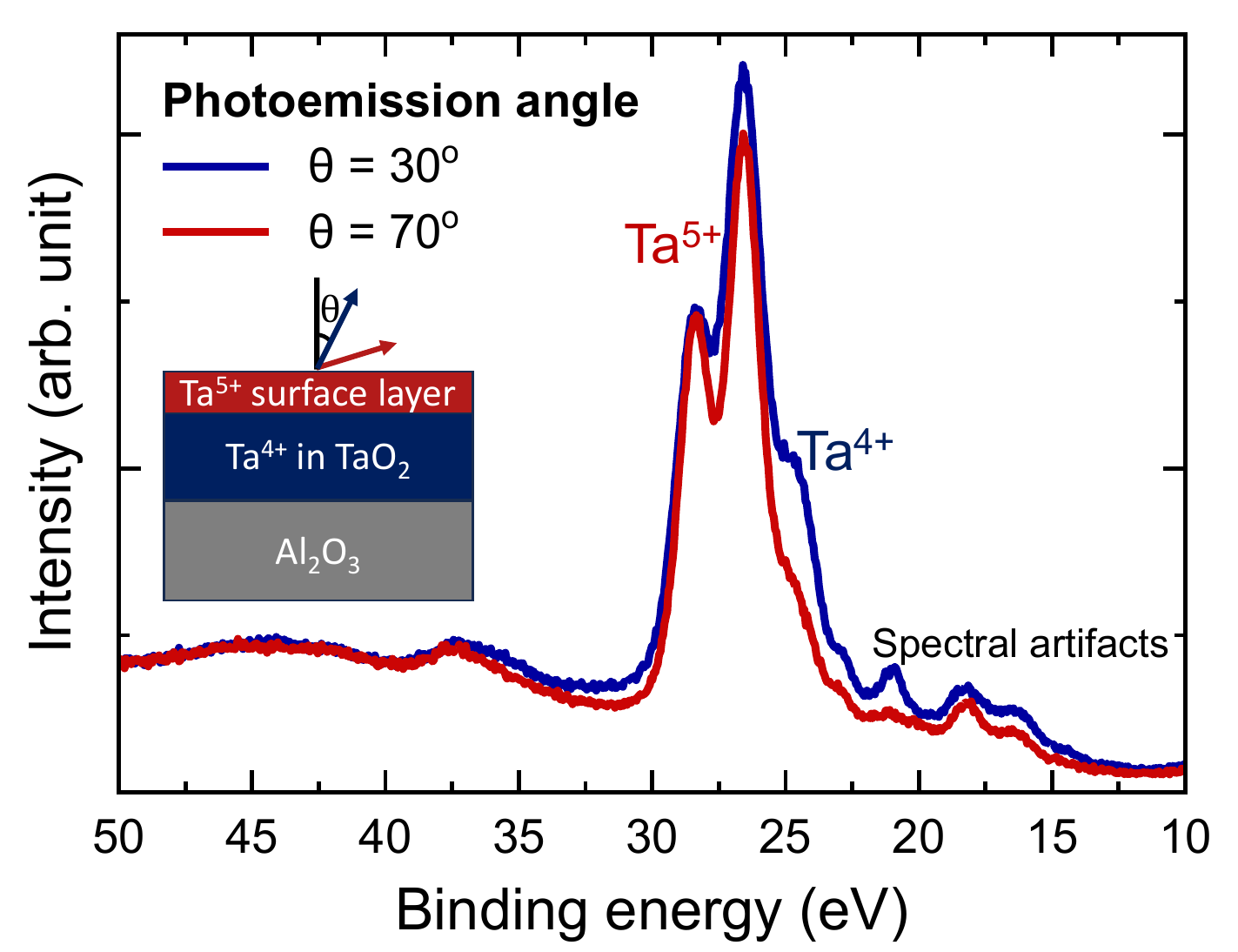}
	\caption{Tantalum $4f$ X-ray photoemission spectroscopy data of MBE-grown \ce{TaO2} collected \textit{in situ} at two different photoemission angles using a non-monochromatic magnesium anode. The inset depicts a cross-sectional model of the sample.}
	\label{fig:XPS-insitu}
\end{figure}

\Cref{fig:XPS-insitu} shows XPS data collected \textit{in situ} on an MBE-grown \ce{TaO2} thin film ($\approx$ \SI{10}{\nano\meter}) before exposure to ambient air. 
Tantalum $4f$ XPS data were collected at two different photoemission angles, providing information from different probing depths. In the more surface-sensitive high photoemission angle geometry, the data looks \ce{Ta^{5+}}-like. At low photoemission angle, however, a strong shoulder appears on the lower binding energy side of the tantalum $4f_{7/2}$ peak, around \SI{24}{\electronvolt}, indicative of a reduced valence state which we assign to \ce{Ta^{4+}}.
In summary, the XPS data points towards the formation of a pentavalent overlayer on top of the tetravalent \ce{TaO2}.
\citet{Muraoka2016} reported a similar result and estimated an overlayer thickness of \SI{3}{\nano\meter} of unintentional \ce{Ta2O5}-like crust. The formation of an oxidized surface layer was previously described for the structurally and electronically similar oxides \ce{VO2} \cite{BirkholzerSotthewes2022, Quackenbush2015} and \ce{NbO2} \cite{Fajardo2021} as well.
We cautiously conclude that the surface oxidation is self-limiting as we observed no aging effects of the samples during the course of two years of exposure to air when comparing XRD data.

To enhance the experimental sensitivity for the buried tetravalent \ce{TaO2} layer, we next turn to hard X-ray photoemission spectroscopy (HAXPES).
Figure \ref{Fig_HAXPES_Ta3d_Ga} shows a comparison of \ce{TaO2} and \ce{Ta2O5} data.
Our \ce{Ta2O5} data are in excellent agreement with HAXPES reference spectra that were recently published by Zheng \textit{et al.}.\cite{Zheng2023} For \ce{TaO2}, we observe the asymmetric peaks of the \ce{Ta^{5+}} doublet with pronounced shoulders on the lower binding energy side, indicative of a reduced valence state compared to \ce{Ta^{5+}}.

We performed operando HAXPES measurements to test the high-temperature stability of \ce{TaO2} thin films in UHV (pressure < \SI{1E-8}{\Torr}).
Figure \ref{Fig_HAXPES_Ta3d_Cr} shows \textit{ex situ} HAXPES data at elevated temperatures.
Up to $\approx$ \SI{700}{\celsius}, the HAXPES spectra look identical to the room-temperature ones. Around \SI{900}{\celsius}, an irreversible oxidation reaction took place, fully transforming \ce{TaO2} into \ce{Ta2O5}. This is evidenced by the disappearance of the small density of states near the Fermi level observed for \ce{TaO2} as well as the \ce{Ta^{4+}} $4f_{7/2}$ peak near \SI{24}{\electronvolt} binding energy.
This oxidation is remarkable as most oxide thin films tend to reduce in UHV conditions at elevated temperatures and emphasizes the challenge of stabilizing tetravalent tantalum in \ce{TaO2}. \footnote{While the temperature and pressure at which oxidation was observed in operando HAXPES was nominally comparable to the MBE growth conditions, calibration differences between different chambers and laboratories likely exist.}

\begin{figure}[t]
	\centering
	\includegraphics[width=\columnwidth]{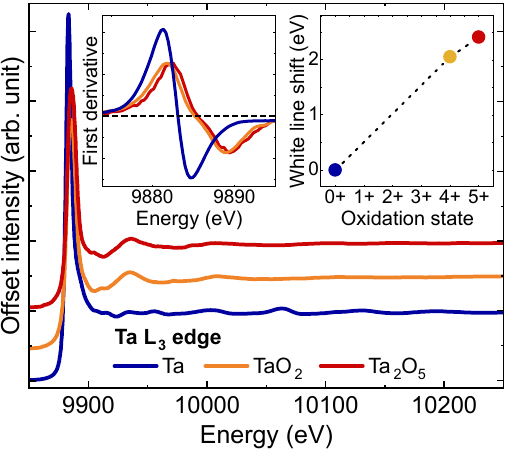}
	\caption{Tantalum $L_3$ X-ray absorption spectroscopy of MBE-grown \ce{TaO2}. The insets show the derivative of the white line (left) and the white-line shift  against the formal oxidation state of tantalum, \ce{TaO2}, and \ce{Ta2O5} (right) as determined from the root of the derivative.}
	\label{Fig_Ta_L3}
\end{figure}

\subsection{X-ray absorption spectroscopy}

The oxygen $K$-edge of our MBE-grown \ce{TaO2} thin films (not shown) resembles spectra of \ce{Ta2O5}. We note that in the TEY mode, the experiment is sensitive only to the top few nanometers of the specimen. This therefore corroborates the presence of a pentavalent tantalum oxide overlayer on the surface of \ce{TaO2} revealed by XPS. 
To obtain more information about the interior of the \ce{TaO2} film, we next turned to a bulk-sensitive hard X-ray measurement, namely the tantalum $L_3$ edge, which probes $2p_{3/2} \rightarrow 5d$ transitions. 
\cref{Fig_Ta_L3} shows a comparison of \ce{TaO2} with tantalum metal and \ce{Ta2O5} absorption spectra.
In rutile \ce{TaO2}, tantalum ions are coordinated octahedrally with oxygen ions. This is also true for most tantalum ions in \ce{Ta2O5} except those that are in a pentagonal bipyramid (\ce{TaO7} unit). This similarity in coordination is the reason why the EXAFS signal of the \ce{TaO2} and \ce{Ta2O5} thin films appear qualitatively very similar.
We acknowledge that our EXAFS data is noisy, as is often the case for thin films measured in the fluorescence yield mode, and therefore we refrain from a detailed analysis. Instead, we focus on the strong white line and its energy shift with the expected oxidation state.
We determine the position of the white line using the roots of the first derivative (see left inset).  The peak position for \ce{TaO2}  falls in between tantalum metal and the pentavalent \ce{Ta2O5} reference samples.

%%%%%%%  RIXS written by Markus
\subsection{Resonant inelastic X-ray scattering}
\begin{figure}[t]
	\centering
	\includegraphics[width=\columnwidth]{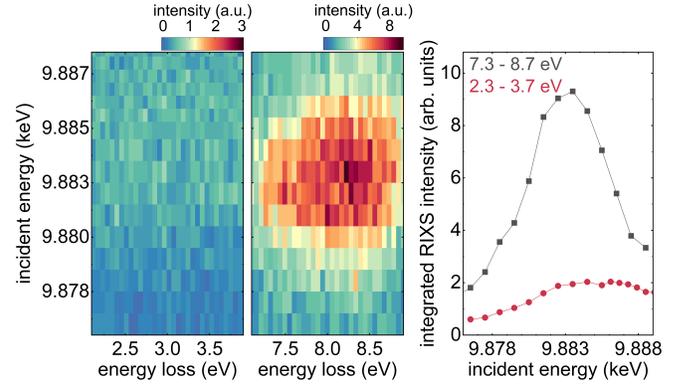}
	\caption{Resonance behavior of tantalum $L_3$-edge RIXS of a TLE-grown \ce{TaO2} film. 
	Left and middle: RIXS intensity as function of the incident energy for energy loss 
	around 3 and 8\,eV, respectively. Note the different color scales.  
	Right: integrated RIXS intensity for the energy windows shown in the other panels. 
    We observe a charge-transfer peak at about 8\,eV energy loss but no spectral feature 
    at lower energies. 	} 
	\label{fig:RIXS}
\end{figure}

To further investigate the elusive $5d^1$ electronic structure, we attempted 
resonant inelastic X-ray scattering (RIXS) at the tantalum $L_3$ edge on a TaO$_2$ sample 
grown by TLE. The RIXS process consists of an initial X-ray absorption step in which a 
$2p$ core electron is promoted into the $5d$ shell. Then, the corresponding $2p$ core hole 
is filled by a valence electron, leaving the system in a low-energy excited state. 
In the case of a Mott insulator with a $5d^1$ configuration, the absorption into the 
$5d$ shell boosts spin and orbital excitations. \cite{Ament2011}
More precisely, one expects to observe the excitation from a local, spin-orbit-driven 
$j$\,=\,3/2 ground state to a $j$\,=\,1/2 excited state at low energy and excitations 
from $t_{2g}$ to $e_g$ orbitals at roughly 3\,eV, 
as reported for Rb$_2$TaCl$_6$ 
with $5d^1$ Ta$^{4+}$ ions in an octahedral ligand cage. \cite{Ishikawa2019}
Therefore, RIXS is an excellent tool to determine the local electronic structure. 
A further motivation was to search for a possible dimerization at low temperature and
in analogy to the case of VO$_2$. Hard X-ray $L$-edge RIXS can detect such a superstructure 
via the corresponding interference pattern in the RIXS intensity, as observed, e.g., for dimers 
in iridates or, at the tantalum L$_3$ edge, for tetrahedra in the lacunar spinel GaTa$_4$Se$_8$.
\cite{Revelli2019,Magnaterra2024}

We studied the resonance behavior, i.e., the RIXS intensity for different incident energies 
$E_{\rm in}$ across the tantalum $L_3$ edge; see Fig.\ \ref{fig:RIXS}. 
We find a pronounced charge-transfer peak at about 8\,eV energy loss which is resonantly 
enhanced at $E_{\rm in}$ about 9.883\,keV.\@ 
Nevertheless, for energy loss between 2 and 4\,eV -- note the different color scale -- 
the data do not show the $t_{2g}$-$e_g$ excitations expected for a $5d^1$ compound. 
This suggests a $5d^0$ state that allows for charge-transfer excitations, 
but not for $d$-$d$ excitations, as reported for $5d^0$ SrLaMgTaO$_6$. \cite{Oh2018}
Unfortunately, the sample immediately degraded under the intense X-ray beam, 
showing clear visual signatures of radiation damage. 
At present, the RIXS result is inconclusive as to whether the sample is indeed 
a $5d^0$ tantalum oxide or whether the elusive $5d^1$ state also rapidly decays under the
beam. There remains hope that samples with a higher structural quality and deliberate capping layers will exhibit greater stability and spectral cleanliness. Success for such an endeavor necessitates 
better lattice matched substrates and is motivation for future work into the design of new substrates for oxides with the rutile structure.

\begin{comment}
RIXS measurements ESRF ID20 09/24
E. Bergamasco, L. Pätzold, and M. Grüninger
ESRF: C. Sahle, B. Detlefs, K. Ruotsalainen
\end{comment}

\subsection{Electron energy-loss spectroscopy}
To spectroscopically investigate the interior of the \ce{TaO2} thin film, we prepared a cross-sectional lamella using focused-ion beam milling. 
Because the tantalum $N_{2,3}$ edges are weak and broad, we focus on the oxygen $K$-edge instead.
The EELS data are presented in Fig. \ref{Fig_O-K_TaO2_Ta2O5}, where they are also compared to a pentavalent tantalum oxide reference.
To qualitatively guide our expectations, we computed the oxygen $K$-edges on the DFT-LDA level.
%% Have Nong and Frank check this part
The key difference between \ce{TaO2} and \ce{Ta2O5} is the ratio between the first two peaks. This part of the oxygen $K$-edge is sensitive to the tantalum $5d$ orbitals due to oxygen $2p$ - tantalum $5d$ hybridization.
Qualitatively, the first peak is expected to be larger for \ce{Ta2O5} because the latter has an empty $5d- t_{2g}$ manifold.
This is in contrast to \ce{TaO2} which has 1 electron in the $5d$ orbital.
See supplementary material for a comparison between the experimental data and DFT-based calculated spectra (see Fig. \ref{Fig_O-K_exp_dft}).

\subsection{Spectroscopic ellipsometry}

\begin{figure}[tb]
	\centering
	\includegraphics[width=\columnwidth]{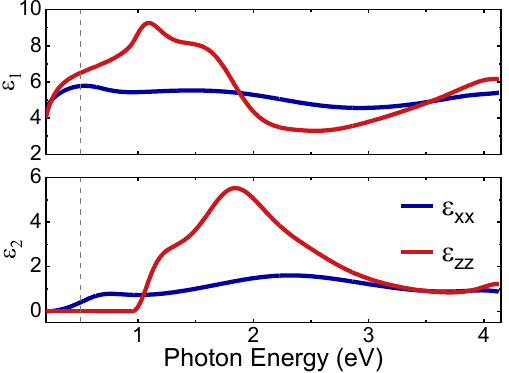}
	\caption{Real ($\epsilon_1$) and imaginary ($\epsilon_2$) parts of the complex dielectric function determined from spectroscopic ellipsometry measurements in the ordinary ($\epsilon_{xx}$) and extraordinary ($\epsilon_{zz}$) directions. The dotted line indicates the transition between two different instruments.}
	\label{fig:Epsilon}
\end{figure}

The directionally dependent complex dielectric function ($\epsilon = \epsilon_1 + i \epsilon_2$) spectra of an \SI{82}{\nano\meter} thick MBE-grown \ce{TaO2} thin film were determined from generalized spectroscopic ellipsometry measurements and are depicted in Fig. \ref{fig:Epsilon}. Measurements on a bare \ce{Al2O3} substrate are shown in Fig. \ref{fig:Al2O3_Ellipsometry}.

Crystals with a rutile structure exhibit uniaxial birefringence, meaning that the optical response depends on the orientation of the incident light’s wavevector $\mathbf{k}$ relative to the crystal’s optic axis, the rutile $c$-axis.
The ordinary response $\epsilon_{xx}$ occurs when the electric field $\mathbf{E}$ is entirely perpendicular to the optic axis ($\mathbf{E} \perp c$); in this case, light experiences the ordinary refractive index $n_o$. The extraordinary response $\epsilon_{zz}$ occurs when $\mathbf{E}$ has a component parallel to the optic axis; in this case, the effective refractive index $n_e$ depends on the angle between the wavevector and the optic axis.
For (101)-oriented rutile films, the optic axis is inclined relative to the surface. In \ce{TaO2}, the angle between the (001) and (101) planes is \SI{33.8}{\degree}, see Fig. \ref{Fig_Cartoon} (b) for illustration.
The strong anisotropy in the complex dielectric function spectra (Fig. \ref{fig:Epsilon}) agrees with the rutile crystal structure. 
Factors that are not captured in the analysis of the ellipsometry data are possible compositional gradients in the film due to the surface oxidation of \ce{TaO2} as well as the anisotropic strain and the resulting defects.

For ease of comparison with the literature, the imaginary part of the permittivity was converted to the real part of the optical conductivity and displayed in Fig. \ref{fig:TaO2_sigma}.
In particular the strength of the peaks below \SI{3}{\electronvolt} are consistent with values reported for other Mott insulators, such as $3d, 4d, 5d$ transition metal compounds.\cite{Goessling2008, Reul2012, Vergara2022, Moon2009, Guiot2013}
In contrast, a clean $d^0$ system would have no optical transitions below the charge-transfer gap. While defects in a $d^0$ compound would yield $d^1$ states with similar energies to the ones reported here, the contribution of such defect states to the optical conductivity is expected to be weaker than the excitations across the Mott-Hubbard gap from $\ket{d^1 d^1}$ to $\ket{d^0 d^2}$, which are typically on the order of a few \SI{100}{\per\ohm\per\centi\meter}. 
For comparison, the bandgap of \ce{Ta2O5} is $\approx$ \SI{4}{\electronvolt} and there are no absorption features at lower energies. \cite{Hur2019, Prato2011}
The absorption coefficient $\alpha$ is provided in Fig. \ref{fig:TaO2_alpha}.
The smallest direct and indirect optical transitions of our \ce{TaO2} film are extracted via Tauc plots of $(\alpha h \nu)^{1/2}$ and $(\alpha h \nu)^2$, respectively (see Figs. \ref{fig:Tauc_indirect} and \ref{fig:Tauc_direct}). 
The transition occuring at the lowest photon energy is observed in the ordinary direction and is considered to be indirect.
We therefore determine the optical bandgap of \ce{TaO2} to be $\approx$ \SI{0.3}{\electronvolt}.

\subsection{Electrical transport}
% Electrical measurements at room temperature were attempted in a linear four-probe configuration with manually applied In contacts as well as sputtered Ti capped with Au using a shadow mask process.

We attempted to measure the DC resistance-temperature relationship of MBE-grown \ce{TaO2} thin films above room temperature.
To this end, we built a custom setup using an evacuated Fine Instruments tube furnace.
The temperature was measured with a K-type thermocouple using a Keithley 740 thermometer.
The resistance was measured using the linear four-probe method using a Keithley 196 digital multimeter.
Kanthal wires were glued to sputtered chromium contacts on the surface of the tested layer with Dupont 6838 high-temperature conductive silver paste. The sample was mounted on an alumina support.
Despite best efforts, the resistance of all tested samples exceeded the measurement range near room temperature and was limited by setup leakage at high temperatures. 

We expected to measure strongly anisotropic electrical transport properties,  given the dislocations and periodic defects in one direction and high degree of structural quality in the commensurately strained direction. In contrast to our expectations, we find the resistance in both directions to exceed our measurement capabilities (>> \SI{}{\giga\ohm}).
%This is a puzzling result given the presence of a narrow optical bandgap.

The oxidized surface layer on top of \ce{TaO2} poses a challenge to the fabrication of ohmic contacts and might act as a wide-gap tunnel barrier.
In an attempt to reduce the effective Schottky barrier effect of the oxidized surface layer, we used a shadow mask process to pattern mm-sized contacts. To this end, we locally sputter-cleaned and gently reduced the surface layer of a \ce{TaO2} thin film before Ti-Pt deposition via magnetron sputtering. Unfortunately, four-probe resistance measurements remained out-of-range.
We therefore speculate that the carriers in \ce{TaO2} are highly localized.

Preliminary attempts to chemically dope \ce{TaO2} with magnesium concentrations of up to \SI{20}{\percent} did not result in a measurable reduction of the electrical resistance and produced no discernible changes in the XRD $\theta$–$2\theta$ scans. Magnesium was selected as a dopant because of its stable valence and because \ce{Mg^{2+}} has an ionic radius within \SI{6}{\percent} of that of octahedrally coordinated \ce{Ta^{4+}}.\footnote{Other ions with a comparable ionic radius to \ce{Ta^{4+}} are \ce{Li^{1+}} and \ce{Sc^{3+}} as well as \ce{W^{5+}} and \ce{Re^{5+}}} Subsequent growth experiments of \ce{TaO2} on \ce{Mg2SiO4} substrates revealed the unintentional formation of \ce{MgTa2O6}, a $d^0$ compound. 
The propensity to form this phase may account for the ineffectiveness of magnesium doping in \ce{TaO2}.

\section{First-principles calculations}
First-principles density-functional theory calculations were used to evaluate the structural, electronic, and dielectric properties of \ce{TaO2}, assuming an infinite, strain-free periodic crystal. The computational details and results are discussed in \cref{sec:DFT} of the supplementary material. Two phases of \ce{TaO2} were thoroughly investigated: the high-symmetry rutile-like $P4_2/mnm$ structure (space group \#136) and a lower-symmetry $I4_1/a$ structure (space group \#88), which had a lower total energy. The transition from $P4_2/mnm$ to $I4_1/a$ is the result of a phonon at the $R$-point condensing into the $P4_2/mnm$ structure and matches with the expectation of the experimentally observed transition in \ce{NbO2}, the 4$d^1$ analog of \ce{TaO2}.\cite{Janninck1966, Belanger1974}
%These were selected based on the experimentally observed high- and low-temperature phases of \ce{NbO2}, the 4$d^1$ analog of \ce{TaO2}.
Our calculations show that whereas the $P4_2/mnm$ phase of \ce{TaO2} is consistently metallic, the properties of the $I4_1/a$ phase are sensitive to the value of an applied Hubbard $U$ term on the Ta-$d$ states. Notably, there exists an energetic barrier between the two phases at $U$ = 0 eV which vanishes as $U$ approaches 4 eV. Additionally, the $I4_1/a$ phase transitions from being conducting at $U$ = 0 eV to insulating as $U$ increases, with  an indirect band gap of $\approx$ \SI{0.7}{\electronvolt} and a direct gap of $\approx$ \SI{1}{\electronvolt}. 
The direct bandgap is in rough agreement with that obtained by spectroscopic ellipsometry ($\approx$ 0.7 and \SI{1.1}{\electronvolt} for the ordinary and extraordinary direction, respectively).
Furthermore, both techniques find a smaller indirect bandgap, albeit with different energies (ellipsometry $\approx$ \SI{0.3}{\electronvolt}, DFT $\approx$ \SI{0.7}{\electronvolt}).

Figures \ref{fig:LOPTIC_U_high}, \ref{fig:LOPTIC_U_low}, \ref{fig:LOPTIC_psuedo_high}, \ref{fig:LOPTIC_psuedo_low} in the supplementary material show relative permittivity spectra computed from first principles using various values of the Hubbard $U$ parameter and three different exchange-correlation functionals for the undistorted rutile polymorph of \ce{TaO2} ($P4_2/mnm$) as well as the \ce{NbO2}-like ($I4_1/a$) polymorph. Although there are some differences between the computational and experimentally measured spectra, the spectra of the distorted $I4_1/a$ polymorph qualitatively match the experimental complex permittivities irrespective of the exact value of $U$ and for all three exchange-correlation functionals. In summary, our computational results for $5d^1$ \ce{TaO2} approximately capture both the bandgap energy as well as the peak energies.

%%%%%%%%%%%%%%%%%%%%%%%%%%%%%%%%%%%%%%%
\section{Conclusion}
In summary, we show that $r$-plane sapphire is a chemically stable substrate for the growth of \ce{TaO2}.
Unlike $c$-plane sapphire, $r$-plane enables the growth of single-domain, untwinned rutile \ce{TaO2} thin films. Using X-ray, optical, and electron spectroscopies, we characterize the metastable $5d^1$ electron configuration of tantalum in the interior of \ce{TaO2} thin films and detect the presence of an oxidized surface layer. 
In contrast to an earlier report, \cite{Muraoka2016} we demonstrate direct heteroepitaxy of \ce{TaO2} without the need for a dedicated template or buffer layer and we show that this result is attainable using two deposition techniques, namely $S$-MBE and TLE.
First-principles calculations support the existence of the rutile phase and reveal a ground state with a \ce{NbO2}-like distortion. While the calculations suggest that the undistorted rutile polymorph would be metallic, all samples in this study are highly electrically insulating with a $\approx$ \SI{0.3}{\electronvolt} optical bandgap in rough agreement with calculated dielectric spectra for the distorted phase. The prediction of two competing phases close in energy motivates further studies of a possible metal-insulator phase transition between them analogous to the archetypic case of \ce{VO2}.

Despite best efforts, experimental signatures of a metal-insulator transition in this $5d^1$ compound have so far remained elusive. We speculate that higher-quality samples with fewer structural defects are needed to unlock the intrinsic properties of notoriously hard to stabilize \ce{TaO2}. This hinges on the availability of suitable substrates with a better lattice match. In our future work, we will explore the growth of trirutile \ce{MgTa2O6} and columbite \ce{MgNb2O6} as candidate substrates. Preliminary tests using olivine \ce{Mg2SiO4} as a candidate substrate for high-temperature rutile film growth led to decomposition and intermixing at the interface and unintentional \ce{MgTa2O6} formation.

%%%%%%%%%%%%%%%%%%%%%%%%%%%%%%%%%%%%%%%
\section{supplementary material}
%% List and summarize all

\begin{acknowledgments}
This work was primarily supported by U.S. Department of Energy, Office of Science, Office of Basic
Energy Sciences, under Contract No. DE-SC0019414 (thin film synthesis: A.S.P., Y.A.B., D.G.S.;
X-ray experiments and interpretation Y.A.B., B.Z.G, A.S.; first-principles theory: J.Z.K. and N.A.B.).

The authors thank Yibin Bu for performing preliminary HAXPES measurements, Mark Pfeifer for technical assistance with the high-temperature XRD, and Steven Button for substrate preparation.
The authors thank Qijun Che, Jelle Ruiter, and Koen M. Draijer for support during synchrotron beamtimes,
and Jochen Mannhart for supporting the TLE.

This work made use of the synthesis and electron microscopy facilities of the Platform for the Accelerated Realization, Analysis, and Discovery of Interface Materials (PARADIM), which are supported by the NSF under Cooperative Agreement No. DMR-2039380.
This work made use of the Cornell Center for Materials Characterization shared instrumentation facility. The FEI Titan Themis 300 was acquired through NSF-MRI-1429155, with additional support from Cornell University, the Weill Institute, and the Kavli Institute at Cornell. The Helios FIB was acquired with support by NSF DMR-2039380.
This work entails research conducted at the Center for High-Energy X-ray Sciences (CHEXS), which is supported by the National Science Foundation (BIO, ENG and MPS Directorates) under award DMR-2342336.
This research was funded, in part, by the Gordon and Betty Moore Foundation’s EPiQS Initiative (Grant nos. GBMF3850 and GBMF9073) to Cornell University.
The authors acknowledge the European Synchrotron Radiation Facility (ESRF) for provision of synchrotron radiation facilities under proposal number BLC-15590 and thank K. Ruotsalainen, B. Detlefs, and C. Sahle for assistance and support in using beamline ID20.
This publication is part of the project \textit{Conductivity on demand -- turning an insulator into a metal} with project number 019.223EN.017 of the research program RUBICON, which is partly financed by the Dutch Research Council (NWO).
Part of this research was carried out at the MESA+ operando HAXPES user facility at the University of Twente. This publication is part of the project ”Lab-based HAXPES facility for operando studies of energy-material interfaces” with file number 175.2019.001 of the research programme NWO-groot 2019/2020 which is partly financed by the Dutch Research Council (NWO).
N.A. thanks Utrecht University for a start-up grant (Dutch Sector Plan) supporting this work.
E.B. and M.G. acknowledge funding from the Deutsche Forschungsgemeinschaft 
(DFG, German Research Foundation) through Project No. 277146847-CRC 1238 (B03). 
The work at AGH University was supported by the National Science Centre, Poland, Grant OPUS No. UMO2021/41/B/ST3/03454.
Computational resources were provided by the Cornell Center for Advanced Computing. The authors thank Turan Birol for helpful discussions regarding the computational results.

\end{acknowledgments}

%%%%%%%%%%%%    AUTHOR CONTRIBUTIONS
\section{Author declarations}
\subsection{Conflict of Interest}
The authors F.V.E.H and D.G.S. have been granted U.S. Patent No. 11,462,402 (4 October 2022) with the title “Suboxide Molecular-Beam Epitaxy and Related Structures.”
The author B.D.F. is an employee of epiray GmbH, a company that sells commercial \ce{CO2} laser heater and TLE systems.

\subsection{Author contributions}
Y.A.B. and A.S.P. contributed equally.
Y.A.B. conceived the project and fabricated the samples together with A.S.P. under the guidance of M.R.B. and D.G.S.
T.S. prepared the \ce{Ta2O5} source rods and fabricated supplementary \ce{TaO2} and \ce{Ta2O5} samples.
F.V.E.H. performed the TLE.
N.S. performed the STEM analysis with guidance from D.A.M.
J.Z.K. performed the DFT calculations with guidance from N.A.B.
Y.A.B. and B.Z.G. performed the synchrotron XRD with guidance from S.S. and A.S.
Y.A.B. and T.A.K. performed and analyzed the AFM.
Y.A.B, I.C.G.B., E.M.K., and C.J.P.  performed the XAS with guidance from F.M.F.G. 
E.B. and M.G. performed the RIXS at ESRF.
Y.A.B and I.C.G.B performed the operando HAXPES with guidance from C.B.
B.D.F. and M.J.W. performed additional XPS and HAXPES.
A.B. and S.C. performed the ellipsometry with guidance from N.J.P.
W.T and W.T. performed the resistance measurements.
N.A. performed the DFT-based XAS calculations.
Y.A.B. wrote the manuscript with input from all authors.

% All authors read and approved of the manuscript

%%%%%%%%%%%%%

\section*{Data Availability Statement}
%The data that support the findings of this study are available from the corresponding author upon reasonable request.
%% Or do we need to upload everything on the PARADIM server?

The data that support the findings of this study are openly available at 
\textcolor{red}{
DOI will be added in production step
}
%\url{https://doi.org/10.34863/zsda-pa72}.
%Replace with actual one!

\bibliography{bib_20260107}% Produces the bibliography via BibTeX.

\newpage
~
\newpage
\onecolumngrid

% ===== Supplementary Information numbering =====
\setcounter{figure}{0}
\setcounter{table}{0}
\setcounter{equation}{0}

\renewcommand{\thefigure}{S\arabic{figure}}
\renewcommand{\thetable}{S\arabic{table}}
\renewcommand{\theequation}{S\arabic{equation}}

\setcounter{section}{0}
\renewcommand{\thesection}{S\arabic{section}}

\preprint{AIP/123-QED}

\title[Supporting Information]{Synthesis of epitaxial \ce{TaO2} by suboxide molecular-beam epitaxy}

\author{Yorick A. Birkhölzer}
\homepage{contributed equally}
\affiliation{Department of Materials Science and Engineering, Cornell University, Ithaca, New York 14853, USA}
\email{y.birkholzer@cornell.edu}

\author{Anna S. Park}
\homepage{contributed equally}
\affiliation{Department of Materials Science and Engineering, Cornell University, Ithaca, New York 14853, USA}
\affiliation{Platform for the Accelerated Realization, Analysis, and Discovery of Interface Materials (PARADIM), Cornell University, Ithaca, New York 14853, USA}
\author{Noah Schnitzer}
\affiliation{Department of Materials Science and Engineering, Cornell University, Ithaca, New York 14853, USA}
\affiliation{Kavli Institute at Cornell for Nanoscale Science, Ithaca, New York 14853, USA}

\author{Jeffrey Z. Kaaret}
\affiliation{School of Applied and Engineering Physics, Cornell University, Ithaca, New York 14853, USA}
\author{Benjamin Z. Gregory}
\affiliation{Department of Materials Science and Engineering, Cornell University, Ithaca, New York 14853, USA}

\author{Tomas A. Kraay}
\affiliation{Department of Materials Science and Engineering, Cornell University, Ithaca, New York 14853, USA}

\author{Tobias Schwaigert}
\affiliation{Department of Materials Science and Engineering, Cornell University, Ithaca, New York 14853, USA}
\affiliation{Platform for the Accelerated Realization, Analysis, and Discovery of Interface Materials (PARADIM), Cornell University, Ithaca, New York 14853, USA}

\author{Matthew R. Barone}
\affiliation{Department of Materials Science and Engineering, Cornell University, Ithaca, New York 14853, USA}
\affiliation{Platform for the Accelerated Realization, Analysis, and Discovery of Interface Materials (PARADIM), Cornell University, Ithaca, New York 14853, USA}

\author{Brendan D. Faeth}
\affiliation{Platform for the Accelerated Realization, Analysis, and Discovery of Interface Materials (PARADIM), Cornell University, Ithaca, New York 14853, USA}

%%%%%%%
\author{Felix V.E. Hensling}
\affiliation{Max Planck Institute for Solid State Research, 70569 Stuttgart, Germany}

\author{Iris C.G. van den Bosch}
\affiliation{MESA+ Institute for Nanotechnology, University of Twente, 7500 AE, Enschede, The Netherlands}
\author{Ellen M. Kiens}
\affiliation{MESA+ Institute for Nanotechnology, University of Twente, 7500 AE, Enschede, The Netherlands}
%\author{Jelle R.H. Ruiters}
%\affiliation{MESA+ Institute for Nanotechnology, University of Twente, 7500 AE, Enschede, The Netherlands}
\author{Christoph Baeumer}
\affiliation{MESA+ Institute for Nanotechnology, University of Twente, 7500 AE, Enschede, The Netherlands}

%

%%%%%
\author{Enrico Bergamasco}
\affiliation{Institute of Physics II, University of Cologne, 50937 Cologne, Germany}

\author{Markus Grüninger}
\affiliation{Institute of Physics II, University of Cologne, 50937 Cologne, Germany}

%%%%%%%

\author{Alexander Bordovalos}
\affiliation{Department of Physics and Astronomy, University of Toledo, Toledo, Ohio 43606, USA}
\affiliation{Wright Center for Photovoltaics Innovation and Commercialization, University of Toledo, Toledo, Ohio 43606, USA}

\author{Suresh Chaulagain}
\affiliation{Department of Physics and Astronomy, University of Toledo, Toledo, Ohio 43606, USA}
\affiliation{Wright Center for Photovoltaics Innovation and Commercialization, University of Toledo, Toledo, Ohio 43606, USA}

\author{Nikolas J. Podraza}
\affiliation{Department of Physics and Astronomy, University of Toledo, Toledo, Ohio 43606, USA}
\affiliation{Wright Center for Photovoltaics Innovation and Commercialization, University of Toledo, Toledo, Ohio 43606, USA}

%%%%%
\author{Waldemar Tokarz}
\affiliation{Faculty of Physics and Applied Computer Science, AGH University of Krakow, 30-059, Krakow, Poland}

\author{Wojciech Tabis}
\affiliation{Faculty of Physics and Applied Computer Science, AGH University of Krakow, 30-059, Krakow, Poland}
%%%%%%

\author{Matthew J. Wahila}
\affiliation{Analytical and Diagnostics Lab, Binghamton University, State University of New York,  Binghamton, New York 13902, USA}
%%%%
\author{Suchismita Sarker}
\affiliation{Cornell High Energy Synchrotron Source, Wilson Laboratory, Cornell University, Ithaca, New York 14853, USA}

\author{Christopher J. Pollock}
\affiliation{Cornell High Energy Synchrotron Source, Wilson Laboratory, Cornell University, Ithaca, New York 14853, USA}
%%%%

\author{Shun-Li Shang}
\affiliation{Department of Materials Science and Engineering, The Pennsylvania State University, University Park, Pennsylvania 16802, USA}

\author{Zi-Kui Liu}
\affiliation{Department of Materials Science and Engineering, The Pennsylvania State University, University Park, Pennsylvania 16802, USA}

\author{Nongnuch Artrith}
\affiliation{Debye Institute for Nanomaterials Science, Utrecht University, 3584 CG, Utrecht, The Netherlands}

\author{Frank M.F. de Groot}
\affiliation{Debye Institute for Nanomaterials Science, Utrecht University, 3584 CG, Utrecht, The Netherlands}

\author{Nicole A. Benedek}
\affiliation{Department of Materials Science and Engineering, Cornell University, Ithaca, New York 14853, USA}

\author{Andrej Singer}
\affiliation{Department of Materials Science and Engineering, Cornell University, Ithaca, New York 14853, USA}

\author{David A. Muller}
\affiliation{Platform for the Accelerated Realization, Analysis, and Discovery of Interface Materials (PARADIM), Cornell University, Ithaca, New York 14853, USA}
\affiliation{Kavli Institute at Cornell for Nanoscale Science, Ithaca, New York 14853, USA}
\affiliation{School of Applied and Engineering Physics, Cornell University, Ithaca, New York 14853, USA}

\author{Darrell G. Schlom}
\email{schlom@cornell.edu}
\affiliation{Department of Materials Science and Engineering, Cornell University, Ithaca, New York 14853, USA}
\affiliation{Platform for the Accelerated Realization, Analysis, and Discovery of Interface Materials (PARADIM), Cornell University, Ithaca, New York 14853, USA}
\affiliation{Kavli Institute at Cornell for Nanoscale Science, Ithaca, New York 14853, USA}
\affiliation{Leibniz-Institut für Kristallzüchtung, Max-Born-Str. 2, 12489 Berlin, Germany}

%Lines 
 
%Lines 

\date{\today}% It is always \today, today,

\maketitle

\end{comment}

%%%%%%%%%%%%%%%%%%%%%%%%%%%%%%%%%%%%%%%

{\Large
\noindent Supplementary Material
\\~\\
\noindent for
\\~\\
\noindent ``Synthesis of epitaxial \ce{TaO2} thin films on \ce{Al2O3} by suboxide molecular-beam epitaxy and thermal laser epitaxy"
}
~ 
\clearpage

\section{Suboxide molecular-beam epitaxy}

\subsection{Vapor pressure}

\begin{figure}[h]
	\centering
	\includegraphics[width=\columnwidth]{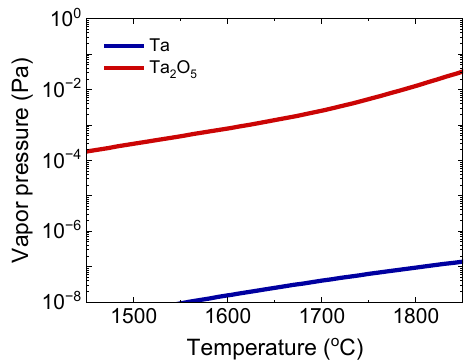}
	\caption{Vapor pressure of tantalum over tantalum metal (blue) and total vapor pressure of species emanating from a \ce{Ta2O5} charge (red) as a function of temperature. The tantalum vapor pressure was computed based on an equation published by \citet{Alcock1984}; the \ce{Ta2O5} vapor pressure data were previously published by \citet{Adkison2020}.}
	\label{Fig_VaporPressure_Ta_Ta2O5}
\end{figure}

\begin{figure}[h]
	\centering
	\includegraphics[width=\columnwidth]{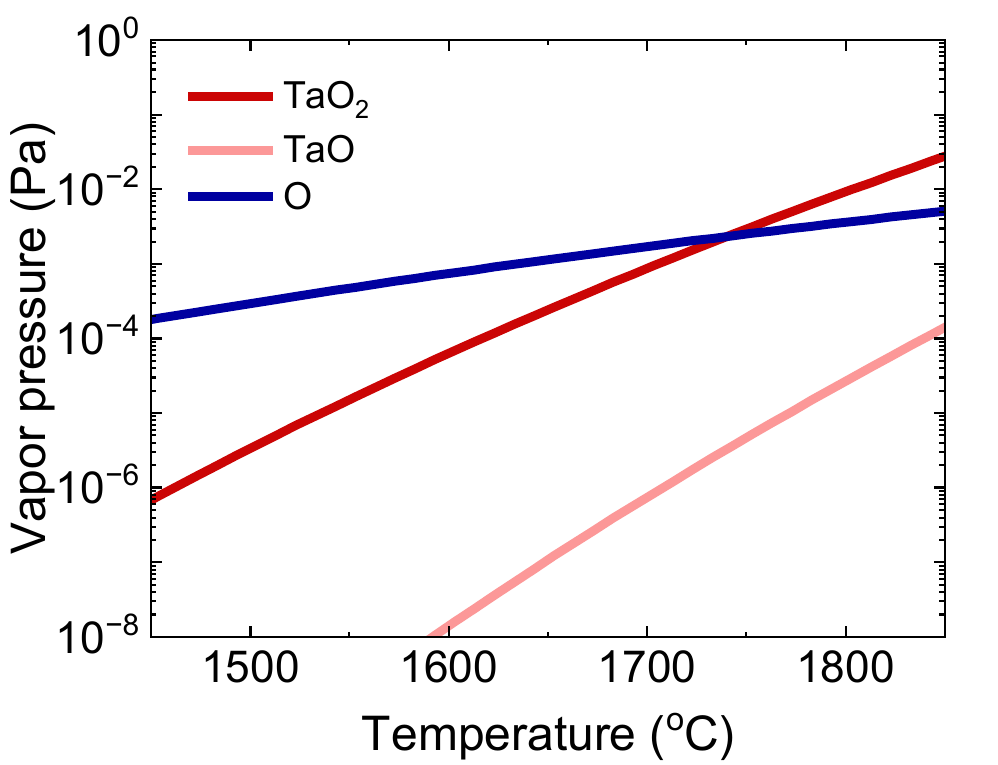}
	\caption{Partial vapor pressure of species emanating from a \ce{Ta2O5} charge as a function of temperature showing the dominance of \ce{TaO2} $\gtrsim$ \SI{1740}{\celsius}. The same data were previously published by \citet{Adkison2020}.}
	\label{Fig_VaporPressureComponents_Ta2O5}
\end{figure}

%\clearpage
%\section{Thermodynamics}
%~

\clearpage
\section{Epitaxial relationship between rutile and $r$-plane sapphire \label{sec:misfit}}
As the only two commercially available rutile substrates \ce{TiO2} and \ce{MgF2} are not stable under the high temperatures (around \SI{1000}{\celsius}) and reducing conditions (below \SI{1E-8}{\Torr}) necessary to crystallize \ce{TaO2}, we  had to search for alternative substrates that may enable heteroepitaxial growth of rutile. We selected $r$-plane sapphire, which shows a similar surface motif to rutile (101).

% Al2O3         TaO2
% 5.13          5.667
% 4.762         4.752

As the cross-sectional schematics in Fig. \ref{Fig_Cartoon} illustrate, the heteroepitaxial interface between \ce{Al2O3} ($1\bar{1}02$) and \ce{TaO2} (101) is highly anisotropic. While the lattice mismatch in one direction is very small (see \cref{Fig_Cartoon} (a)) resulting in a commensurately strained epitaxial film, the mismatch is so big in the orthogonal direction (see \cref{Fig_Cartoon} (b)) that it leads to periodic defects. 
The commensurate in-plane direction is \ce{TaO2} [010] ($b_\text{rutile}$), which is parallel to \ce{Al2O3} [$11\bar{2}0$].
The incommensurate in-plane direction is \ce{TaO2} [$\bar{1}01$], which is parallel to \ce{Al2O3} [$\bar{1}101$]. The misfit strains\cite{Freund2004} in the two directions are:

\[
\setlength{\jot}{12pt}  % or try 2pt, 4pt, etc.
\begin{aligned}
    \epsilon &= \frac{a_\text{substrate} - a_\text{film}}{a_\text{film}} \\
    \epsilon_\text{com}   &= \frac{4.762 - 4.752}{4.752} &&\approx \phantom{-}0.2 \% \\
    \epsilon_\text{incom} &= \frac{5.130 - 5.667}{5.667} &&\approx -9.5 \%
\end{aligned}
\]

where $a_\text{film}$ is the stress-free lattice parameter of the film material and $a_\text{substrate}$ is the lattice parameter of the substrate.
For \ce{TaO2}, we use the values published by \citet{Syono1983} (PDF number 01-079-9603).

\begin{figure}[h]
	\centering
	\includegraphics[width=0.6\columnwidth]{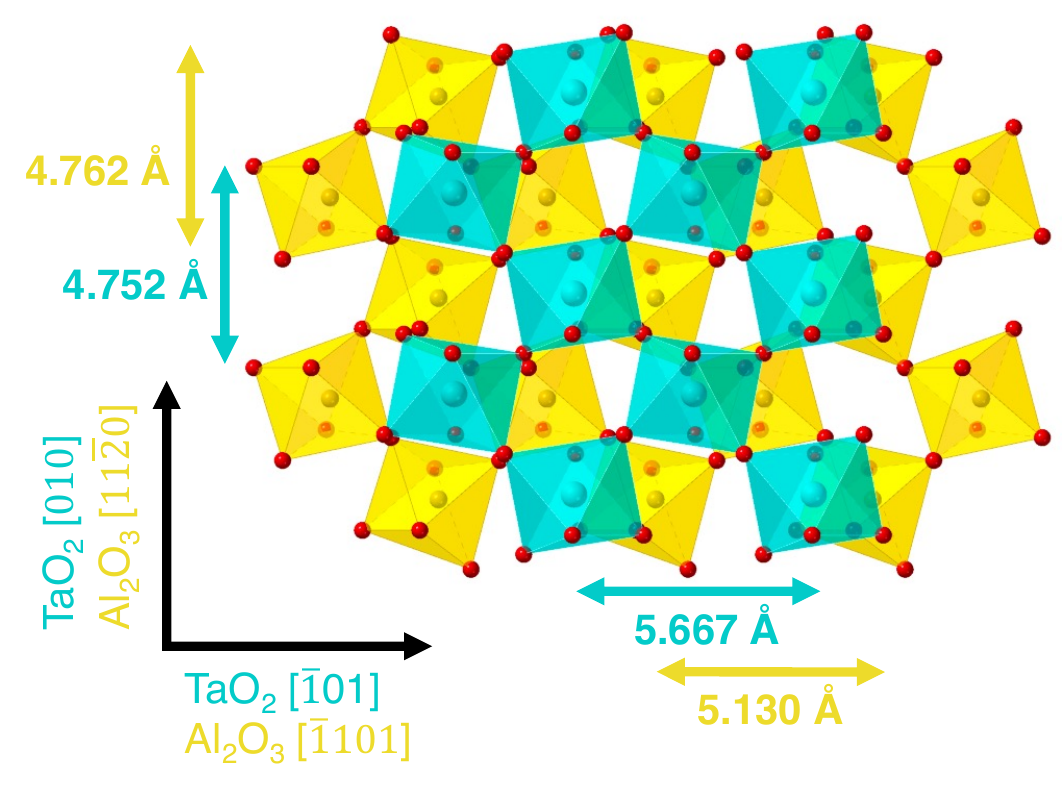}
	\caption{Schematic top view of \ce{TaO2} (101) on \ce{Al2O3} ($1\bar{1}02$). For clarity, only one monolayer of the substrate and the film are depicted. The indicated lattice parameters are unstrained bulk values.}
	\label{Fig_Cartoon_Topview}
\end{figure}

The experimental data confirming this anisotropic strain states are shown in \cref{Fig_RSM_Lab,Fig_RSM_overlay,Fig_STEM}.
A schematic top-view of the interfacial alignment of \ce{TaO2} (101) on \ce{Al2O3} ($1\bar{1}02$) is shown in \cref{Fig_Cartoon_Topview}.

~
\clearpage
\section{Atomic Force Microscopy}

\begin{figure}[tbh]
	\centering
	\includegraphics[width=\columnwidth]{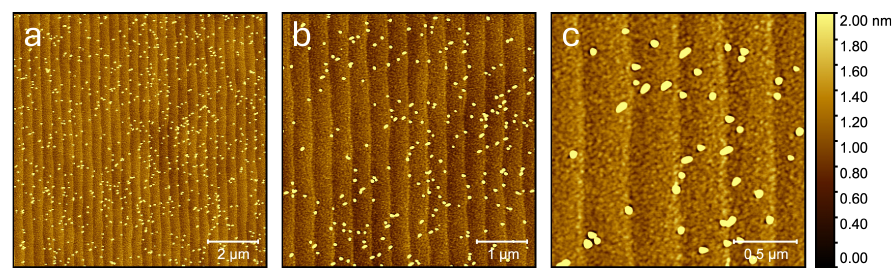}
	\caption{AFM of MBE-grown \ce{TaO2} (101) on \ce{Al2O3} ($1\bar{1}02$). Field of view 10x10 (a), 5x5 (b), and 2x\SI{2}{\micro\meter\squared} (c).}
	\label{fig:AFM}
\end{figure}
AFM topography images reveal that the step-and-terrace structure of the underlying \ce{Al2O3} (1$\bar{1}$02) substrate is preserved during \ce{TaO2} film growth by $S$-MBE. The nominal thickness of the \ce{TaO2} film is \SI{11.4}{\nano\meter}.
The surface appears smooth, except for a few nanometer-sized particles.

%%%%%% XRD
~
\clearpage
\section{X-ray diffraction of TLE-grown \ce{TaO2} thin films on $c$-plane sapphire}

\begin{figure}[h]
    \centering
    \includegraphics[width=0.9\linewidth]{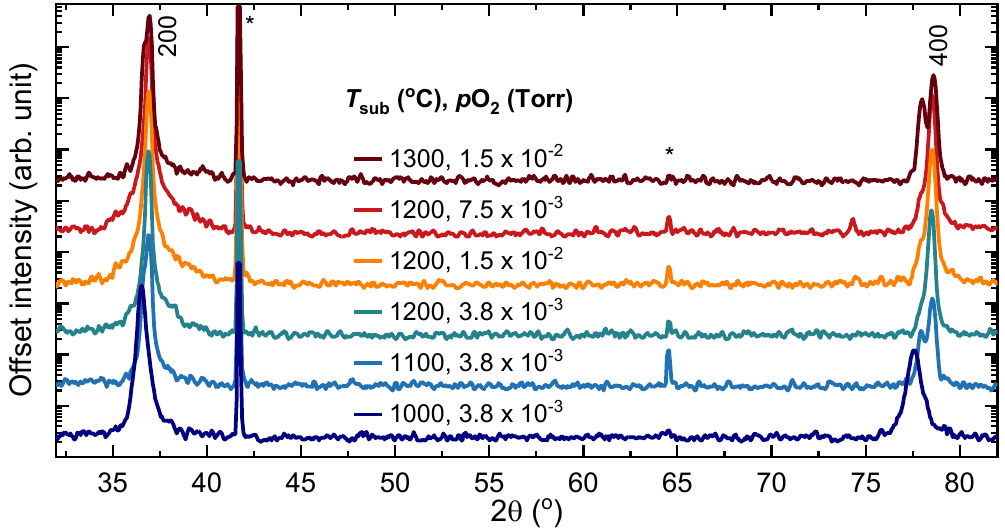}
    \caption{XRD $\theta-2\theta$ scans of TLE-grown \ce{TaO2} (100) on \ce{Al2O3} (0001) confirming a wide temperature and pressure window for the formation of the rutile phase. Asterisks indicate the allowed 0006 and the forbidden 0009 peaks of the sapphire substrate.}
    \label{fig:XRD_TLE_c-plane}
\end{figure}

\section{X-ray diffraction of MBE- and TLE-grown \ce{TaO2} thin films on $r$-plane sapphire}
\begin{figure}[h]
    \centering
    \includegraphics[width=0.9\linewidth]{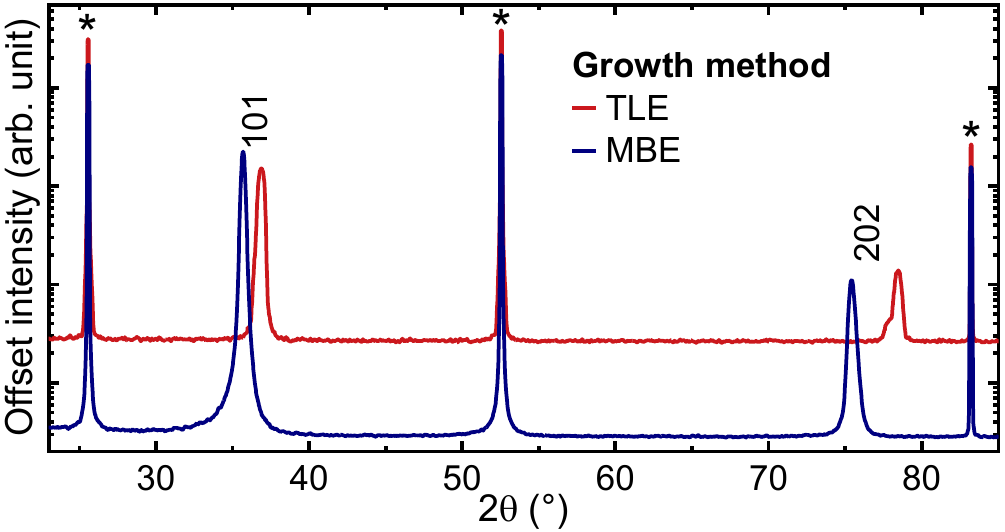}
    \caption{XRD $\theta-2\theta$ scans of MBE- and TLE-grown \ce{TaO2} (101) on \ce{Al2O3} (1$\bar{1}$02). Asterisks indicate the substrate peaks. The TLE growth was not optimized on this substrate orientation. This TLE sample was grown at \SI{1100}{\celsius} and \SI{3.8e-3}{\Torr}.}
    \label{fig:XRD_MBE_TLE}
\end{figure}

~
\section{X-ray diffraction of MBE-grown \ce{TaO2} thin films of various thicknesses on $r$-plane sapphire}

\begin{figure}[h]
    \centering
    \includegraphics[width=\linewidth]{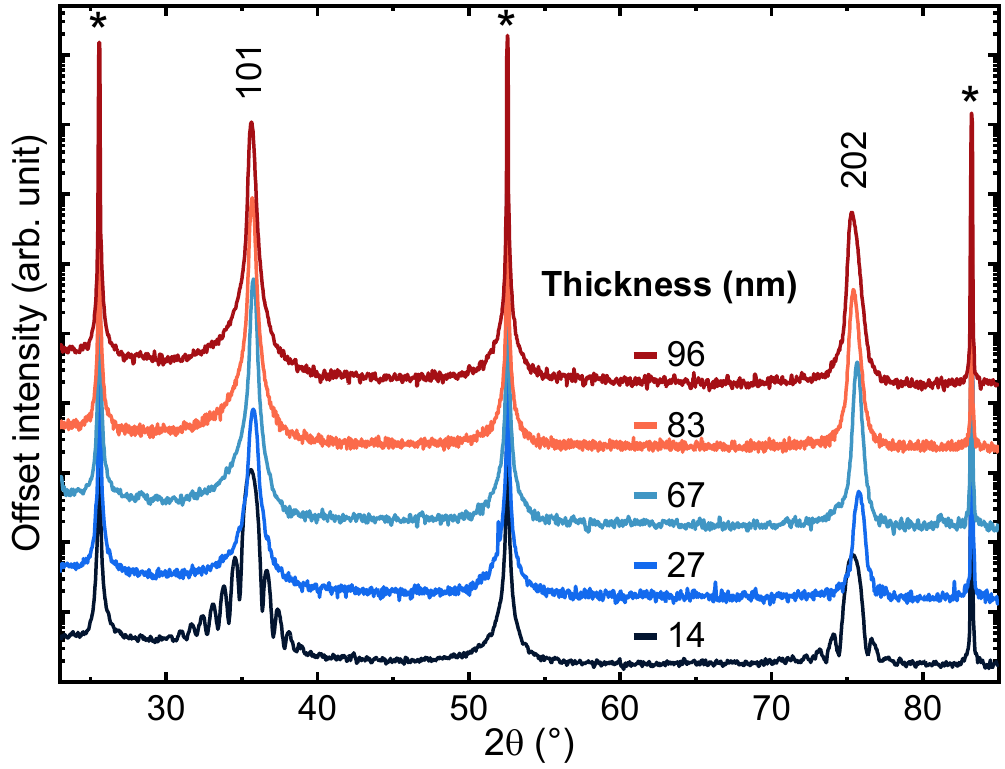}
    \caption{XRD $\theta-2\theta$ scans of MBE-grown \ce{TaO2} (101) of various thicknesses on \ce{Al2O3} (1$\bar{1}$02) grown at $T_\text{sub} \approx$ 1050 - \SI{1100}{\celsius}.}
    \label{fig:XRD_Thicknesses}
\end{figure}

~
\clearpage
\section{X-ray diffraction phi scans}
~
\begin{figure}[h]
    \centering
    \includegraphics[width=0.5\linewidth]{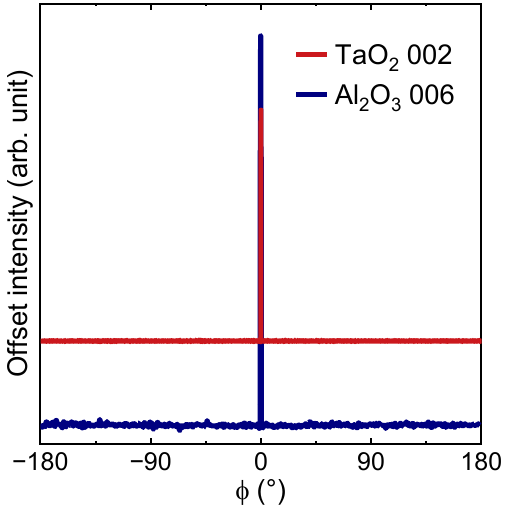}
    \caption{XRD $\phi$-scans of MBE-grown \ce{TaO2} (101) on \ce{Al2O3} (1$\bar{1}$02) confirming the untwinned nature of the film. The rutile and corundum $c$-axes point in the same in-plane direction.}
    \label{fig:PhiScans}
\end{figure}

\section{RHEED}
\begin{figure}[h]
	    \centering
	    \includegraphics[width=0.5\linewidth]{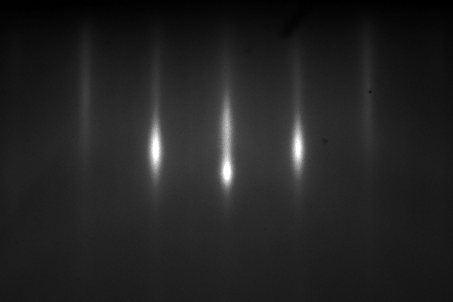}
	    \caption{Reflective high-energy electron diffraction (RHEED) image of MBE-grown \ce{TaO2} taken at room temperature along the [010] direction of the film.}
	    \label{fig:RHEED}
	\end{figure}

\clearpage
\section{Reciprocal space maps}
\begin{figure}[th]
	\centering
	\includegraphics[width=\columnwidth]{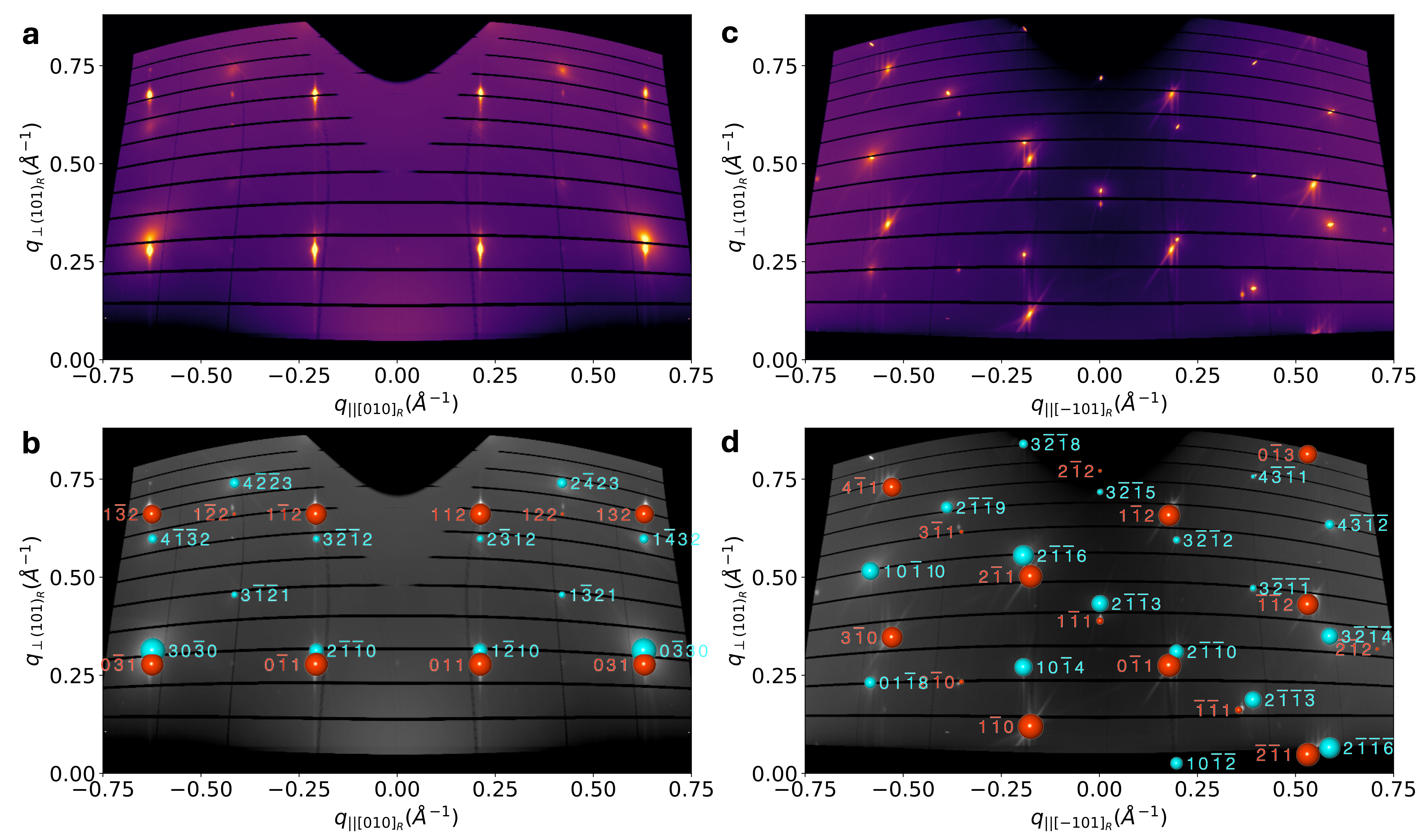}
	\caption{Synchrotron X-ray reciprocal space maps on a false-color scale (a, c) and grayscale (b, d). Lower panels show superimposed single-crystal diffraction simulations. \ce{TaO2} peaks are depicted and labeled in red, \ce{Al2O3} peaks in cyan.}
	\label{Fig_RSM_overlay}
\end{figure}

To verify the long-range order and orientational alignment inferred from STEM imaging and laboratory-source XRD measurement, we performed additional synchrotron X-ray measurements, mapping a wide range of three-dimensional reciprocal space.
Figure \ref{Fig_RSM_overlay} shows two orthogonal slices of the 3D dataset. Panels (a, b) depict a $K-L$ plot at $H=1$ (index referring to the rutile \ce{TaO2} film), panels (c, d) depict a $H-L$ plot at $K=1$.
To aid navigation in these feature-rich plots, we performed single crystal diffraction simulations using the program SingleCrystal (version 5.1.1) by CrystalMaker Software Ltd.

The fact that the simulations can uniquely index every experimentally observed peak proves the untwinned nature of the \ce{TaO2} film. The reason for this is the unique epitaxial match between rutile and $r$-plane sapphire, which lacks rotational symmetry at the interface that would lead to twinning.

For panels (a, b) in \cref{Fig_RSM_overlay}, the view direction is parallel to the \ce{TaO2} lattice vector $[\bar{1}01]$ while the vertical direction is normal to the (101) plane. The substrate view direction is parallel to the \ce{Al2O3} lattice vector $[1\bar{1}0\bar{1}]$ while the vertical direction is normal to the ($1\bar{1}02$) plane.

For panels (c, d) in \cref{Fig_RSM_overlay}, the view direction is parallel to the \ce{TaO2} lattice vector $[0\bar{1}0]$ while the vertical direction is normal to the (101) plane. The substrate view direction is parallel to the \ce{Al2O3} lattice vector [$11\bar{2}0$] while the vertical direction is normal to the ($1\bar{1}02$) plane.

~
\clearpage
\section{High-temperature X-ray diffraction}
\begin{figure}[h]
	\centering
	\includegraphics[width=0.8\columnwidth]{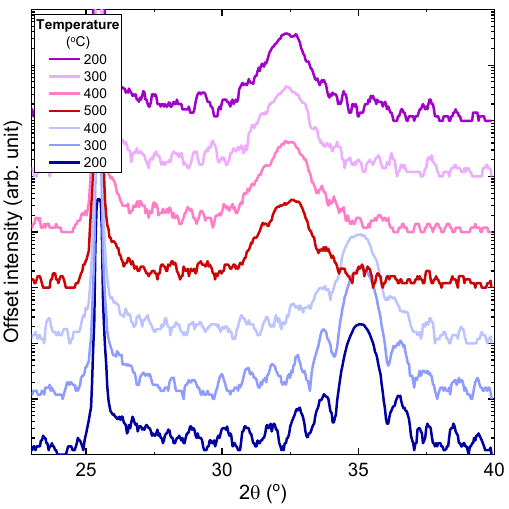}
	\caption{XRD at elevated temperature shows the irreversible oxidation of \ce{TaO2} to \ce{Ta2O5} in air above \SI{400}{\celsius}. From bottom to top, the figure shows scans during heating and subsequent cooling measured with a Rigaku SmartLab diffractometer equipped with an Anton Paar DHS 1100 domed hot stage and a monochromated copper anode.}
	\label{Fig_XRD_Heating}
\end{figure}

To test the stability of \ce{TaO2} in air at elevated temperatures, we performed \textit{operando} XRD measurements using a temperature-controlled hot stage. The results are summarized in Fig. \ref{Fig_XRD_Heating}. The data shows that above \SI{300}{\celsius}, \ce{TaO2} oxidizes to \ce{Ta2O5}. This reaction is irreversible and the sample remains \ce{Ta2O5} upon subsequent cooling.

\begin{comment}
\clearpage
\section{XRR}

\begin{figure}[h]
	\centering
%	\includegraphics[width=0.75\columnwidth]{XRR_TaO2_SiOx.pdf}
	\caption{X-ray reflectivity data (red circles) with overlaid fit (blue line) of the sample used for ellipsometry. See table \ref{tab:XRR_Fit} for details.}
	\label{Fig_XRR}
\end{figure}

\begin{table}[h]
    \caption{X-ray reflectivity fitting results of a capped film used for ellipsometry measurements. Fits were performed using PANalytical AMASS 2.0 software. The density of the substrate was fixed during the fitting procedure.}
    \label{tab:XRR_Fit}
    \centering
    \setlength{\tabcolsep}{12pt}
    \begin{tabular}{lccc}
        \toprule
        Layer & Thickness & Roughness & Density \\
        ~ &  (nm) &  (nm) &  (\SI{}{\gram\per\cubic\centi\meter}) \\
        \midrule
        \ce{SiO_x} capping layer     & \SI{10.64(17)}{} & \SI{0.91(11)}{}  & \SI{2.26(13)}{}  \\
        \ce{TaO2} film               & \SI{30.93(11)}{} & \SI{0.40(1)}{}    & \SI{9.60(12)}{}  \\
        \ce{Al2O3} substrate         & $\infty$         & \SI{0.41(3)}{}    & 3.989  \\
        \bottomrule
    \end{tabular}
\end{table}

\end{comment}

~
\clearpage

\begin{comment}
\clearpage
\section{XPS}
While most metal oxides would reduce in the ultra-high vacuum chamber at elevated temperatures, \ce{TaO2} instead oxidizes to \ce{Ta2O5}.
The latter is consistent with thermodynamic data available for bulk Ta on the equilibrium between the metal and its stable pentavalent oxide. 

Fig. \ref{Fig_Ellingham_Ta2O5_SiO2_MgO} shows 
Ellingham diagram for the reaction

$$ \frac{4}{3} \ce{Ta} + \ce{O_2} \rightarrow \frac{2}{5} \ce{Ta_2O_5}$$
\end{comment}

\begin{comment}
\clearpage
\section{\ce{Mg2SiO4} as a novel substrate for \ce{TaO2}}

To select a novel substrate for \ce{TaO2}, we looked for an oxide compound that consists of ions that are stable in reducing conditions, i.e., in an ultrahigh vacuum chamber at elevated temperatures.
To guide our selection, we examined binary oxide Ellingham diagrams.

We noticed that the olivine structure bears some resemblance to rutile. 
\ce{Mg2SiO4} can be thought to consist of MgO and \ce{SiO2}. 
However, as the experiment would reveal, this approach based on thermodynamic data on binary oxide constituents is an oversimplification.
As the STEM and XPS data revealed, \ce{Mg2SiO4} decomposed and instead of a rutile \ce{TaO2} film, trirutile \ce{MgTa2O6} formed instead.

\begin{figure}[h]
	\centering
%	\includegraphics[width=\columnwidth]{Ta_Si_Mg_Ellingham.png}
	\caption{Ellingham diagrams of the stable oxides of tantalum, silicon, and magnesium}
	\label{Fig_Ellingham_Ta2O5_SiO2_MgO}
\end{figure}

\end{comment}

\clearpage
\section{HAXPES}
\begin{figure}[h]
	\centering
	\includegraphics[width=0.62\columnwidth]{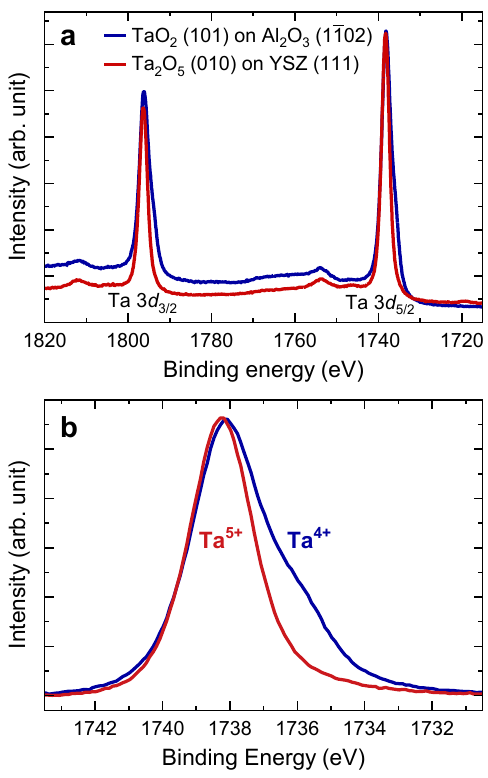}
	\caption{Tantalum \textit{3d} HAXPES data collected using a Scienta Omicron system with a monochromated gallium liquid metal jet source operating at \SI{9.25}{\kilo\electronvolt}. The \ce{TaO2} data shows asymmetric peaks originating from the pentavalent surface layer with shoulders on the low binding energy side which we assign to the tetravalent interior of the film. The \ce{Ta2O5} thin film was fabricated via suboxide MBE while applying a flux of ozone gas. Both samples were exposed to air for several months prior to the measurements explaining the strong \ce{Ta^{5+}} signal of the \ce{TaO2} sample despite the increased information depth of HAXPES compared to soft X-ray photoelectron spectroscopy. Panel (b) shows a magnification of the tantalum $3d_{5/2}$ region after Shirley-background \cite{Shirley1972} subtraction.}
	\label{Fig_HAXPES_Ta3d_Ga}
\end{figure}

%\textcolor{blue}{Variable-temperature data from Twente}

\begin{figure}[h]
	\centering
	\includegraphics[width=\columnwidth]{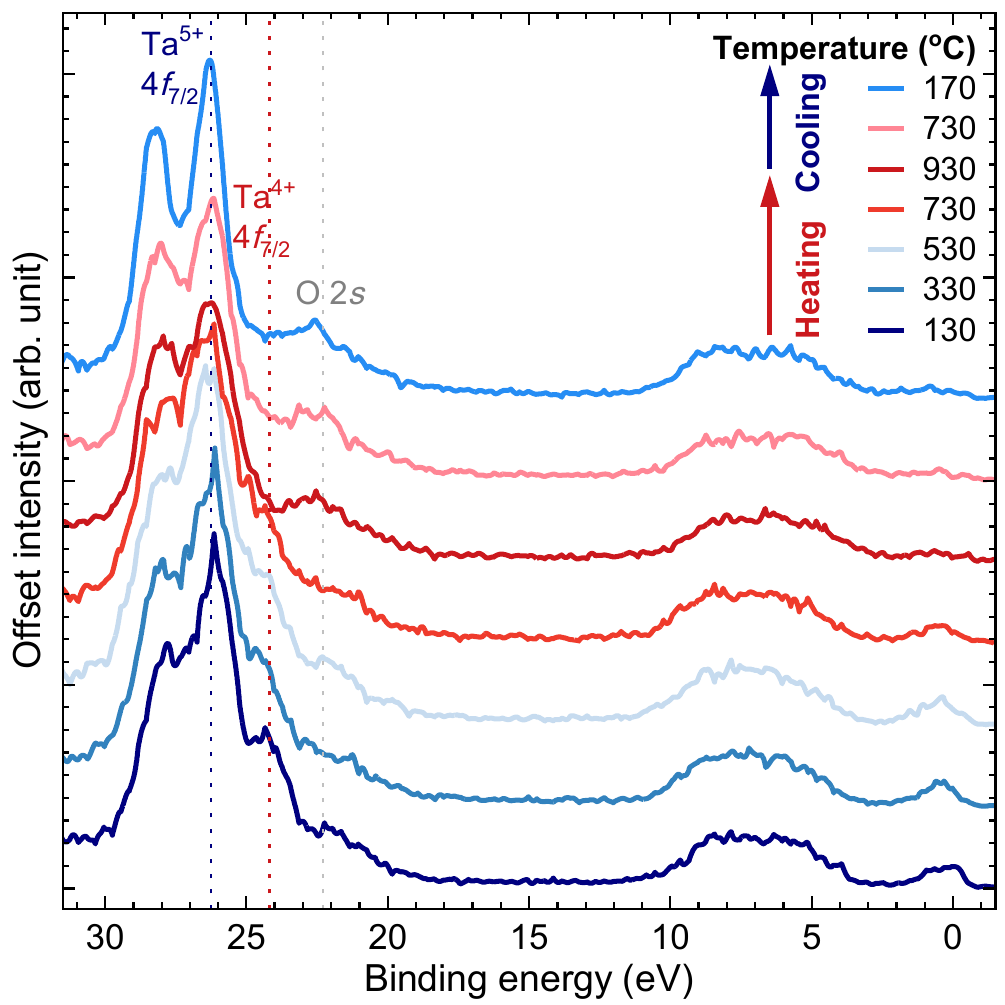}
	\caption{Variable-temperature \textit{operando} HAXPES. Tantalum \textit{4f}, oxygen \textit{2s} and valence band data collected using a SPECS system operating at \SI{5.4}{\kilo\electronvolt} using a chromium anode. The sample was first heated and then cooled as spectra were measured. Dashed vertical lines are guides to the eye. Around \SI{930}{\celsius}, \ce{TaO2} irreversibly oxidized to \ce{Ta2O5} in UHV conditions. Residual gas analysis indicated $p$\ce{H2O} and $p$\ce{O2} $<$ \SI{1E-8}{\Torr}. The sample was exposed to air for several months prior to the measurements. Even at room temperature, the sample showed a strong \ce{Ta^{5+}} contribution due to the surface oxidation of \ce{TaO2}.}
	\label{Fig_HAXPES_Ta3d_Cr}
\end{figure}

\clearpage
\section{Oxygen K-edge EELS}

\begin{figure}[h]
	\centering
	\includegraphics[width=0.75\columnwidth]{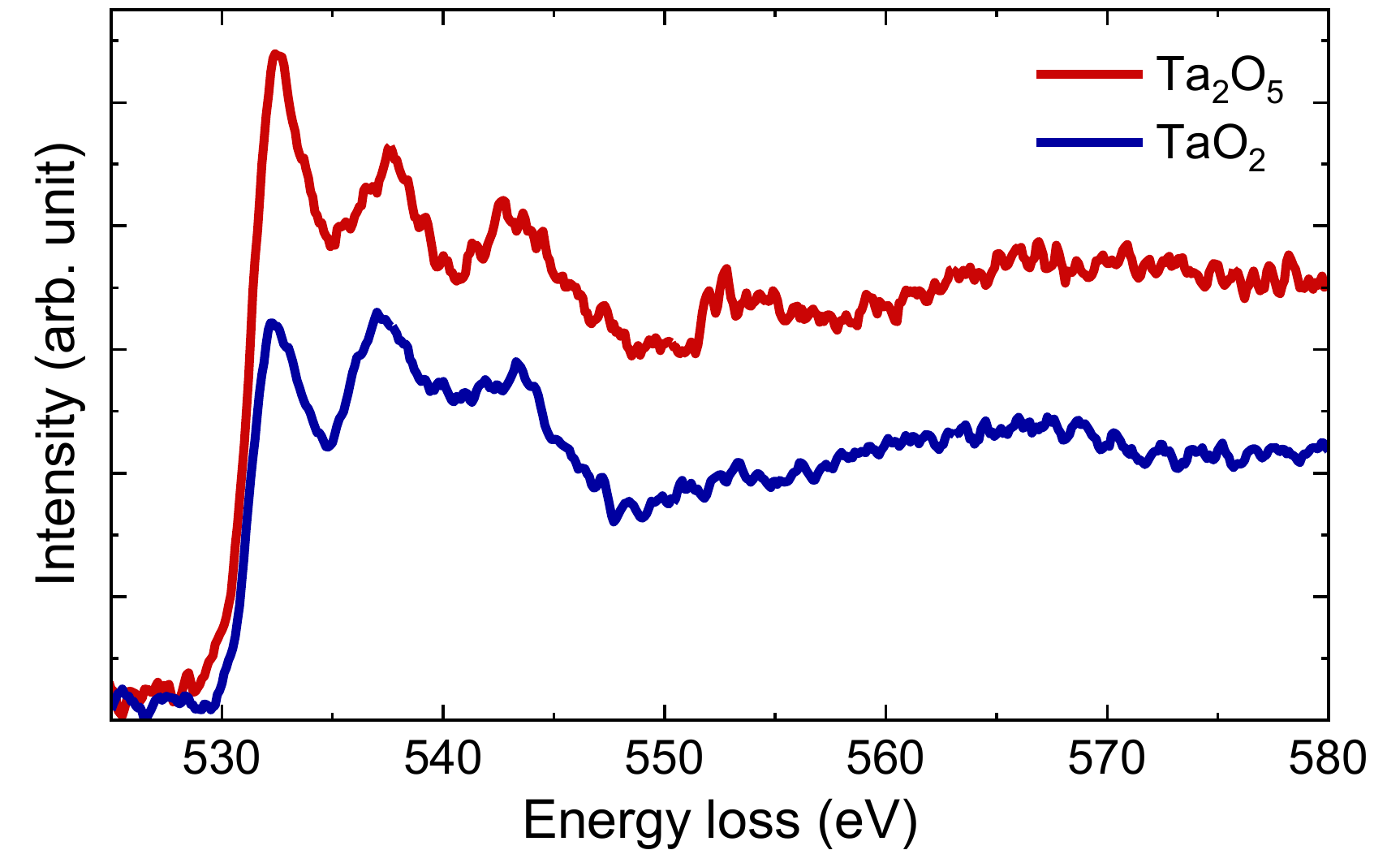}
	\caption{Oxygen $K$-edge electron energy-loss spectroscopy comparing \ce{TaO2} and \ce{Ta2O5}. The key difference is the intensity of the first peak, commonly ascribed to be indicative of the $t_{2g}$ of the metal $d$-band filling.}
	\label{Fig_O-K_TaO2_Ta2O5}
\end{figure}

\section{Oxygen K-edge simulation}
X-ray absorption spectroscopy (XAS) simulations were carried out using density-functional theory (DFT) \cite{Hohenberg1964, Kohn1965} within the supercell core-hole approach,\cite{Karsai2018} as implemented in the Vienna ab initio simulation package (VASP) \cite{Kresse1996} with projector-augmented wave (PAW) pseudopotentials.\cite{Blochl1994}
Supercells of \ce{TaO2} (\ce{Ta72O144}) and \ce{Ta2O5} (\ce{Ta72O180}) were used in all calculations. 
Approximately 4000 bands were included in the core-hole calculations. 
A plane-wave energy cutoff of \SI{400}{\electronvolt} was employed, and the Brillouin zone was sampled using a $k$-point grid with a density of 6000 divided by the number of atoms in the supercell.
The XAS spectra were obtained from the imaginary part of the frequency-dependent dielectric function. To account for instrumental resolution and core-hole lifetime effects, the calculated spectra were convoluted with a Gaussian function with a full width at half maximum (FWHM) of \SI{0.3}{\electronvolt} and a Lorentzian function with a linearly increasing width from 0.0 to \SI{0.7}{\electronvolt}. Several exchange–correlation functionals were tested; results presented here were obtained using the local density approximation (LDA) \cite{Ceperly1980, Perdew1981} with spin polarization.

\begin{figure}[h]
	\centering
	\includegraphics[width=0.75\columnwidth]{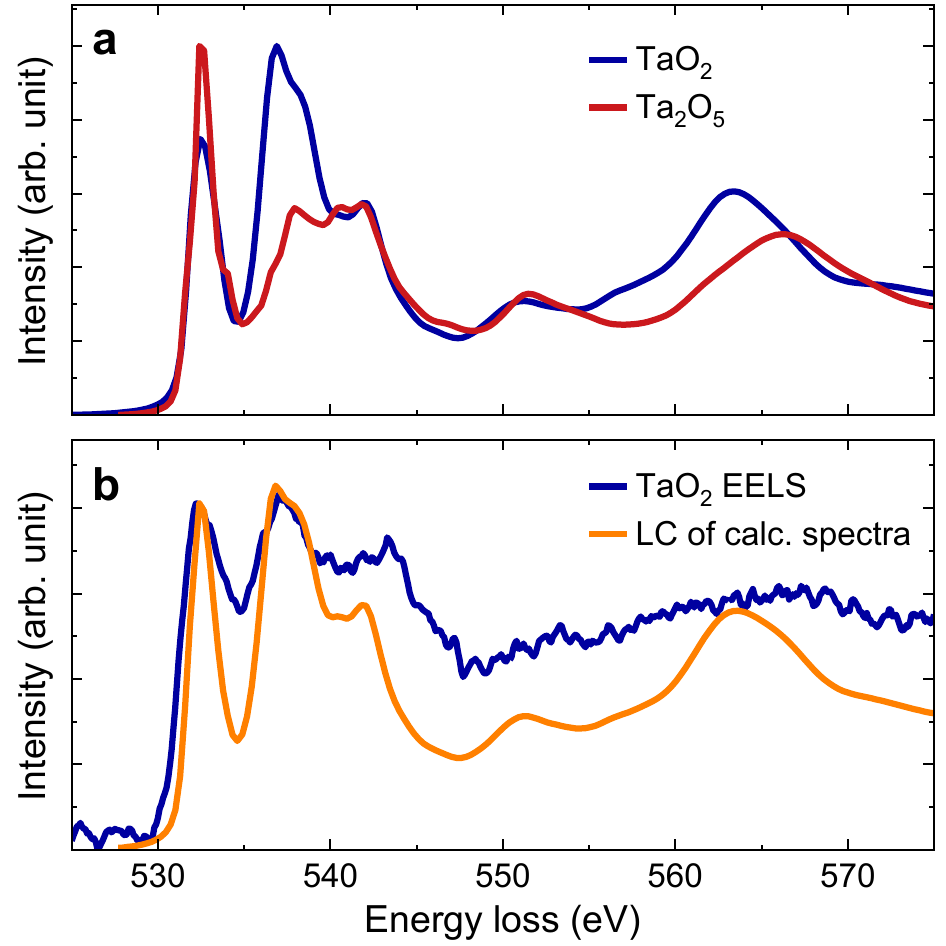}
	\caption{Calculated oxygen $K$-edge electron energy-loss spectra for \ce{TaO2} and \ce{Ta2O5} (a), comparison of experimental EELS data (as shown in Fig. \ref{Fig_O-K_TaO2_Ta2O5}) with a linear combination (LC) of calculated spectra. The LC consists of 75 \% \ce{TaO2} and 25 \% \ce{Ta2O5} to mimic the unpreventable surface oxidation of \ce{TaO2}.}
	\label{Fig_O-K_exp_dft}
\end{figure}

\Cref{Fig_O-K_exp_dft} (a) shows a comparison of simulated XAS (or EELS) spectra for \ce{TaO2} (in blue) and \ce{Ta2O5} (in red). The key difference in the simulations -- as for the experimental data shown in Fig. \ref{Fig_O-K_TaO2_Ta2O5} -- is the ratio of the first peak (around \SI{532}{\electronvolt}) compared to the second peak (around \SI{537}{\electronvolt}).
\cref{Fig_O-K_exp_dft} (b) shows a comparison of the experimental EELS data of a \ce{TaO2} lamella with a linear combination of \ce{TaO2} and \ce{Ta2O5} simulations. The latter combination was simulated to mimic the oxidation of \ce{TaO2} in air.
While the ratio of \ce{TaO2} and \ce{Ta2O5} was chosen to match the ratio of the first two peaks, the remainder of the O $K$-edge spectrum was not optimized.

~
\clearpage
\section{Spectroscopic ellipsometry \label{sec:Ellipsometry}}
\subsection{$r$-plane \ce{Al2O3} substrate}
\begin{figure}[h]
    \centering
    \includegraphics[width=0.75\linewidth]{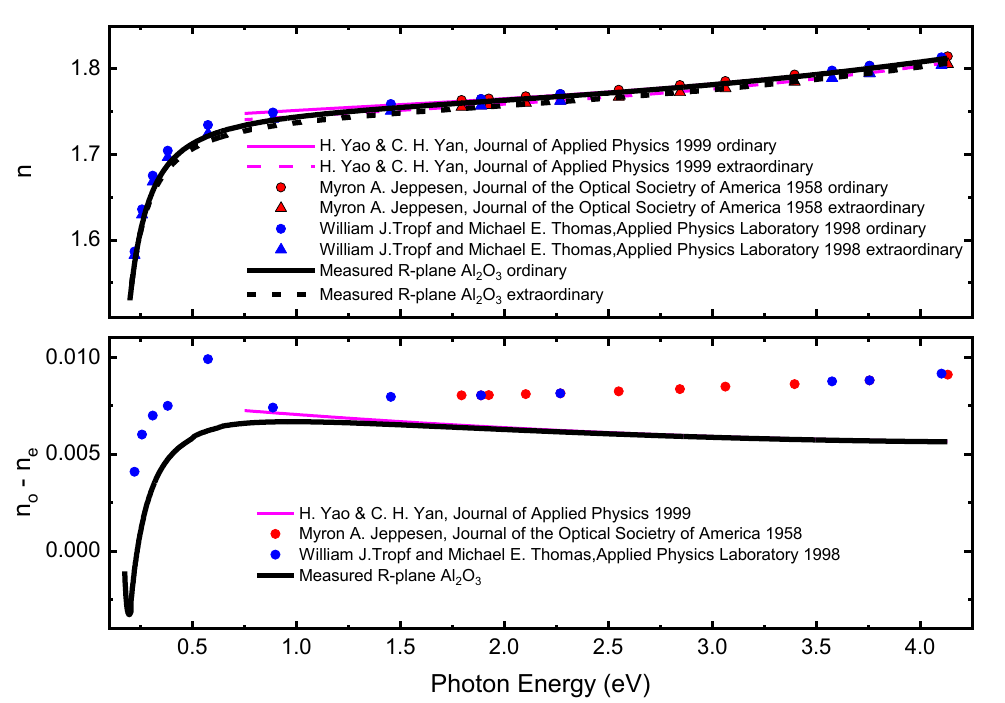}
    \caption{Index of refraction ($n$) for \ce{Al2O3} in the ordinary ($n_o$) and extraordinary ($n_e$) direction, and birefringence ($n_o - n_e$) spectra obtained in this work compared to that reported in literature.}
    \label{fig:Al2O3_Ellipsometry}
\end{figure}

For the generalized ellipsometric spectra collected for the $r$-plane \ce{Al2O3} substrate, all four $\phi$ rotations are analyzed simultaneously \cite{Amonette2023, Jellison2022} to determine surface layer thickness and \ce{Al2O3} complex optical properties by fitting a structural and optical model to experimental ellipsometric spectra with an iterative least square regression to minimize the mean square error (MSE). \cite{Johs2008}
The structural model for the \ce{Al2O3} substrate consists of a semi-infinite single crystal substrate / surface layer / air ambient with the optical response of the surface layer described by an anisotropic Bruggeman effective medium approximation (EMA)\cite{Aspnes1982, Schmidt2013}
consisting of equal parts \ce{Al2O3} and void, resulting in a surface layer thickness of \SI{1.80\pm0.03}{\nano\meter}. The polar Euler angle ($\theta$) is fixed to \SI{57.6}{\degree} which is the angle the $c$-axis relative to the $r$-plane surface. The sample has been rotated by nominally \SI{45}{\degree} increments with the azimuthal Euler angle $\phi$.
The parametric model to describe the optical response of \ce{Al2O3} in both the ordinary and extraordinary directions consists of the parameterization developed by \citet{YaoYan1999}
with the addition of a Sellmeier \cite{Tompkins2005} 
expression for both the ordinary and extra ordinary directions to account for contributions from phonon modes at photon energies below the measured spectral range.\cite{Harman1994}
The anisotropic Bruggeman EMA utilizes the same directionally dependent complex optical properties and Euler angles as the bulk film.\cite{Schmidt2013}
The model is used to fit the experimentally measured ellipsometric spectra using a least squares regression with an unweighted MSE. A good quality of fit is obtained with a low MSE = 8 × $10^{-3}$. The experimentally determined optical properties are in reasonable agreement with literature, with the index of refraction ($n = \epsilon_1^{1/2}$ when $\epsilon_2 = 0$) for electric fields oscillating parallel to the ordinary and extraordinary directions and birefringence shown in Fig. \ref{fig:Al2O3_Ellipsometry}.\cite{Tropf1997, YaoYan1999, Jeppesen1958} 

\subsection{\ce{TaO2} film}
Generalized ellipsometric spectra were collected at four different sample orientations with both ellipsometers (see \cref{sec:Characterization}) to determine the structural and optical properties of the \ce{TaO2} film deposited on an $r$-plane \ce{Al2O3} substrates. The optical properties of \ce{Al2O3} were fixed from measurements of an uncoated $r$-plane \ce{Al2O3} substrate. \ce{TaO2} was modeled with a rutile crystal structure with the polar Euler angle $\theta$ fixed to \SI{33.8}{\degree}, corresponding to the angle between the $c$-axis and the $(101)$ sample surface plane.

Divided spectral range analysis \cite{Gautam2014,Adhikari2016,Amonette2023,Ghimire2015,Barone2022} was applied to determine the structural model and spectra of the dielectric function $\epsilon$ for electric fields oscillating parallel to the ordinary and extraordinary directions. Divided spectral range analysis involves simultaneously modeling the transparent spectral region and the highly absorbing spectral region of the material, each using a physically realistic parametric model to describe $\epsilon$ within that region while keeping the structural model common and excluding the weakly absorbing region. This approach avoids model-dependent bias in the weakly absorbing region near the bandgap.

The transparent region was determined to span \SIrange{0.20}{0.25}{\electronvolt}, while the highly absorbing region was selected to cover \SIrange{1.15}{4.13}{\electronvolt}. The transparent region was measured using the FTIR-VASE ellipsometer, whereas the highly absorbing region was measured using the V-VASE ellipsometer. For both the ordinary and extraordinary directions, the parametric model describing $\epsilon$ for \ce{TaO2} in the transparent region consisted of a Sellmeier expression and a constant additive term to the real part of the complex dielectric function, $\epsilon_1$, denoted $\epsilon_{\infty}$. In the highly absorbing region, the parametric model consisted of a Sellmeier expression, $\epsilon_{\infty}$, and a series of Lorentz oscillators \cite{Tompkins2005} to account for above-bandgap electronic transitions, with the number of oscillators depending on the film and crystallographic direction.

The structural model consisted of a \ce{Al2O3} substrate / interfacial layer / \ce{TaO2} / air ambient stack. The interfacial layer shared the same $\theta$ as the \ce{Al2O3} substrate and had a fixed thickness of \SI{1.8}{\nano\meter}, consistent with the surface layer established from the bare substrate measurement. The optical properties of the interfacial layer were described using an EMA \cite{Aspnes1982,Schmidt2013} consisting of equal fractions of \ce{Al2O3} and bulk \ce{TaO2}.

The \ce{TaO2} sample had a nominal thickness of \SI{82}{\nano\meter} as measured by X-ray reflectivity (XRR). Generalized spectroscopic ellipsometry fitting yielded a thicknesses of $\SI{81.7 \pm 0.1}{\nano\meter}$ for the same sample, in excellent agreement with the XRR results. Deviations from the nominal angle of incidence were allowed to vary collectively and were fit to $-\,\SI{0.11 \pm 0.01}{\degree}$  for the \SI{82}{\nano\meter} film in the FTIR measurements to account for differences in angle of incidence between the two ellipsometers. The azimuthal Euler angles $\phi_m$ were treated similarly to those of the \ce{Al2O3} substrate. The samples were rotated by nominal \SI{45}{\degree} increments between measurements, and each sample was placed in nominally identical orientations for each instrument.

\begin{figure}[h]
    \centering
    \includegraphics[width=0.65\linewidth]{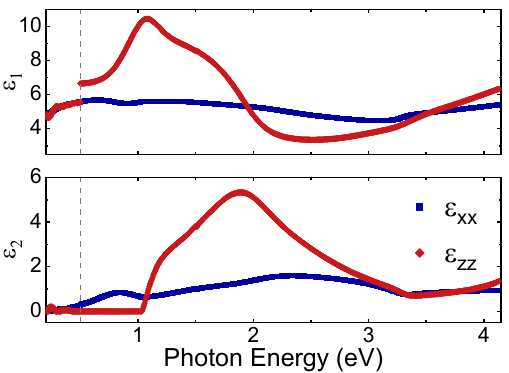}
    \caption{Real ($\epsilon_1$) and imaginary ($\epsilon_2$) parts of the complex dielectric function determined by numerical inversion of spectroscopic ellipsometry measurements in the ordinary ($\epsilon_{xx}$) and extraordinary ($\epsilon_{zz}$) directions. The dotted line indicates the transition between two different instruments.}
    \label{fig:Epsilon_numerically_inverted}
\end{figure}

\begin{figure}[h]
    \centering
    \includegraphics[width=0.65\linewidth]{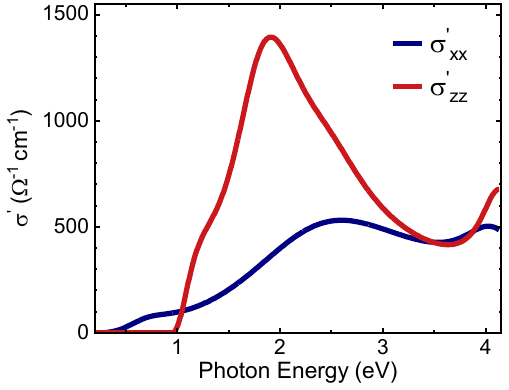}
    \caption{Real part ($\sigma^\prime$) of the optical conductivity $\sigma = \sigma^\prime + i \sigma^{\prime\prime}$ for an MBE-grown \SI{82}{\nano\meter} thick \ce{TaO2} film on $r$-plane sapphire in both the ordinary ($\sigma^\prime_{xx}$) and the extraordinary ($\sigma^\prime_{zz}$) directions. The optical conductivity was obtained from the ellipsometry data shown in Fig. \ref{fig:Epsilon}.}
    \label{fig:TaO2_sigma}
\end{figure}

\begin{figure}[h]
    \centering
    \includegraphics[width=0.65\linewidth]{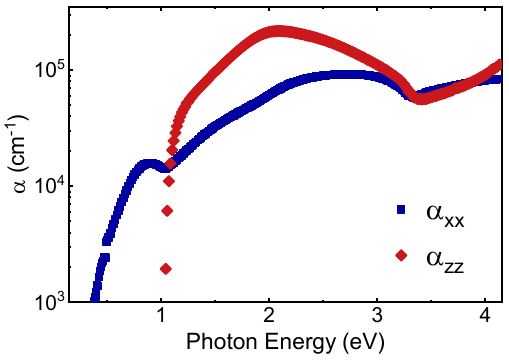}
    \caption{Absorption coefficient $\alpha$ for an MBE-grown \SI{82}{\nano\meter} thick \ce{TaO2} film on $r$-plane sapphire in both the ordinary ($\alpha_{xx}$) and the extraordinary ($\alpha_{zz}$) directions.} % as determined from numerically inverted $\epsilon$.}
    \label{fig:TaO2_alpha}
\end{figure}

The structural parameters and Euler angles were subsequently fixed and used as inputs for numerical inversion \cite{Oldham1969} across the entire spectral range, including the weakly absorbing region near the bandgap, to determine spectra of $\epsilon$ in both the ordinary and extraordinary directions. 
In the ordinary direction, a well-defined feature in $\epsilon_2$ was observed in the \SIrange{0.75}{0.85}{\electronvolt} range, along with a broader peak spanning \SIrange{2.1}{2.3}{\electronvolt}. The absorption onset in the ordinary direction occurred in the \SIrange{0.2}{0.3}{\electronvolt} range. In the extraordinary direction, features in $\epsilon_2$ were observed near \SI{1.2}{\electronvolt} and \SI{1.9}{\electronvolt}, with absorption onset occurring near \SI{1}{\electronvolt}.
These features are evident in the parametric and numerically inverted $\epsilon$ shown in \cref{fig:Epsilon} and \cref{fig:Epsilon_numerically_inverted}, the real part of the parametric optical conductivity ($\sigma'$) shown in \cref{fig:TaO2_sigma}, and the numerical inverted absorption coefficient ($\alpha$) shown in \cref{fig:TaO2_alpha}.

\begin{comment}
The dielectric functions obtained from numerical inversion were not Kramers--Kronig consistent. Possible contributing factors include deviations from the assumed polar Euler angle $\theta$, strain, compositional gradients, or artifacts arising from combining spectra collected with two different ellipsometers.     
\end{comment}

\clearpage
A series of Tauc--Lorentz oscillators \cite{Jellison1996} were fit to the measured ellipsometric spectra with the structural model fixed to values obtained from the divided spectral range analysis. The resulting $\epsilon$ spectra are shown in Fig. \ref{fig:Epsilon} in the main text and reproduce qualitatively similar features to the numerically inverted dielectric function while enforcing Kramers--Kronig consistency. The fitted parameters are summarized in Table \ref{tab:Tauc-Lorentz}.

The MSE values for the numerically inverted fit was $14 \times 10^{-3}$. Using the Tauc--Lorentz oscillator model yielded a nearly identical MSE value of $15 \times 10^{-3}$ for the same film.
%The small differences in MSE indicate limited sensitivity to the features responsible for the lack of Kramers--Kronig consistency in the numerical inversion results.

\begin{figure}[h]
    \centering
    \includegraphics[width=0.7\linewidth]{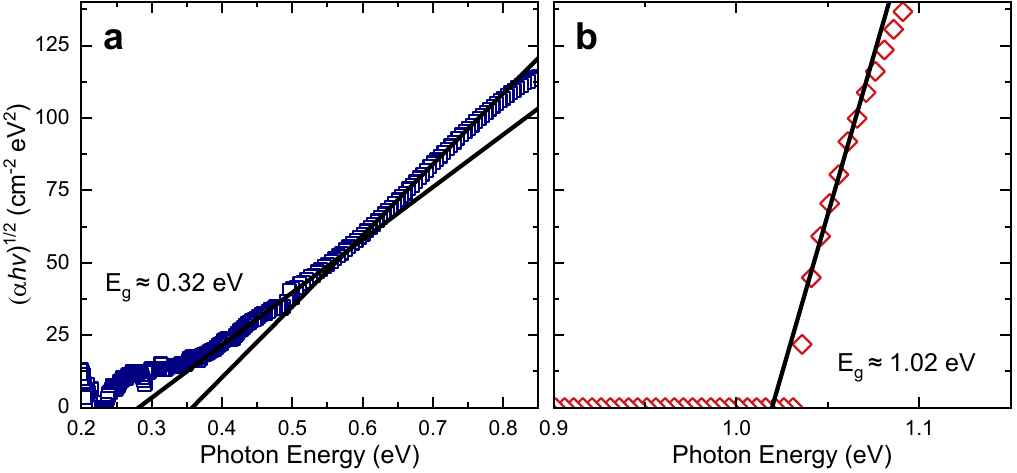}
    \caption{Tauc plots for the extraction of the lowest indirect transition energies. The linear extrapolation of $(\alpha h \nu)^{1/2}$ to zero for the \SI{82}{\nano\meter}-thick \ce{TaO2} film in the ordinary (a) direction indicates the smallest optical transition, which is interpreted as the bandgap. There are two distinct slopes in the ordinary direction indicating phonon-assisted absorption and the presence of an indirect transition, and the intercept of both slopes are averaged to produce the bandgap energy. Panel (b) shows the same fit for the extraordinary direction, but with only a single slope indicating a lack of phonon-assisted absorption. The intercept for this fit is \SI{1.02}{\electronvolt}.}
    \label{fig:Tauc_indirect}
\end{figure}

\begin{figure}[h]
    \centering
    \includegraphics[width=0.7\linewidth]{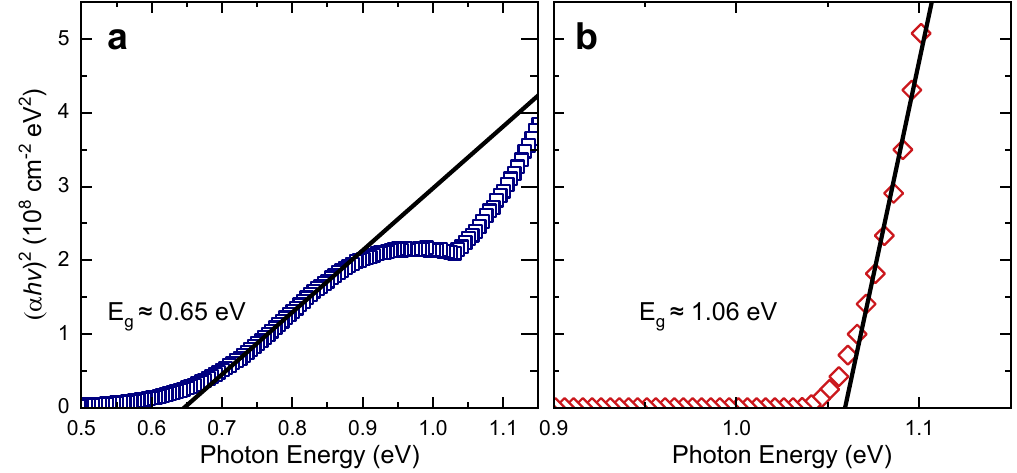}
    \caption{Linear extrapolation of $(\alpha h \nu)^2$ to zero for the \SI{82}{\nano\meter} thick \ce{TaO2} film in the ordinary (a) and extraordinary (b) directions. The intercepts identify the lowest direct transitions.}
    \label{fig:Tauc_direct}
\end{figure}

The absorption coefficient $\alpha$ was determined from the numerically inverted $\epsilon$ spectra for both crystallographic directions, as shown in Fig. \ref{fig:TaO2_alpha}. 
Analysis of the $(\alpha h\nu)^{1/2}$ spectra revealed two distinct slopes in the ordinary direction, indicating phonon-assisted absorption and the presence of an indirect bandgap \cite{Pankove1975}. \Cref{fig:Tauc_indirect} (a) shows linear extrapolations of $(\alpha h\nu)^{1/2} = 0$ for both slopes, with the indirect bandgap defined as the average of the intercepts. The \SI{82}{\nano\meter} thick film exhibited an indirect gap of $\SI{0.32}{\electronvolt}$. This indirect gap represents the lowest-energy optical transition and is therefore taken as the bandgap of the material.
When using the Tauc--Lorentz oscillator model (Table \ref{tab:Tauc-Lorentz}), the Tauc gap parameter in the ordinary direction was fit to $\SI{0.23}{\electronvolt}$ for the \SI{82}{\nano\meter} thick film. 
The $(\alpha h\nu)^2$ spectra were examined in both directions to identify the lowest direct transitions. \Cref{fig:Tauc_direct} (a) shows linear extrapolations of $(\alpha h\nu)^2 = 0$ in the ordinary direction, dominated by absorption peaks in the \SIrange{0.75}{0.85}{\electronvolt} range. The lowest direct transitions were determined to be \SI{0.65}{\electronvolt}.
\Cref{fig:Tauc_direct} (b) shows similar analysis for the extraordinary direction, yielding lowest direct transitions of $\SI{1.06}{\electronvolt}$.
Similar spectroscopic ellipsometry data were obtained for a \SI{96}{\nano\meter} thick sample (not shown).
The estimated uncertainty of the transition energies is on the order of \SI{0.1}{\electronvolt}, arising from the choice of range for linear extrapolations.

\begin{table}[h]
\centering
\caption{The Tauc-Lorentz amplitude ($A$), broadening ($\Gamma$), resonance energy ($E_0$) and bandgap energy ($E_g$) as well as the Sellmeier UV pole amplitude ($A_{UV}$), IR pole amplitude ($A_{IR}$), and the constant additive term ($\varepsilon_\infty$) that describe spectra in $\epsilon$ for the \ce{TaO2} in Fig. \ref{fig:Epsilon}. The UV pole has a fixed energy of \SI{6}{\electronvolt} and the IR pole has a fixed energy of \SI{0}{\electronvolt}. \label{tab:Tauc-Lorentz}}
\begin{tabular}{lcc}
\toprule
Parameter & Ordinary & Extraordinary \\
\midrule
$\varepsilon_{\infty}$            & $2.11 \pm 0.03$       & $0.78 \pm 0.05$ \\
$A_{\mathrm{UV}}$ (eV$^2$) & $72.5 \pm 0.9$        & $117 \pm 1$ \\
$A_{\mathrm{IR}}$ (eV$^2 \times 10^4$) & $466 \pm 8$          & $1070 \pm 10$ \\
$E_g$ (eV)              & $0.225 \pm 0.009$     & $0.95 \pm 0.01$ \\
$A_1$ (eV)              & $0.70 \pm 0.06$       & $33 \pm 8$ \\
$\Gamma_1$ (eV)         & $0.60 \pm 0.03$       & $0.42 \pm 0.03$ \\
$E_{01}$ (eV)           & $0.705 \pm 0.008$     & $1.06 \pm 0.03$ \\
$A_2$ (eV)              & $4.25 \pm 0.07$       & $16 \pm 1$ \\
$\Gamma_2$ (eV)         & $2.33 \pm 0.04$       & $0.85 \pm 0.04$ \\
$E_{02}$ (eV)           & $2.53 \pm 0.01$       & $1.812 \pm 0.007$ \\
$A_3$ (eV)              & $0.37 \pm 0.04$       & $1.1 \pm 0.3$ \\
$\Gamma_3$ (eV)         & $0.761 \pm 0.06$      & $0.84 \pm 0.09$ \\
$E_{03}$ (eV)           & $4.07 \pm 0.02$       & $2.50 \pm 0.03$ \\
$A_4$ (eV)              & --                    & $0.58 \pm 0.05$ \\
$\Gamma_4$ (eV)         & --                    & $0.47 \pm 0.03$ \\
$E_{04}$ (eV)           & --                    & $4.14 \pm 0.02$ \\
\bottomrule
\end{tabular}
\end{table}

\clearpage

\section{First-Principles Calculations}\label{sec:DFT}
\subsection{Computational Details}\label{sec:DFT_details}
In order to investigate the properties of TaO$_2$, first-principles density functional theory calculations were performed using VASP 6.2.0, \cite{VASP1,VASP2,VASP3} using the projector augmented-wave (PAW) method. We explored the properties with the following exchange-correlation functionals: local-density approximation (LDA), Perdew-Burke-Ernzerhof (PBE), and PBEsol, \cite{VASP_PAW} and found that the general features of TaO$_2$ were consistent regardless of the choice of functional (Sec. \ref{sec:DFT_functionals}). In contrast to experimental conditions, our calculations assume an infinite periodic crystal with zero epitaxial strain. Our calculations show that the (rutile-like) $P4_2/mnm$ high symmetry structure is metallic and that there exists a lower energy, lower symmetry $I4_1/a$ structure, which is insulating. The band gap in the $I4_1/a$ phase is sensitive to the value of the Hubbard $U$, see Sec. \ref{sec:DFT_U}. Our results are similar to the findings of Ref. \cite{Zhu_2015}, however these previous authors claim the lower symmetry structure has space group $P\bar{1}$ as opposed to our identification of $I4_1/a$.

The following states were included in the valence of the PAW potentials:  $5p^65d^46s^1$ for tantalum and $2s^22p^4$ for oxygen. A force convergence tolerance of $5\times{10^{-4}}$ eV/\AA\ was used for all calculations with a plane-wave energy cutoff of 900 eV. A 16$\times$16$\times$27 Monkhorst-Pack \textbf{k}-point grid was used for calculations of the $P4_2/mnm$ structure using the conventional 6-atom unit cell and an 11$\times$11$\times$11 \textbf{k}-point grid was used for the $I4_1/a$ structure in the primitive 24-atom unit cell. These values were chosen to converge the lattice constants of the $P4_2/mnm$ phase (which requires more stringent parameters as it is metallic) to within 0.5 picometers when compared to incrementing the \textbf{k}-point grid to 20$\times$20$\times$40 and the energy cutoff to 1900 eV.  Gaussian smearing was used in our calculations on the $P4_2/mnm$ phase, with a smearing width of 0.09 eV such that the entropy term was between 1 and 2 meV/atom (LDA 1.08 meV/atom, PBE 1.14 meV/atom, PBEsol 1.08 meV/atom). Calculations on the $I4_1/a$ phase were also performed with Gaussian smearing. 

\begin{figure}[h]
	\centering
	\includegraphics[width=.9\columnwidth]{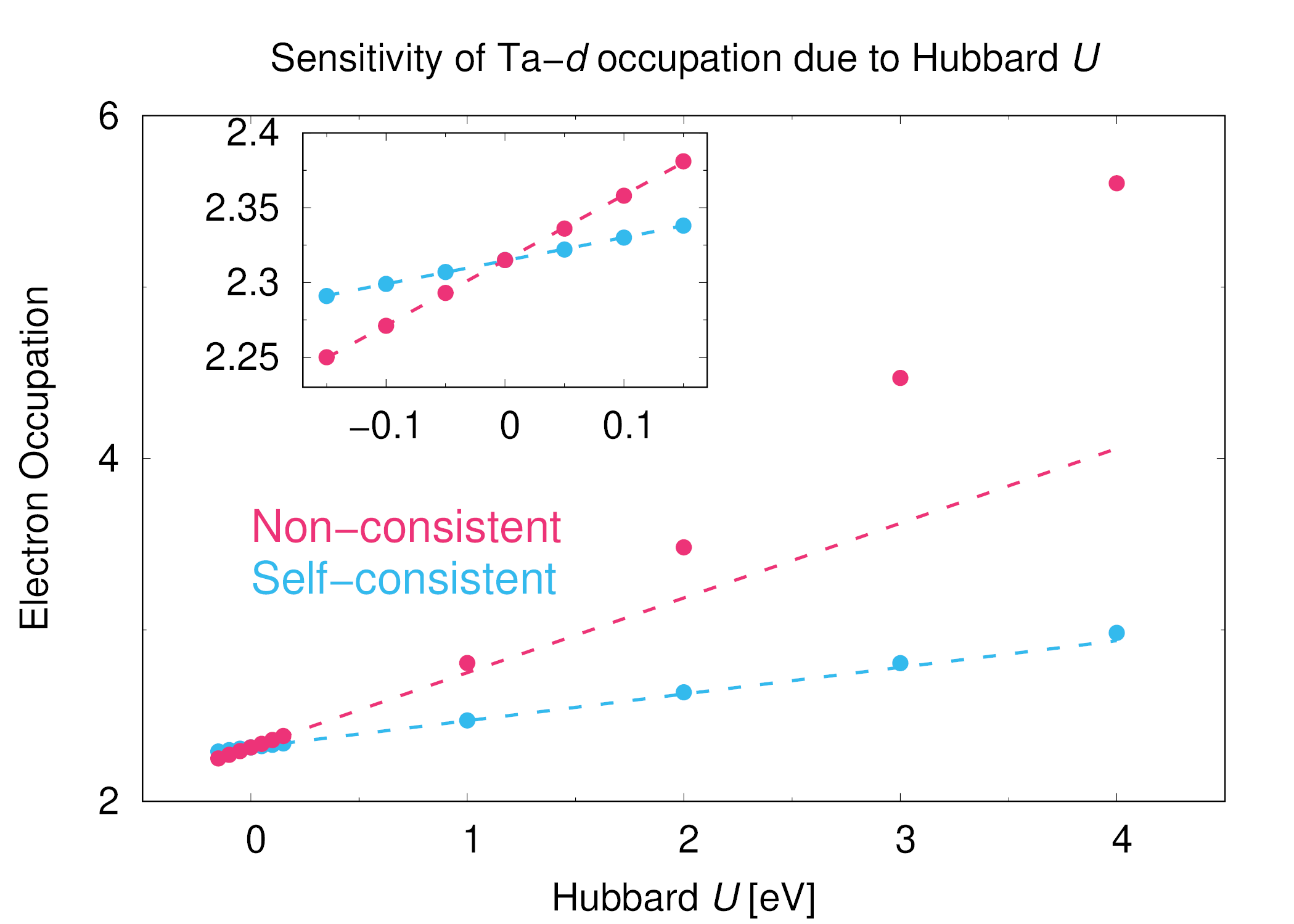}
	\caption{Dependence of the Ta-$d$ occupation as a function of the Hubbard $U$ for the $I4_1/a$ structure using the LDA functional. The red data points correspond to the non-consistent results, whereas the blue data points correspond to the self-consistent results. The inset shows the data points close to zero which were used to fit the linear slopes to calculate the self-consistent value of $U$ (4.127 eV). We note that a second order term becomes comparable to the linear term for the non-consistent results as $U$ approaches 4 eV.   }
	\label{HubbardU}
\end{figure}

We found that the occupation of the tantalum-$d$ orbital was sensitive to the application of an on-site Coulomb interaction (Hubbard $U$)  using LDA+U/GGA+U as implemented by Dudarev. \cite{Dudarev1998LDAUU} The values of the on-site term ($U-J$) for each of the exchange correlation-functionals were found using the linear response approach defined in Ref. \cite{Cococcioni2005}. The $U-J$ values extracted from this method for the low-symmetry structures (LDA 4.13 eV, PBE 3.95 eV, and PBEsol 3.97 eV) were used for calculations involving the high-symmetry $P4_2/mnm$ structure as they did not differ significantly from the self-consistent $U-J$ values calculated in the high-symmetry structure (LDA: 4.22 eV, PBE 4.17 eV, and PBEsol 3.97 eV). Our converged lattice constants using these $U$ values can be found in Table \ref{tab:Lattice_Constants_Pseudo}. It should be noted that while the on-site occupation had a linear response to the on-site value for small $U-J$, a second order term becomes appreciable as the $U-J$ term approaches 4 eV (Fig. \ref{HubbardU}), which is concerning when assuming a linear response. We explore the sensitivity of properties due to the applied Hubbard $U$ in further detail in Sec. \ref{sec:DFT_U}.

We used a \textbf{k}-point grid of 32$\times$32$\times$54 for the $P4_2/mnm$ phase and 19$\times$19$\times$19 for the $I4_1/a$ phase when calculating the density of states. The frequency-dependent dielectric response was calculated using the built-in LOPTIC method in VASP \cite{Gajdos2006}. The current-based results are reported for the $P4_2/mnm$ phase as it is metallic and the direct-based results are reported for the $I4_1/a$ phase, as recommended by Ref. \cite{Sangalli2017}. For all of the density of states, electronic band structure, and frequency-dependent dielectric response calculations the number of electronic bands which were included were 84 for the $P4_2/mnm$ phase and 336 for the $I4_1/a$ phase. Below are our fully relaxed structures ($P4_2/mnm$ and $I4_1/a$) for the LDA functional with a Hubbard $U$ value of 4.127 eV.

\begin{verbatim}
TaO2 - P4_2/mnm
   1.00000000000000
     4.9793858837262732    0.0000000000000000    0.0000000000000000
     0.0000000000000000    4.9793858837262732    0.0000000000000000
     0.0000000000000000    0.0000000000000000    2.8575942530325609
   Ta   O
     2     4
Direct
 -0.0000000000000000 -0.0000000000000000 -0.0000000000000000
  0.5000000000000000  0.5000000000000000  0.5000000000000000
  0.2847778905047560  0.2847778905047560  0.0000000000000000
  0.7152221094952438  0.7152221094952438 -0.0000000000000000
  0.2152221094952439  0.7847778905047562  0.5000000000000000
  0.7847778905047562  0.2152221094952439  0.5000000000000000
\end{verbatim}

\clearpage
\begin{verbatim}
TaO2 - I4_1/a
   1.00000000000000
    -4.8757749036197104    4.8757749050696697    2.9700177379407555
     4.8757749036197104   -4.8757749050696697    2.9700177379407555
     4.8757749050696697    4.8757749036197104   -2.9700177379407555
   Ta   O
     8    16
Direct
  0.6408907722002998  0.8879626861270091  0.7339791835272080
  0.1091092277997002  0.3620373138729907  0.7660208164727920
  0.1539835025998012  0.9069115886730916  0.2660208164727919
  0.5960164974001989  0.3430884113269082  0.2339791835272080
  0.0930884113269082  0.3591092277997002  0.2470719139267094
  0.6569115886730916  0.8908907722002998  0.2529280860732906
  0.6379626861270091  0.4039835025998012  0.7470719139267094
  0.1120373138729908  0.8460164974001989  0.7529280860732906
  0.7547370189934782  0.0015142385332747  0.0249475590730972
  0.2796315625974120  0.5317837384311869  0.0527699818039243
  0.2709862433727373  0.0231384193065140  0.5527699818039243
  0.7734333205398225  0.5202105399796104  0.5249475590730970
  0.9952629810065147  0.2484857614667254  0.4750524409269028
  0.4703684374025880  0.7182162615688131  0.4472300181960757
  0.4790137566272627  0.2268615806934860  0.9472300181960757
  0.9765666794601775  0.7297894600203896  0.9750524409269030
  0.2702105399796103  0.2452629810065148  0.2467772195397895
  0.7731384193065139  0.7203684374025879  0.2521521759337697
  0.7515142385332748  0.2265666794601776  0.7467772195397895
  0.2817837384311871  0.7290137566272626  0.7521521759337695
  0.9768615806934861  0.5296315625974121  0.2478478240662305
  0.4797894600203897  0.0047370189934852  0.2532227804602105
  0.4682162615688129  0.5209862433727374  0.7478478240662305
  0.9984857614667252  0.0234333205398225  0.7532227804602105
\end{verbatim}

\clearpage
\subsection{Ta-$d$ Hubbard $U$ Sensitivity}\label{sec:DFT_U}
\begin{figure}[h]
	\centering
	\includegraphics[width=\columnwidth]{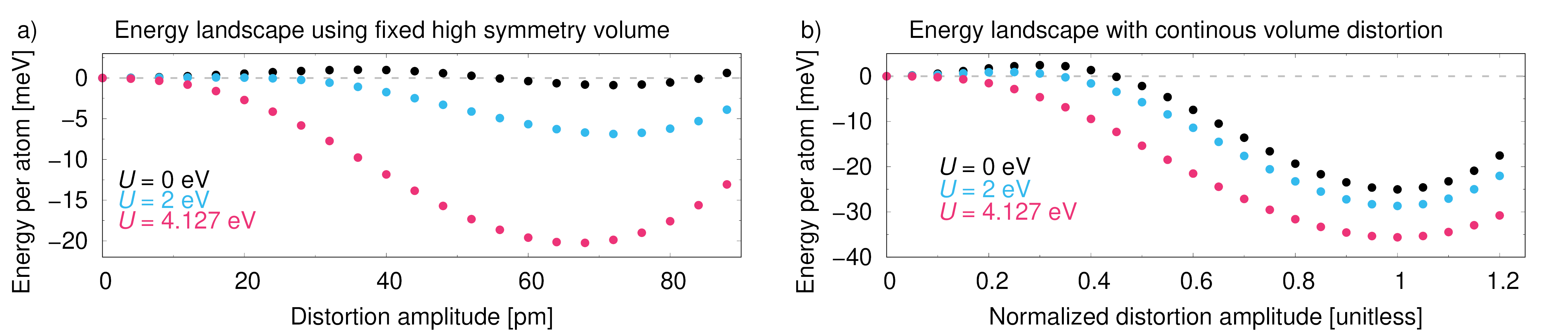}
	\caption{ Energy landscapes from our first-principle calculations from the high-symmetry $P4_2/mnm$ structure to the low-symmetry $I4_1/a$ structure. a) The continuous distortion of atoms from the $P4_2/mnm$ structure to the relaxed $I4_1/a$, with a fixed  $P4_2/mnm$ volume. Each value of $U$ had its own relaxed $P4_2/mnm$ and $I4_1/a$ structures, and the distortions were defined as the difference in atomic positions between the two.   b) Continuous distortion of atoms along with a continuous strain deformation associated with the transition between $P4_2/mnm$ (at 0) and $I4_1/a$ (at 1), for the three different $U$ values. The computational cell for both of these panels was a 48 atom 2$\times$2$\times$2 supercell of the 6 atom conventional cell, and an 8$\times$8$\times$14 $\mathbf{k}$-point grid was used. }
	\label{fig:Barrier_U}
\end{figure}

Our calculations suggest that the properties of the $I4_1/a$ phase of TaO$_2$ are sensitive to the value of the Hubbard $U$ parameter for the Ta-$d$ states. In this section we show how the properties compare with the LDA exchange correlation functional for $U$ values of 0 eV, 2 eV, and 4.127 eV. The structural instability observed in Fig. \ref{fig:Barrier_U} corresponds to displacements of the $R_1^+$ mode of $P4_2/mnm$, which in turn brings $\Gamma_1^+$, $\Gamma_3^+$ and $M_5^+$ displacements into the fully relaxed structure via higher-order coupling terms. There are a number of possible high-symmetry configurations of the 4-dimensional $R_1^+$ distortion. We found that the [a,a,-a,a] configuration (which takes the $P4_2/mnm$ structure to $I4_1/a$) had the lowest total energy compared to the other possible combinations. Most configurations raised the total energy compared to $P4_2/mnm$, the different configurations can be found using the ISOSUBGROUP tool. \cite{ISOTROPYGeneral, Stokes2016} A peculiar feature of the transition from $P4_2/mnm$ to $I4_1/a$ is that the energy landscape has a barrier which depends on the Hubbard $U$ (Fig. \ref{fig:Barrier_U}). There is a finite barrier for $U$ = 0 eV which must be overcome before the system lowers in energy, this could potentially be why this lower-energy phase was missed in prior explorations \cite{birol2013ab}. Increasing $U$ to 2 eV lowers the barrier, but it is still non-zero. When $U$ is set to 4.127 eV the barrier is gone and the transition is purely downhill. The structure also changes volume as the $P4_2/mnm$ structure relaxes to $I4_1/a$. Fig. \ref{fig:Barrier_U}b shows that there still is an energy barrier even as the volume changes for the $U$ = 0 eV and 2 eV cases, and again is absent in the $U$ = 4.127 eV case. Applying the volume deformations without the associated atomic displacements only increased the total energy. The lattice constants of the fully relaxed cells for these different $U$ values can be found in Table \ref{tab:Lattice_Constants_U}. The $a$ and $c$-axes grow as $U$ increases in the $P4_2/mnm$ structure, and only the $a$-axis grows in the $I4_1/a$ structure while the $c$-axis appears to be relatively insensitive to $U$.

\begin{table}[h]
    \caption{Lattice constants after volume relaxation with the LDA exchange-correlation functional, for the different values of $U$. The $I4_1/a$ lattice constants have been converted to reference the 6-atom conventional cell.}
    \label{tab:Lattice_Constants_U}
    \centering
    \begin{tabular}{lccc}
        \toprule
        Hubbard $U$ [eV] & 0  & 2  & 4.127  \\
        \midrule
        $P4_2/mnm$  $a$-axis  [\AA]       & 4.919 & 4.947 & 4.979  \\
        $P4_2/mnm$  $c$-axis  [\AA]       & 2.843 & 2.850 & 2.858  \\
        $I4_1/a$    $a$-axis  [\AA]       & 4.774 & 4.816 & 4.876  \\
        $I4_1/a$    $c$-axis  [\AA]       & 2.977 & 2.978 & 2.970  \\
        % $\epsilon_{xx}^0$           & 7.676 & 7.178  & 6.216  \\
        % $\epsilon_{zz}^0$           & 10.188 & 9.865 & 9.414  \\
        \bottomrule
    \end{tabular}
\end{table}

\begin{figure}[h]
	\centering
	\includegraphics[width=\columnwidth]{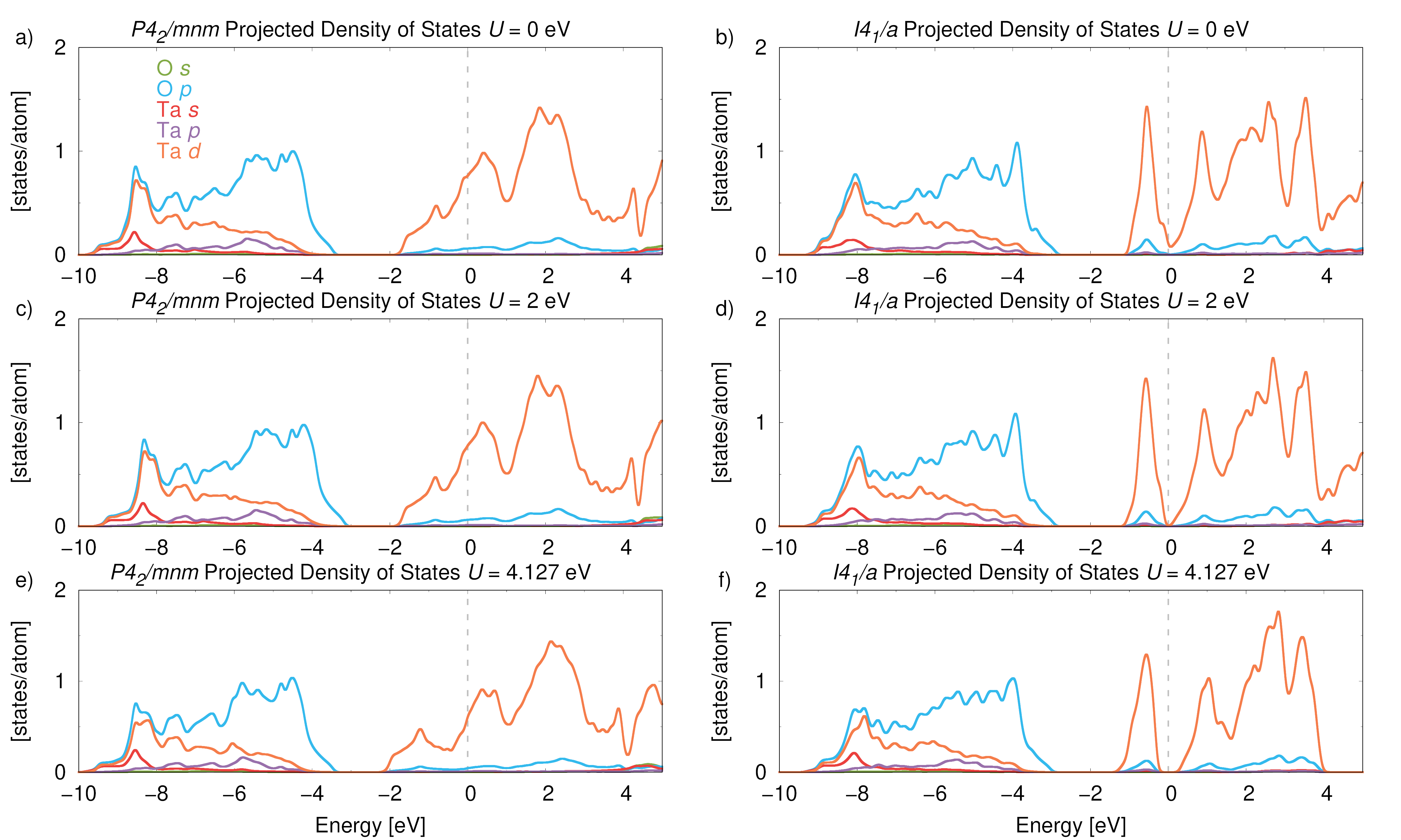}
	\caption{First-principle calculation of the density of states in the $P4_2/mnm$ (panels a, c, e) and $I4_1/a$ (panels b, d, f) phases with the LDA functional for various Hubbard $U$ values. The $P4_2/mnm$ phase appears relatively insensitive to the Hubbard $U$, and most of the  electronic states at the Fermi energy are from Ta-$d$. The $I4_1/a$ phase goes from having no gap at $U$ = 0 eV, to having a band gap which grows as $U$ increases.  }
	\label{fig:DOS_U}
\end{figure}

\begin{figure}[h]
	\centering
	\includegraphics[width=\columnwidth]{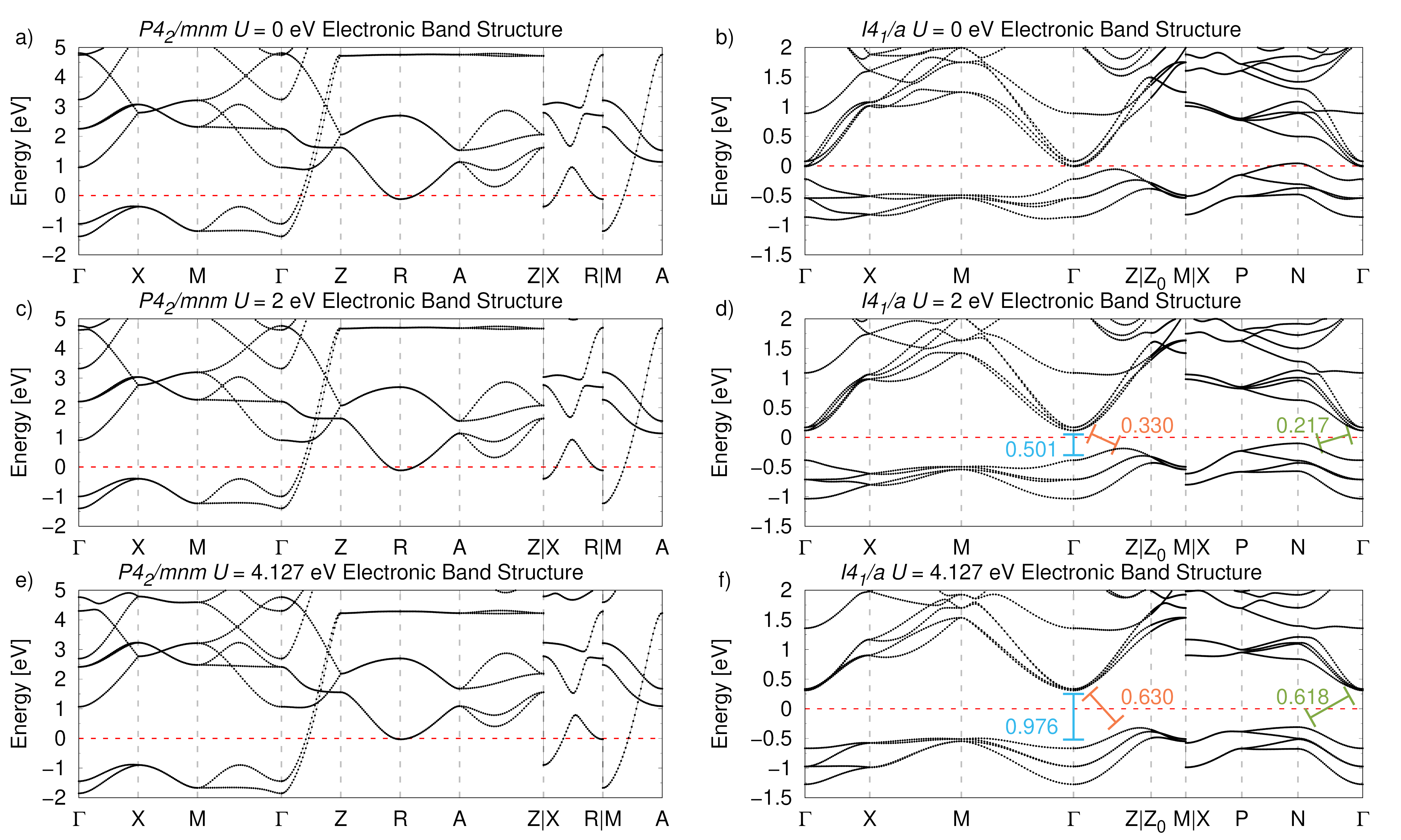}
	\caption{First-principles electronic band structures for the  $P4_2/mnm$ (panels a, c, and e) and $I4_1/a$ phases (panels b, d, and f) for different values of the Hubbard $U$. The $P4_2/mnm$ phase seems relatively insensitive to the Hubbard $U$, and has states at the Fermi energy indicating that the high-symmetry phase is metallic. The $I4_1/a$ phase starts with states at the fermi energy for $U$ = 0 eV, but a gap forms as the Hubbard $U$ increases.}
	\label{fig:Bandstructure_U}
\end{figure}

In addition to the existence of an energy barrier between the $P4_2/mnm$ and $I4_1/a$ phases, the electronic properties of the $I4_1/a$ phase of TaO$_2$ are also sensitive to the value of $U$. The density of states in Fig. \ref{fig:DOS_U} show that while the $P4_2/mnm$ phase is relatively insensitive to the Hubbard $U$ value the $I4_1/a$ phase becomes an insulator as the Hubbard $U$ value is increased from zero, with the gap increasing for larger values of $U$. Applying the Hubbard $U$ on the Ta-$d$ states is pushing them away from the Fermi energy in the $I4_1/a$ phase. The electronic band structure calculations in Fig. \ref{fig:Bandstructure_U} also show electronic states at the Fermi energy separate as the value of $U$ increases in the $I4_1/a$ phase.

\begin{figure}[h]
	\centering
	\includegraphics[width=\columnwidth]{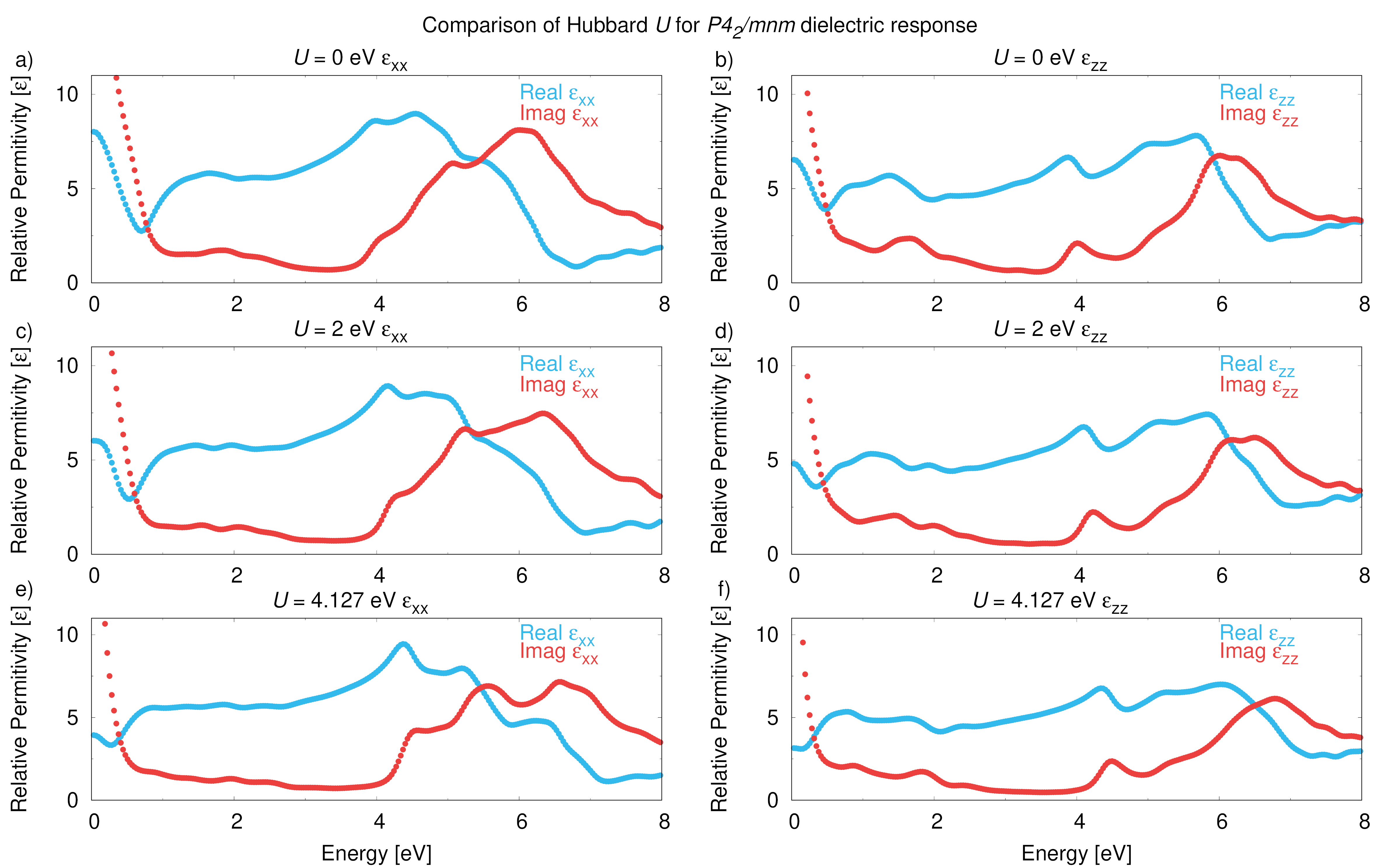}
	\caption{First-principle results of the frequency-dependent relative permittivity in the $P4_2/mnm$ phase for the LDA functional. The left column of panels (a, c, and e) show the real and imaginary components along the $x$-direction (symmetry equivalent to $y$), while the right column of panels (b, d, and f) shows the real and imaginary components along the $z$-axis.  }
	\label{fig:LOPTIC_U_high}
\end{figure}

\begin{figure}[h]
	\centering
	\includegraphics[width=\columnwidth]{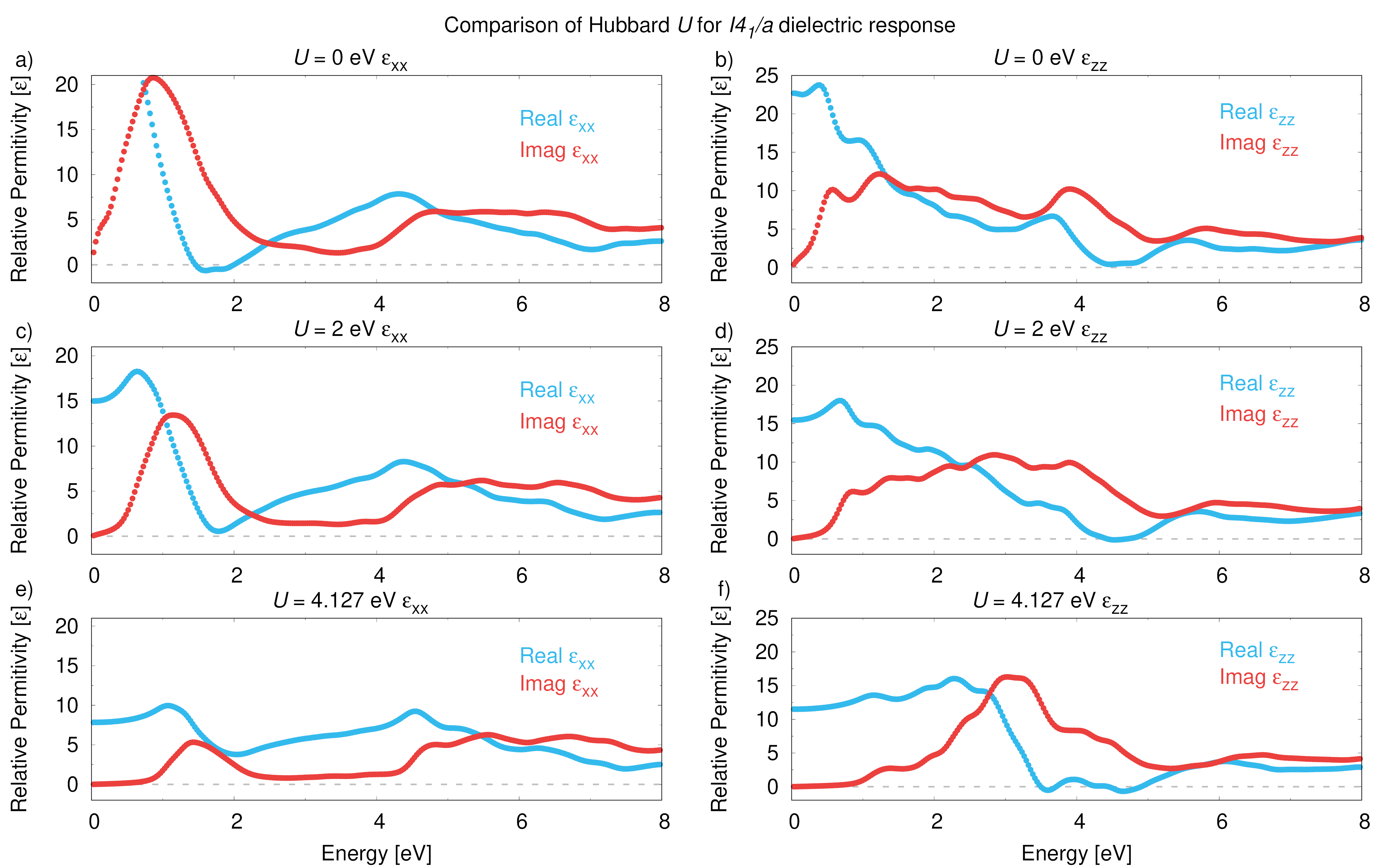}
	\caption{First-principle results of the frequency-dependent relative permittivity in the $I4_1/a$ phase for the LDA functional. The left column of panels (a, c, and e) show the real and imaginary components along the $x$-direction (symmetry equivalent to $y$), while the right column of panels (b, d, and f) shows the real and imaginary components along the $z$-axis.  }
	\label{fig:LOPTIC_U_low}
\end{figure}

Figures \ref{fig:LOPTIC_U_high} and \ref{fig:LOPTIC_U_low} show how $U$ alters the frequency-dependent dielectric response in the $P4_2/mnm$ and $I4_1/a$ phases respectively. The low-frequency imaginary component in the $P4_2/mnm$ phase goes to infinity because it is a metal, and the position of the resonant absorption peaks do not change much as a function of $U$. The $I4_1/a$ phase on the other-hand exhibits significant changes to the absorption spectrum with $U$, particularly along the $z$-axis. 
As we have demonstrated repeatedly throughout this section, the $I4_1/a$ phase of TaO$_2$ is quite sensitive to the value of the Hubbard $U$ and warrants further exploration.

\clearpage
\subsection{Comparing Functionals}\label{sec:DFT_functionals}
The general features from our first-principles calculations of TaO$_2$ seem consistent for the variety of exchange-correlation functionals that we checked. The lattice constants in Table \ref{tab:Lattice_Constants_Pseudo} are the shortest in LDA, the longest in PBE, and in-between for PBEsol, which is the expected trend since LDA has a tendency to `over-bond.' All three functionals feature an elongation of the $c$-axis and a contraction of the $a/b$-axes as the system transitions from the $P4_2/mnm$ phase to the $I4_1/a$ phase.

\begin{figure}[h]
	\centering
	\includegraphics[width=\columnwidth]{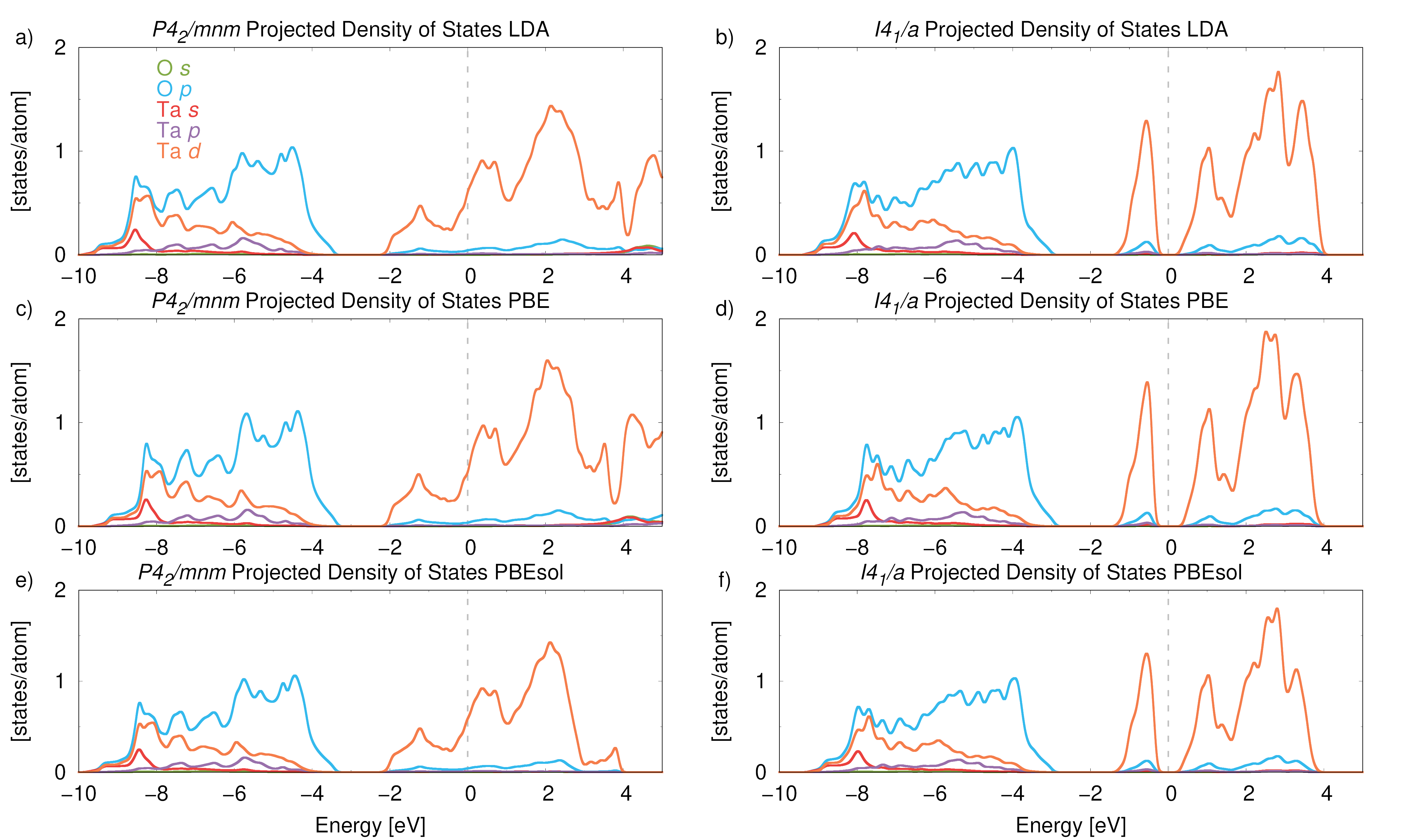}
	\caption{First-principle calculation of the density of states of the $P4_2/mnm$ (panels: a, c, and e) and $I4_1/a$ (panels: b, d, f) phases for each of the LDA (a \& b), PBE (c \& d), and PBEsol (e \& f) functionals. The respective Hubbard $U$ for each function is: 4.127 eV for LDA, 3.955 eV for PBE, and  3.966 eV for PBEsol.   }
	\label{fig:DOS_pseudo}
\end{figure}

\begin{figure}[h]
	\centering
	\includegraphics[width=\columnwidth]{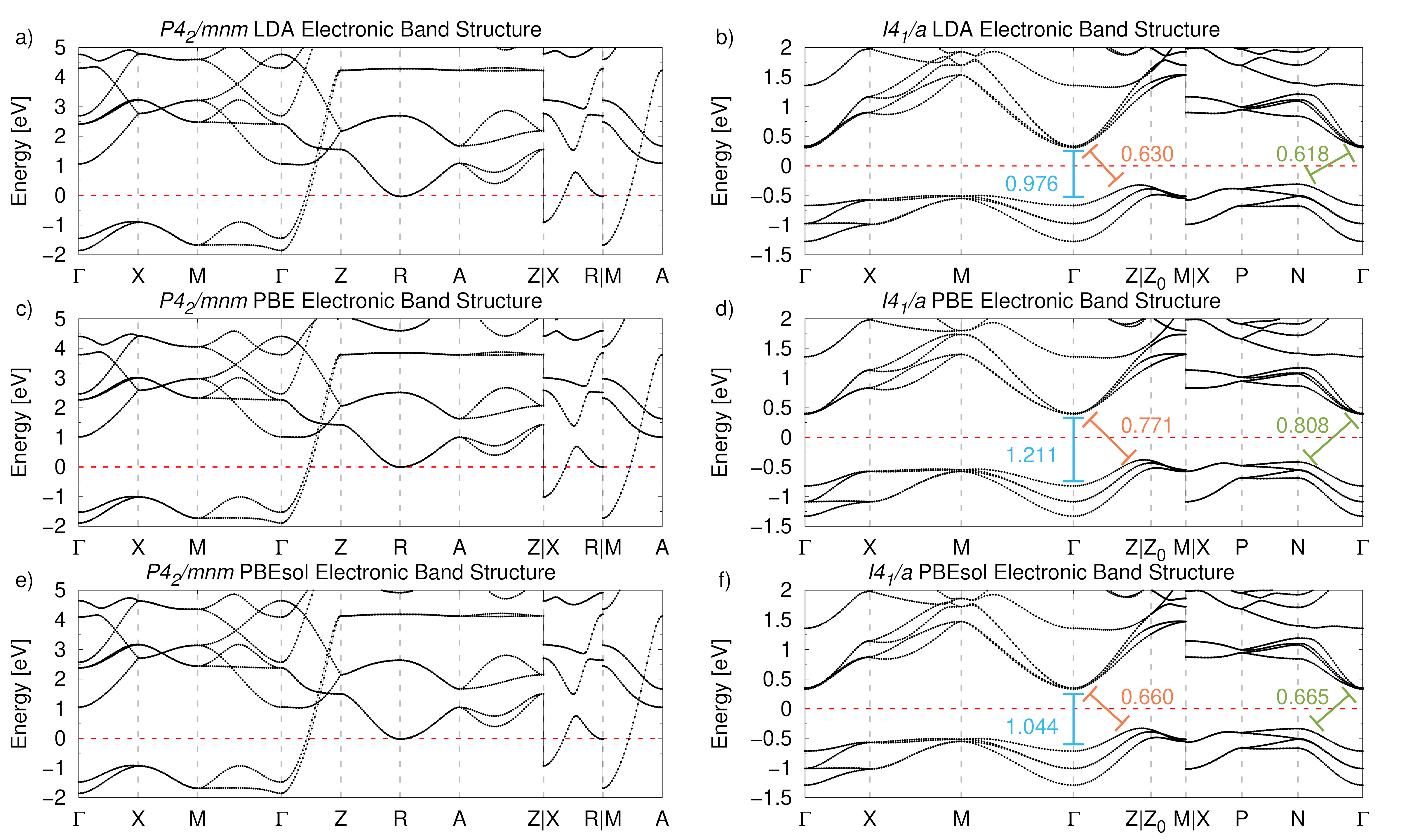}
	\caption{First-principles electronic band structures for each functional for the   $P4_2/mnm$ (panels a, c, and e) and $I4_1/a$ phases (panels b, d, and f). In the $P4_2/mnm$ phase, each functional has electronic states at the Fermi energy, indicating that the high symmetry phase is metallic. In the $I4_1/a$ phase, each functional has a direct band gap of $\sim$1 eV and an indirect gap of  $\sim$0.7 eV. The respective Hubbard $U$ for each function is: 4.127 eV for LDA, 3.955 eV for PBE, and  3.966 eV for PBEsol.}
	\label{fig:Electronic_Bandstructure}
\end{figure}

\begin{figure}[h]
	\centering
	\includegraphics[width=\columnwidth]{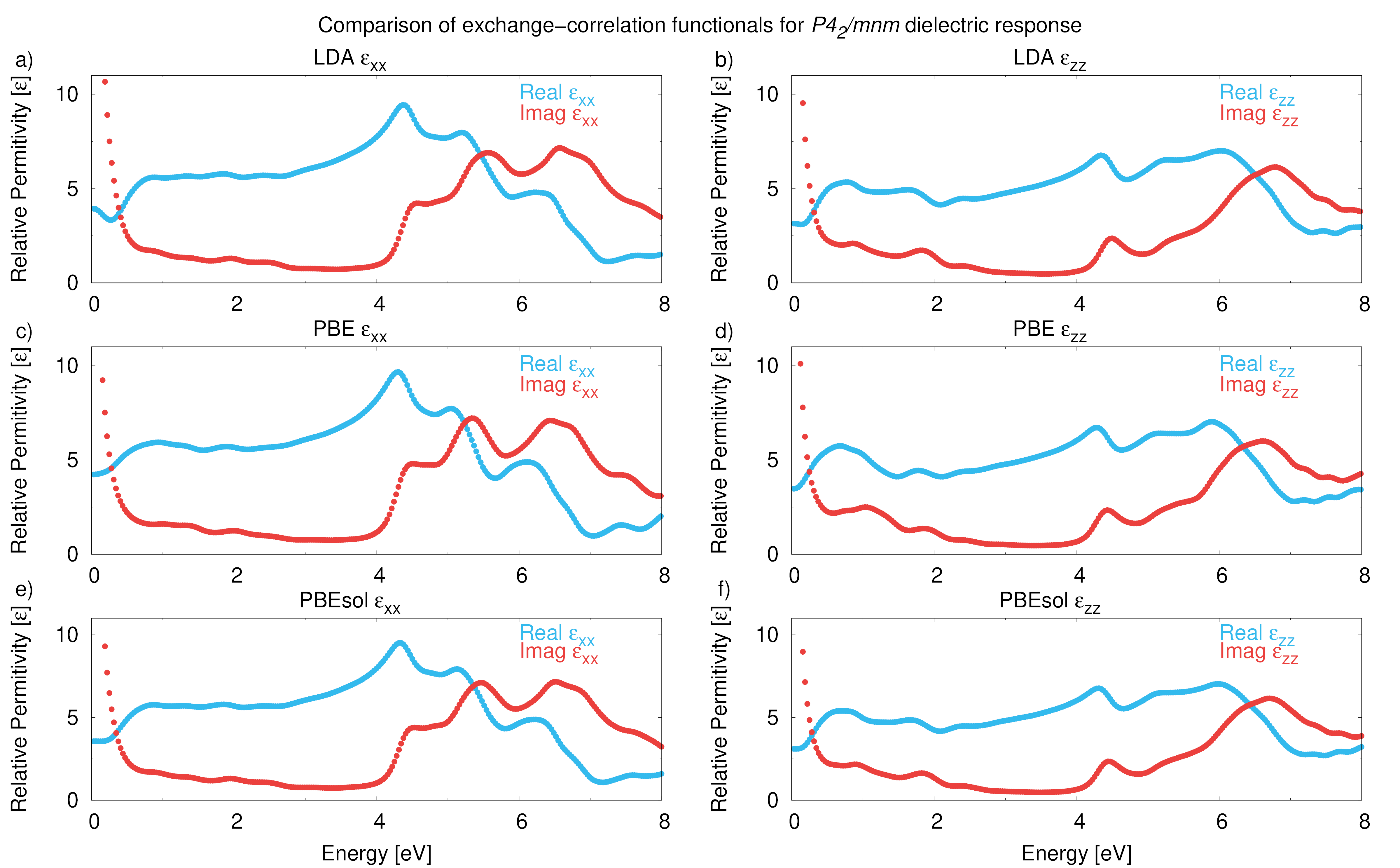}
	\caption{First-principle results of the frequency dependent relative permittivity in the $P4_2/mnm$ phase for the LDA (a, b), PBE (c, d), and PBEsol (e, f) functionals. The left column of panels (a, c, and e) show the real and imaginary components along the $x$-direction (symmetry equivalent to $y$), while the right column of panels (b, d, and f) show the real and imaginary components along the $z$-axis. The imaginary component goes to infinity at low frequency because the phase is a metal. The respective Hubbard $U$ for each function is: 4.127 eV for LDA, 3.955 eV for PBE, and  3.966 eV for PBEsol. }
	\label{fig:LOPTIC_psuedo_high}
\end{figure}

\begin{figure}[h]
	\centering
	\includegraphics[width=\columnwidth]{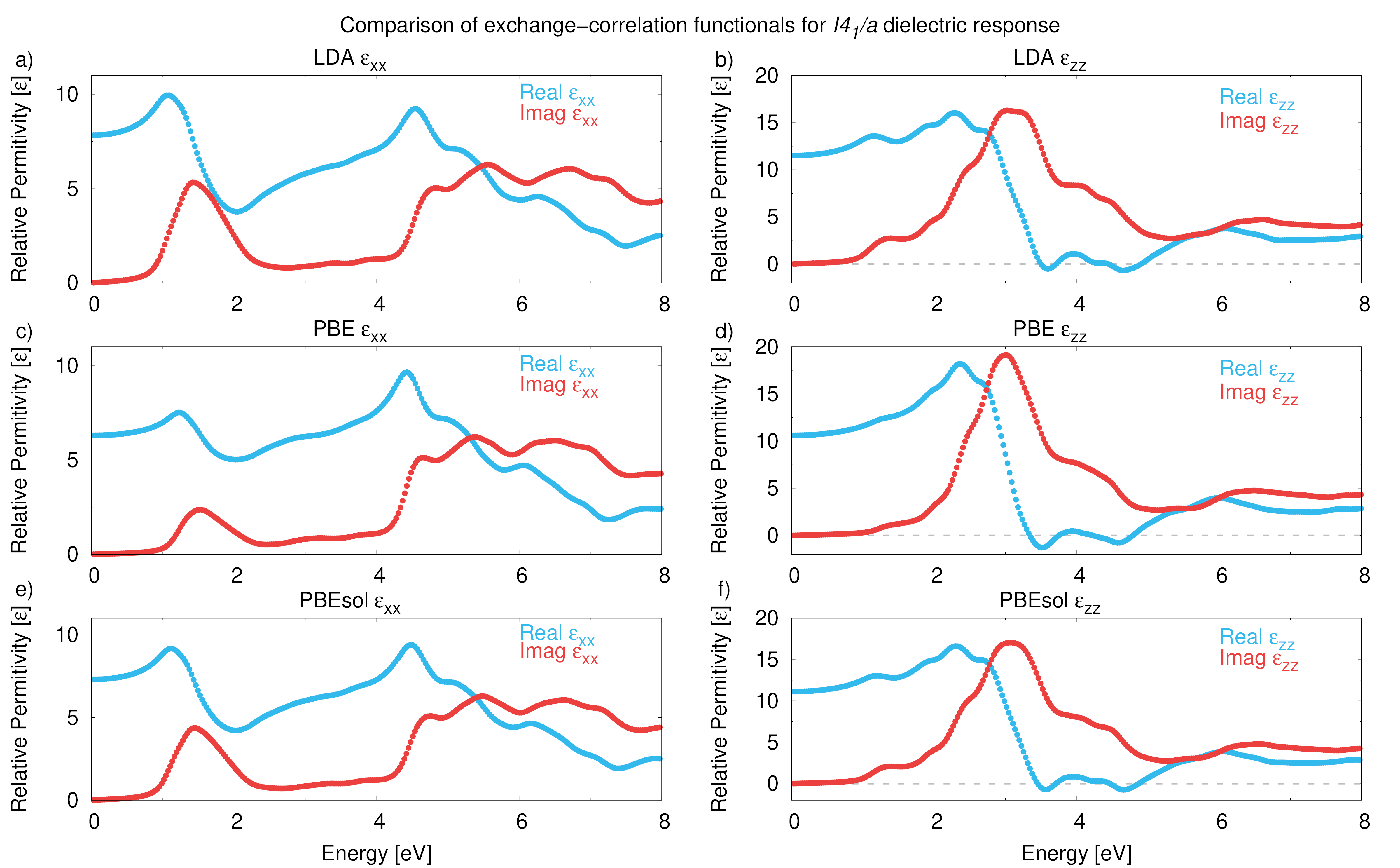}
	\caption{First-principle results of the frequency dependent relative permittivity in the $I4_1/a$ phase for the LDA (a, b), PBE (c, d), and PBEsol (e, f) functionals. The left column of panels (a, c, and e) show the real and imaginary components along the $x$-direction (symmetry equivalent to $y$), while the right column of panels (b, d, and f) show the real and imaginary components along the $z$-axis. The respective Hubbard $U$ for each function is: 4.127 eV for LDA, 3.955 eV for PBE, and  3.966 eV for PBEsol.  }
	\label{fig:LOPTIC_psuedo_low}
\end{figure}

The projected density of states of the $P4_2/mnm$ and $I4_1/a$ phases are visually consistent between the different functionals (Fig. \ref{fig:DOS_pseudo}). The $P4_2/mnm$ phase has Ta-$d$ states at the Fermi energy, indicating the system is metallic. This is in contrast to the $I4_1/a$ phase, which has a band gap at the Fermi energy. The band structure calculations (Fig. \ref{fig:Electronic_Bandstructure}) suggest the gap is smallest for an indirect transition, while the smallest direct gap is roughly 1 eV. The frequency-dependent dielectric response (Figs. \ref{fig:LOPTIC_psuedo_high} and \ref{fig:LOPTIC_psuedo_low}) have fairly consistent spectra across the different functionals. 

\begin{table}[h]
    \caption{First-principles relaxed lattice constants, for the various exchange-correlation functionals. The $I4_1/a$ lattice constants have been converted to match the 6-atom conventional cell.}
    \label{tab:Lattice_Constants_Pseudo}
    \centering
    \begin{tabular}{lccc}
        \toprule
        Functional & LDA & PBE & PBEsol \\
        \midrule
        Ta-$d$ Hubbard $U$ [eV]                         & 4.127 & 3.955 & 3.966   \\
        $P4_2/mnm$  $a$-axis  [\AA]       & 4.979 & 5.054 & 5.012  \\
        $P4_2/mnm$  $c$-axis  [\AA]       & 2.858 & 2.902 & 2.872  \\
        $I4_1/a$    $a$-axis  [\AA]       & 4.876 & 4.971 & 4.915  \\
        $I4_1/a$    $c$-axis  [\AA]       & 2.970 & 3.000 & 2.981  \\
        % $\epsilon_{xx}^0$           & 7.676 & 7.178  & 6.216  \\
        % $\epsilon_{zz}^0$           & 10.188 & 9.865 & 9.414  \\
        \bottomrule
    \end{tabular}
\end{table}

% \begin{figure}[h]
% 	\centering
% 	\includegraphics[width=.7\columnwidth]{DFT_images/dielectric/TaO2_PBEsol_lowsym_DIEL.pdf}
% 	\caption{First-principle results of the frequency dependent relative permittivity in the $I4_1/a$ phase for the PBEsol functional. The top panel shows the real and imaginary components along the $x$-direction (symmetry equivalent to $y$), while the bottom panel shows the real and imaginary components along the $z$-axis.  }
% 	\label{PBEsol_lowsym_DIEL}
% \end{figure}

%\end{document}

\end{document}